\documentclass[aps,prd,superscriptaddress,notitlepage,11pt,final,longbibliography]{revtex4-1}

\usepackage[]{hyperref}

\usepackage{amsmath,amssymb,mathrsfs}

\usepackage{subfigure}
\usepackage{epsf}
\usepackage{epsfig}
\usepackage[usenames,dvipsnames]{xcolor}
\usepackage{bbm}
\usepackage{color}
\usepackage{comment}
\usepackage{cleveref}
\usepackage{caption}

\usepackage{natbib}

\usepackage{physics}

\makeatletter
\newcommand{\figcaption}[1]{\def\@captype{figure}\caption{#1}}
\newcommand{\tblcaption}[1]{\def\@captype{table}\caption{#1}}
\makeatother

\begin{document}

\title{Nodal compact $Q$-ball/$Q$-shell in the $\mathbb{C}P^N$ nonlinear sigma model}

\author{P. Klimas~}
\email{pawel.klimas@ufsc.br}
\affiliation{Departamento de F\'isica, Universidade Federal de Santa Catarina, Campus Trindade, 88040-900, Florian\'opolis-SC, Brazil}

\author{N. Sawado~}
\email{sawadoph@rs.tus.ac.jp}
\affiliation{Department of Physics, Tokyo University of Science, Noda, Chiba 278-8510, Japan}

\author{S. Yanai}
\email{phyana0513@gmail.com}
\affiliation{Department of Physics, Tokyo University of Science, Noda, Chiba 278-8510, Japan}

\begin{abstract}
Nodal, excited compactons in the $\mathbb{C}P^N$ models with V-shaped potentials are analyzed. 
It is shown that the solutions exist as compact $Q$-balls and $Q$-shells. The solutions have a discontinuity in the second derivative associated with the character of the potential, however, their energy and charge densities are both continuous. The excited $Q$-balls and $Q$-shells are analyzed as electrically neutral and electrically charged objects.

\end{abstract}

\maketitle

\section{Introduction}
\label{sec:intro}

Compactons, i.e. field configurations that exist on finite size supports ("compact supports"), 
possess a distinct character among other solutions of standard field theory models.  
Namely, the field takes its vacuum values outside this support and the energy as well as the 
charge are always concentrated on the compact support.  
Initial study of such exotic configurations concerning the complex signum-Gordon model 
was presented in ~\cite{Arodz:2008jk,Arodz:2008nm}. Some preceding results involving real 
scalar fields were presented in \cite{Arodz:2002yt,Arodz:2003mx,Arodz:2005gz}. 
Such models possess standard kinetic terms and special V-shaped potential that gives rise to compact solutions. 
The ``V-shaped character'' of the potential means that the potential has at least one local minimum in the form of a spike. 
The first-order side derivatives at the minimum do not vanish at this point. Moreover, they are not equal to each other (frequently, e.g. for the signum-Gordon model, these derivatives differ from each other just by the overall sign).
	
A complex scalar field theory with some self-interactions has stationary
soliton solutions called $Q$-balls~\cite{Friedberg:1976me,Coleman:1985ki}. 
$Q$-balls have attracted much attention in the studies of 
evolution of the early Universe~\cite{Friedberg:1986tq,Lee:1986ts}.
There is also a certain possibility that they survive the early phase of the Universe and constitute a major ingredient of dark matter
~\cite{Kusenko:1997zq,Kusenko:1997si,Kusenko:1997vp}.
The compact $Q$-balls in the $\mathbb{C}P^N$ model were presented in~\cite{Klimas:2017eft}. Moreover, this model also supports the  $Q$-shell compactons, even in the absence of an electromagnetic  field (the case $N>3$).
The gravitating compact $Q$-balls, {\it i.e.}, the compact boson stars can harbor 
a Schwarzschild and a Reissner-Nordstr\"om type black hole~\cite{Kleihaus:2010ep,Kumar:2014kna,Kumar:2015sia,Kumar:2016sxx}. 

In the last few years, we made some efforts in the study of compact $Q$-balls in the nonlinear sigma model on a target space $\mathbb{C}P^N$, 
and also the boson stars corresponding to the model~\cite{Klimas:2018ywv, Yanai:2019wpv, Sawado:2020ncc, Sawado:2021rsc, Klimas:2021eue}.	
In this paper, we explore excited compacton states, i.e. the nodal solutions of this model that differ from the standard solutions by the form of the radial function. 
The excited states of $Q$-balls or the boson stars are very important for theoretical study and also 
for astrophysical observations
~\cite{Brihaye:2008cg,Bernal:2009zy,Collodel:2017biu,Alcubierre:2018ahf,Alcubierre:2019qnh,Loginov:2020xoj,Loginov:2020lwg, Almumin:2021gax}. 
The multi-state boson stars, which are superposed the ground and excited states of the boson star solutions, are considered 
for obtaining realistic rotation curves of spiral galaxies~\cite{Bernal:2009zy}. 
We explore the multi-nodal, excited $Q$-ball solutions on the compact support. The point is that we impose the condition allowing to get 
second or more node points and then, formally, we are able to get the multi-nodal compacton with arbitrary node numbers.  It requires that the radial function can change the sign at the node points.
Our results have somewhat similarities with \cite{Loginov:2020xoj,Loginov:2020lwg} which both are radially excited.
A major difference is that our solutions are compactons that are nontrivial only on a certain compact support. 

The discontinuous nature of the potential derivative at the minimum is responsible for the appearance of  the signum function in the radial field equation.  Consequently, the solution exhibits  discontinuity of the second derivative at the node points. The first derivative of the radial function is continuous, however, not smooth at the node points. However, it turns out that  the energy density of such field configurations  is free from any difficulties. 
As a result, we obtain a large number of multi-nodal solutions for both the non-gauged and the gauged models. 

The paper is organized as follows. In Section II we shall describe the model. The ansatz for the parametrization of the $\mathbb{C}P^N$ field is given in this section. Section III presents the solutions of the model. We give further analysis and discussion in Section IV.  
Conclusions and remarks are presented in the last Section.

\section{The model}
\label{model}

The action of our model has the following form
\begin{align}
&S = \int \sqrt{-g} d^4x \Bigl[-\frac{1}{4}F_{\mu \nu}F^{\mu \nu} + 4M^2 g^{\mu \nu} \frac{D_{\mu} u^{\dagger}\cdot D_{\nu}u}{1+u^{\dagger} \cdot u}-4M^2g^{\mu \nu}\frac{\left(D_{\mu} u^{\dagger} \cdot u\right) \left(u^{\dagger}\cdot D_{\nu}u\right)}{\left(1+u^{\dagger}\cdot u\right)^2} -\mu^2 V\Bigr]. \label{action}
\end{align}
$F_{\mu\nu}$ is the standard electromagnetic field tensor and the complex fields $u_i$ also are minimally coupled to the Abelian gauge fields $A_\mu$ through $D_\mu=\partial_\mu-ieA_\mu$. 
We employ the `V-shaped' potential which is of the form 
\[
V = \sqrt{\frac{u^{\dagger}\cdot u}{1+u^{\dagger}\cdot u}}.
\]
It is convenient to introduce the dimensionless coordinates 
\begin{align}
x_\mu \to \frac{\mu}{M}x_\mu
\end{align}
and also $A_\mu \to \mu/M A_\mu$. 
We also restrict $N$ to be odd, i.e., $N:=2n+1$.  The solutions with vanishing magnetic field can be obtained within the ansatz 
\begin{align}
&u_m(t,r,\theta,\varphi)=\sqrt{\frac{4\pi}{2n+1}}f(r)Y_{nm}(\theta,\varphi)e^{i\omega t}\,, 
\label{ansatzcpn}\\
&A_\mu(t,r,\theta,\varphi)dx^\mu=A_t(r)dt
\label{ansatzgauge}
\end{align}
which allows for reduction of the partial differential equations to the system of ordinary radial  differential equations.
$Y_{nm}, -n\leq m \leq n$ are standard spherical harmonics and $f(r)$ is the matter profile function. 
Each $2n+1$ field $u=(u_m)=(u_{-n},u_{-n+1},\cdots,u_{n-1},u_n)$ is associated with one of $2n+1$ spherical harmonics for given $n$. 
The relation 
$\sum_{m=-n}^n Y_{nm}^*(\theta,\varphi)Y_{nm}(\theta,\varphi)=\dfrac{2n+1}{4\pi}$ is very useful for obtaining an explicit form of many inner products. 
For convenience, we introduce a new gauge field containing the gauge field $A_{t}$ and the constant $\omega$
\begin{align}
b(r):=\omega-eA_t(r)\,.
\label{ansatzgaugew}
\end{align}
Applying the ansatz \eqref{ansatzcpn} and \eqref{ansatzgauge} we get the dimensionless reduced Lagrangian of the $\mathbb{C}P^N$ model in the form 
\begin{align}
&\tilde{\mathcal{L}}_{\mathbb{C}P^N}=\frac{\kappa b'^2}{2e^2}+\frac{4b^2f^2}{(1+f^2)^2}-\frac{4f'^2}{(1+f^2)^2}-\frac{4n(n+1)f^2}{r^2(1+f^2)}-V
\label{effectivelag}
\end{align}  
where for convenience we have introduced a dimensionless constant $\kappa:=\mu^2/M^4$. The  potential simplifies to the following one
\begin{equation}
V=\frac{|f|}{\sqrt{1+f^2}}.\label{potreduces}
\end{equation}
The potential \eqref{potreduces} in the limit of small amplitude fields behaves as the potential of the signum-Gordon model i.e. $V\sim |f|$. Note that, in contradiction to our preceding publications, in this paper we do not restrict the sign of $f(r)$ to non-negative values -- it can be any real number.
The reduced matter field equation and Maxwell's equations take the form of two coupled ordinary differential equations
\begin{align}
&f'' + \frac{2}{r}f' - \frac{n(n+1)f}{r^2} + \frac{1-f^2}{1+f^2}b^2f-\frac{2ff'^2}{1+f^2} -\frac{1}{8}{\rm Sign}(f) \sqrt{1+f^2} =0\,,\label{eq:f} \\
&\kappa b''+\frac{2}{r}\kappa b' - \frac{8e^2bf^2}{(1+f^2)^2} =0\,.\label{eq:b}
\end{align}
The presence of ${\rm Sign}(f)$ in the field equation is a direct consequence of the V-shaped character of the potential. 
In further part of the paper we solve these two coupled equations with fixed $\kappa$ (for simplicity we set $\kappa =1$).

The dimensionless (reduced) Hamiltonian of the model reds
\begin{align}
\mathcal{H}_{\mathbb{C}P^N}=\frac{4 b^2 f^2}{(1+f^2)^2}+	\frac{4f'^2}{(1+f^2)^2}+	 \frac{4n(n+1)f^2}{r^2(1+f^2)}+ \frac{\kappa b'^2}{2e^2}+V\,
\end{align}
and it has interpretation of radial profile function of the energy density. Integrating the energy density over whole space one gets the energy of the compacon
\begin{align}
&E=4\pi\int r^2dr
\biggl[
\frac{4b^2f^2}{(1+f^2)^2}+\frac{4f'^2}{(1+f^2)^2}+\frac{4n(n+1)f^2}{r^2(1+f^2)}+	 \frac{\kappa b'^2}{2e^2}+V\biggr]\,.
\label{energy}
\end{align}

Now we shall look at the question of Noether charges. The action with the covariant derivative \eqref{action} is invariant under the following local $U(1)^N$ symmetry
\begin{align}
&A_\mu(x) \to A_\mu(x)+e^{-1}\partial_\mu\Lambda(x) \nonumber \\
&u_i\to \exp[iq_i \Lambda(x)]u_i,~~~~i=1,\cdots,N\,.
\label{gtransformation}
\end{align}	
where $q_i$ are some real numbers. 
The following Noether current is associated with the invariance of the action (\ref{action})  
under transformations (\ref{gtransformation})
\begin{align}
J^{(i)}_\mu=-i\frac{4M^2}{\left(1+u^{\dagger}\cdot u\right)^2}\sum_{j=1}^N\Big[u_i^* \Delta_{ij}^2 D_\mu u_j-D_\mu u_j^* \Delta_{ji}^2u_i\Big]\,.
\end{align}
Making use of the ansatz (\ref{ansatzcpn}),(\ref{ansatzgauge}) we find the following form of the Noether currents
\begin{align}
J_t^{(m)}(r,\theta)= \frac{(n-m)!}{(n+m)!}\frac{8bf^2}{(1+f^2)^2}\bigl(P^m_n(\cos\theta)\bigr)^2\,,
\label{current0} \\
J_\varphi^{(m)}(r,\theta)=\frac{(n-m)!}{(n+m)!}\frac{8mf^2}{(1+f^2)^2}\bigl(P^m_n(\cos\theta)\bigr)^2
\label{currentp}
\end{align}
and $J_r^{(m)}=J_\theta^{(m)}=0$ for $m=-n,-n+1,\cdots,n-1,n$. Note that both non vanishing currents do not depend on variables $t$ and $\varphi$.
Hence, the conservation of currents is explicit after writing the continuity equation in the form
\begin{align}
\frac{1}{\sqrt{-g}}\partial_\mu \Big(\sqrt{-g}g^{\mu\nu}J_\nu^{(m)}\Big)=\partial_t J_t^{(m)}
+\frac{1}{r^2\sin^2\theta}\partial_\varphi J_\varphi^{(m)}=0.
\end{align}	
The  corresponding conserved Noether charge reads
\begin{align}
Q^{(m)}&:=\frac{1}{2}\int_{\mathbb{R}^3} d^3x \sqrt{-g} J^{(m)}_t(x) \nonumber \\
&=\frac{16\pi}{2n+1}\int r^2dr\frac{bf^2}{(1+f^2)^2}\,.
\end{align}
Owing to our ansatz, the charge does not depend on index $m$, which means that the symmetry of the solutions 
is reduced to the $U(1)$ symmetry. However we shall keep the index $m$ for completeness. 

Compactons in our model are extended objects with the spherical shape surface (the border of the compacton), such that the matter profile function satisfies the boundary conditions
\begin{align}
f(R)=0,~~f'(R)=0,~~~~r=R \textrm{~:~the~compacton~radius}
\label{compact}
\end{align}
Looking at typical numerical solutions $f(r)$ one can see that in many cases it oscillates. Usually, {\it the first zero} of the function $f$ is chosen as the  radius of the compacton border. In such a case the first zero corresponds with the local minimum of $f(r)$.
Mathematically, it is not obvious that the choice of the first zero as the compacton radius is a unique possibility.

Indeed, for an oscillating function the choice of some other minima as the compacton border leads to certain solutions with the profile function having some extra zeros. The number of such zeros can label different types of compact solutions. 
The purpose of this paper is to find the plausible answer for two questions. First, whether the solution exists for the choice of {\it the second zero} (minimum of maximum) of $f(r)$? Second, if a solution exists, what is his form and properties? We shall answer these questions in the following part of the paper.

\begin{figure*}[t]
  \begin{center}
\subfigure[]{\includegraphics[width=0.45\textwidth, angle =0]{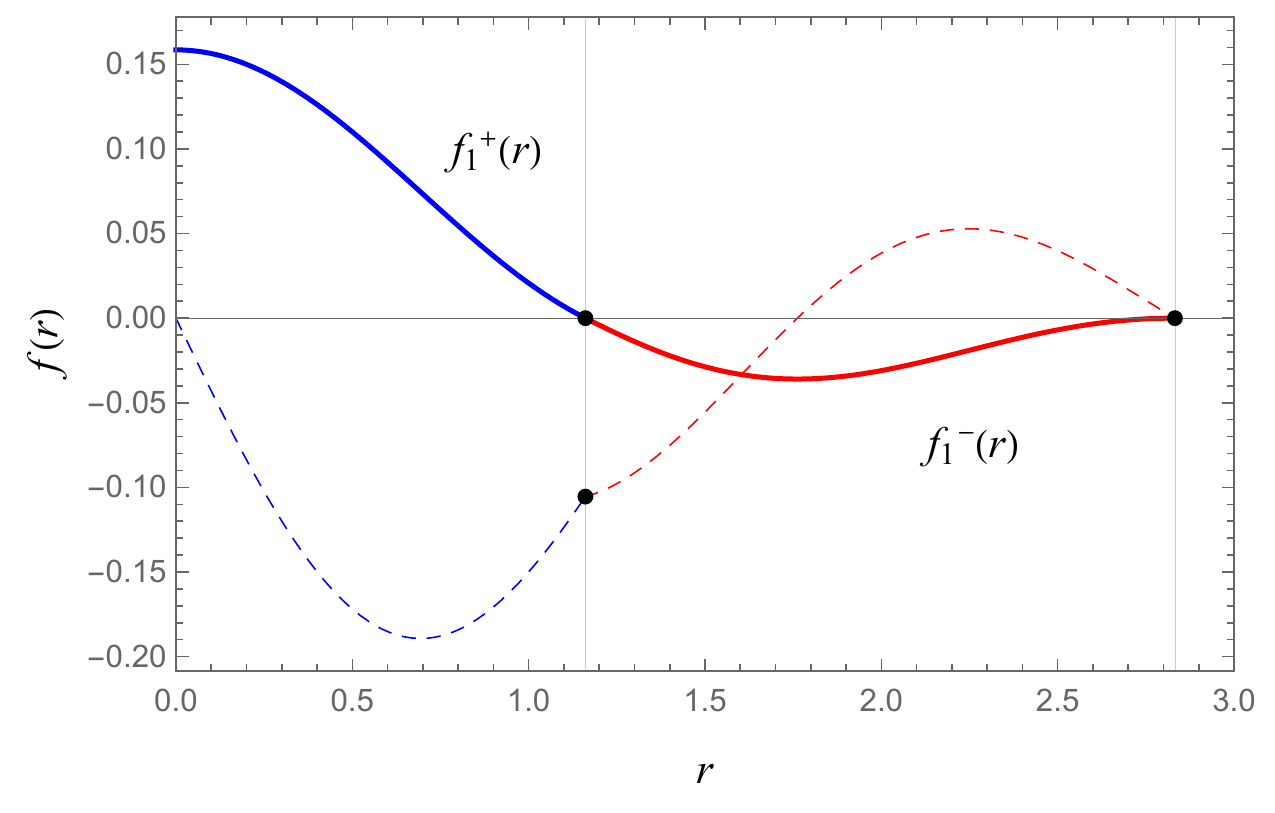}} \hskip0.5cm
\subfigure[]{\includegraphics[width=0.45\textwidth, angle =0]{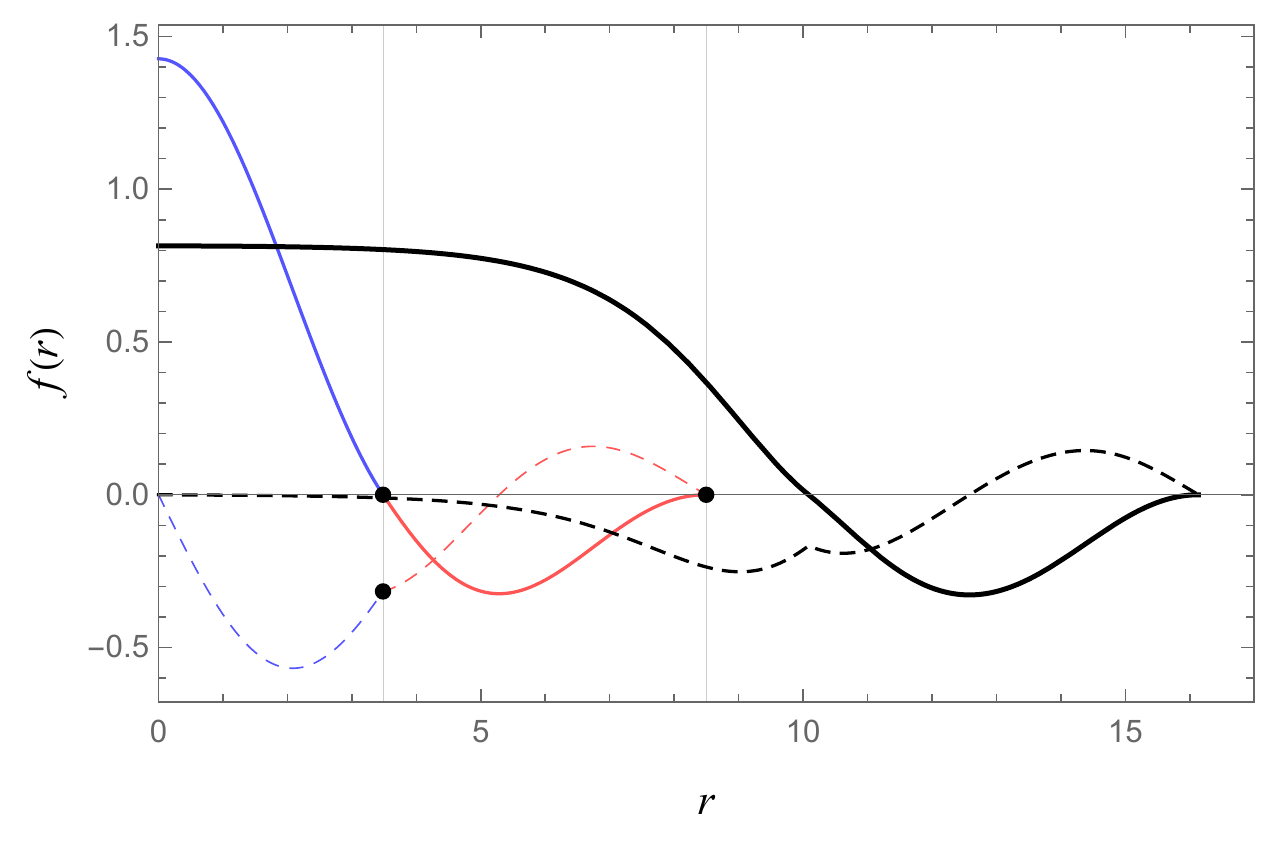}}\\
\subfigure[]{\includegraphics[width=0.45\textwidth, angle =0]{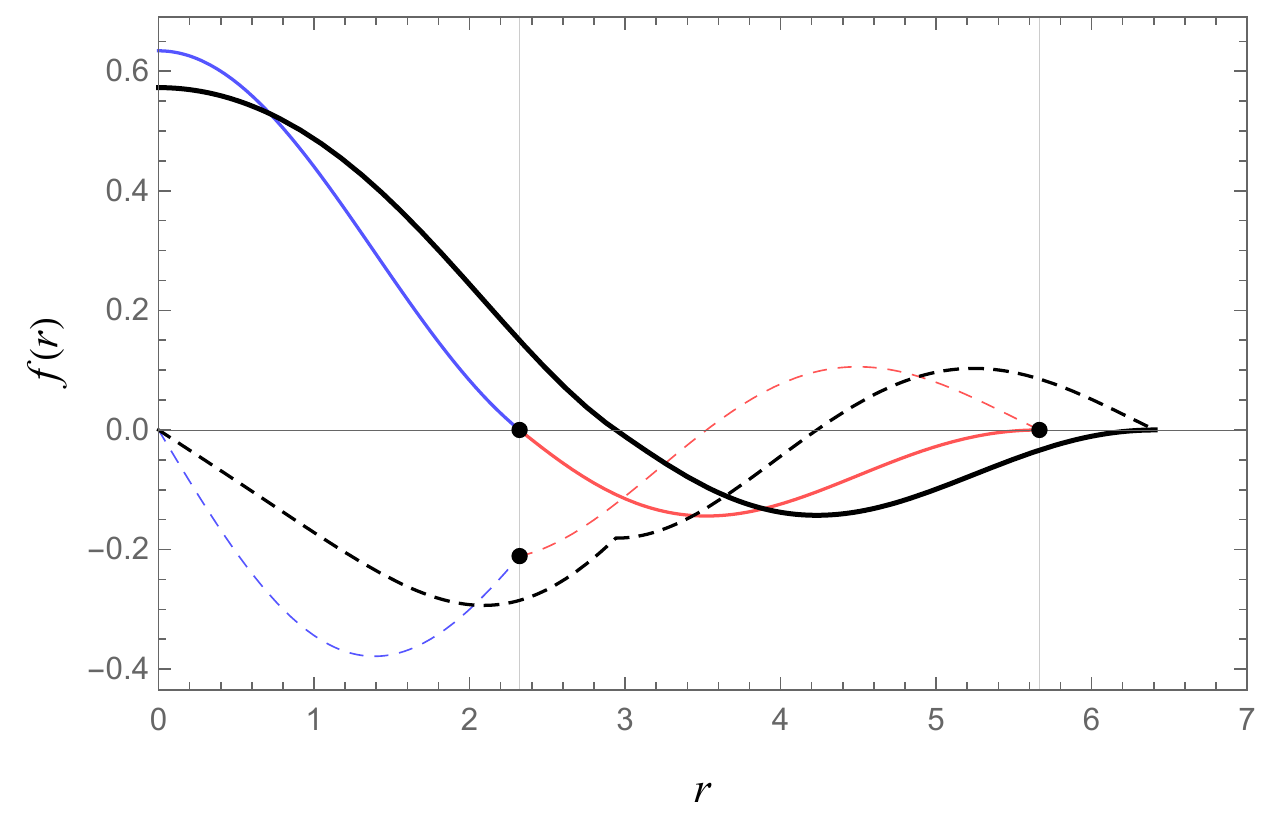}} \hskip0.5cm
\subfigure[]{\includegraphics[width=0.45\textwidth, angle =0]{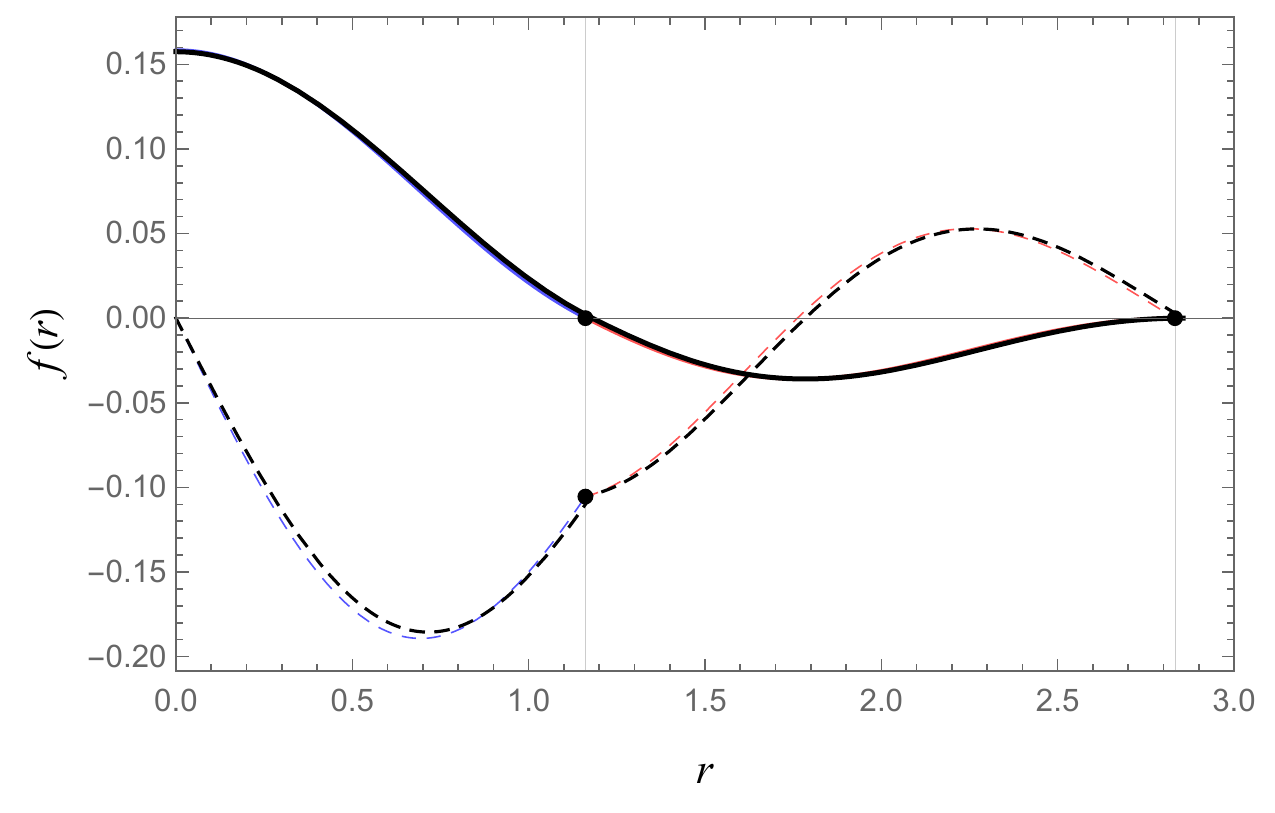}}

   \caption{\label{ANw} The single node analytical solution for signum-Gordon model and the numerical  $\mathbb{C}P^1$ solution. 
The  plot  shows solutions with $\lambda=1/8$. (a) The 1-node analytical solution for the signum-Gordon model with $\omega=3.00$ 
and (b) the numerical solutions (black curves) for the $\mathbb{C}P^1$ model with $\omega=1.00$,
(c) $\omega=1.50$ and (d) $\omega=3.00$. The dashed lines indicate the derivative of $f(r)$. }
  \end{center}
\end{figure*}

\begin{figure*}[t]
  \begin{center}
\subfigure[]{\includegraphics[width=0.45\textwidth, angle =0]{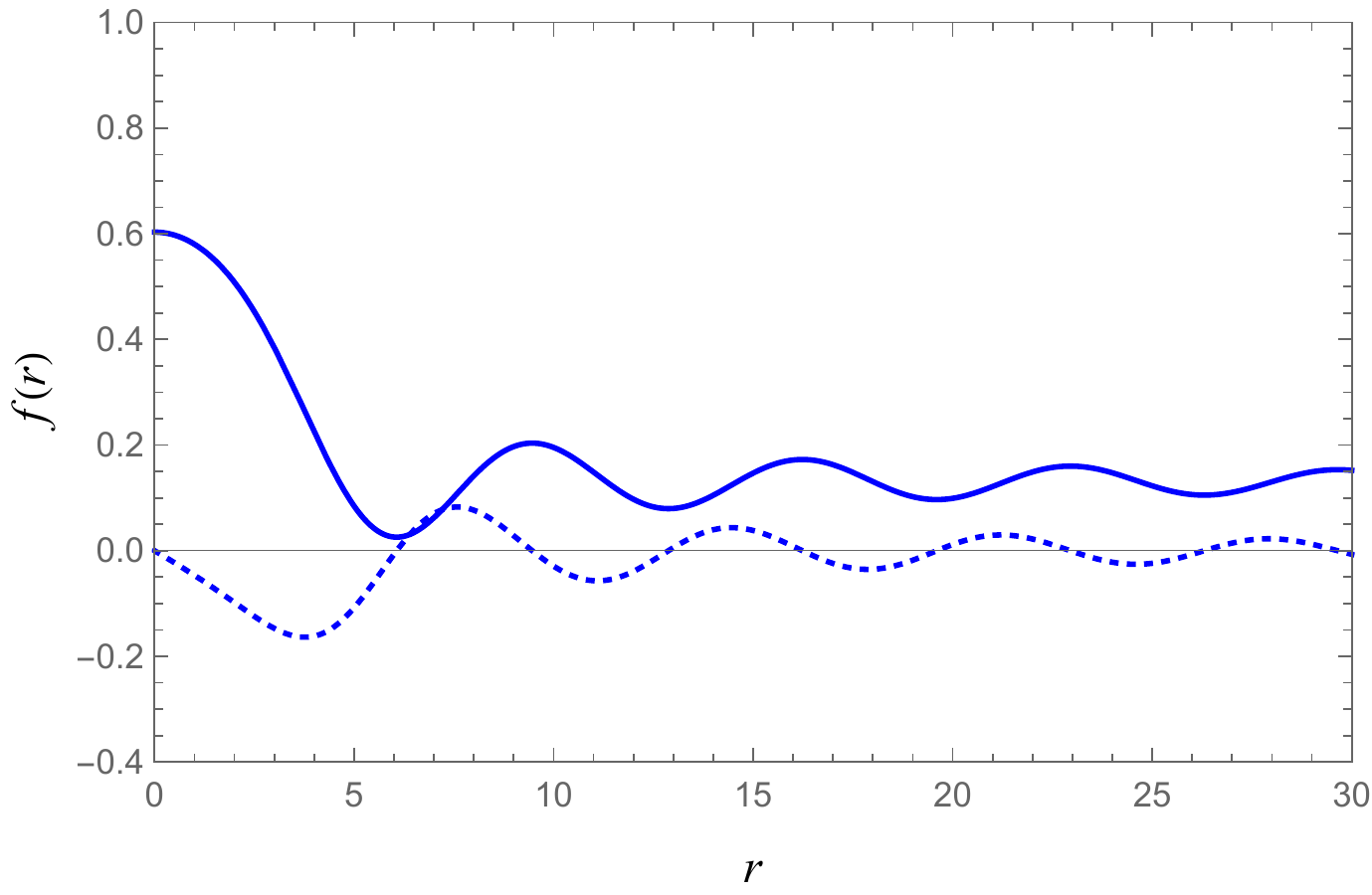}}\hskip0.5cm
\subfigure[]{\includegraphics[width=0.45\textwidth, angle =0]{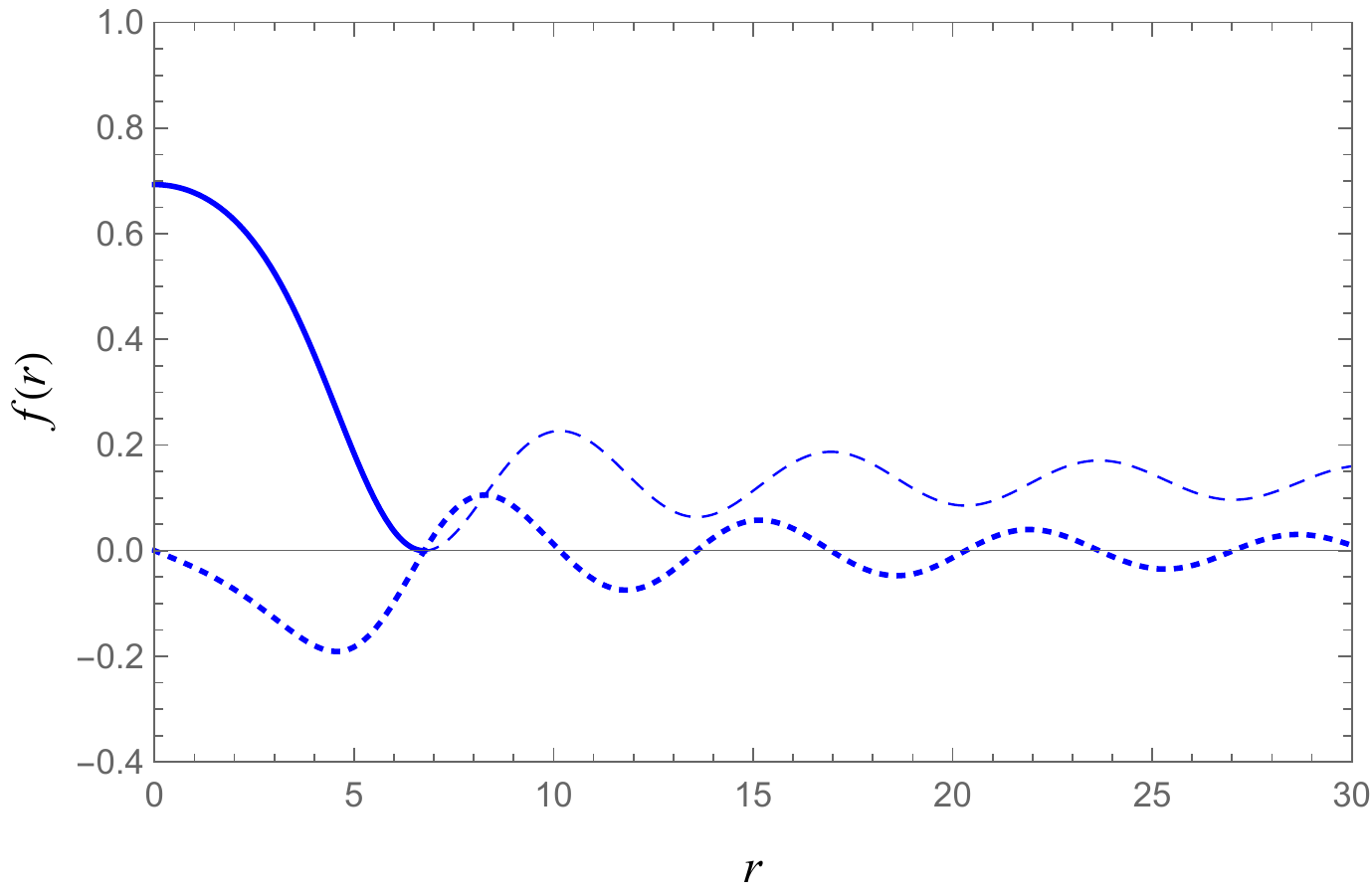}} \\
\subfigure[]{\includegraphics[width=0.45\textwidth, angle =0]{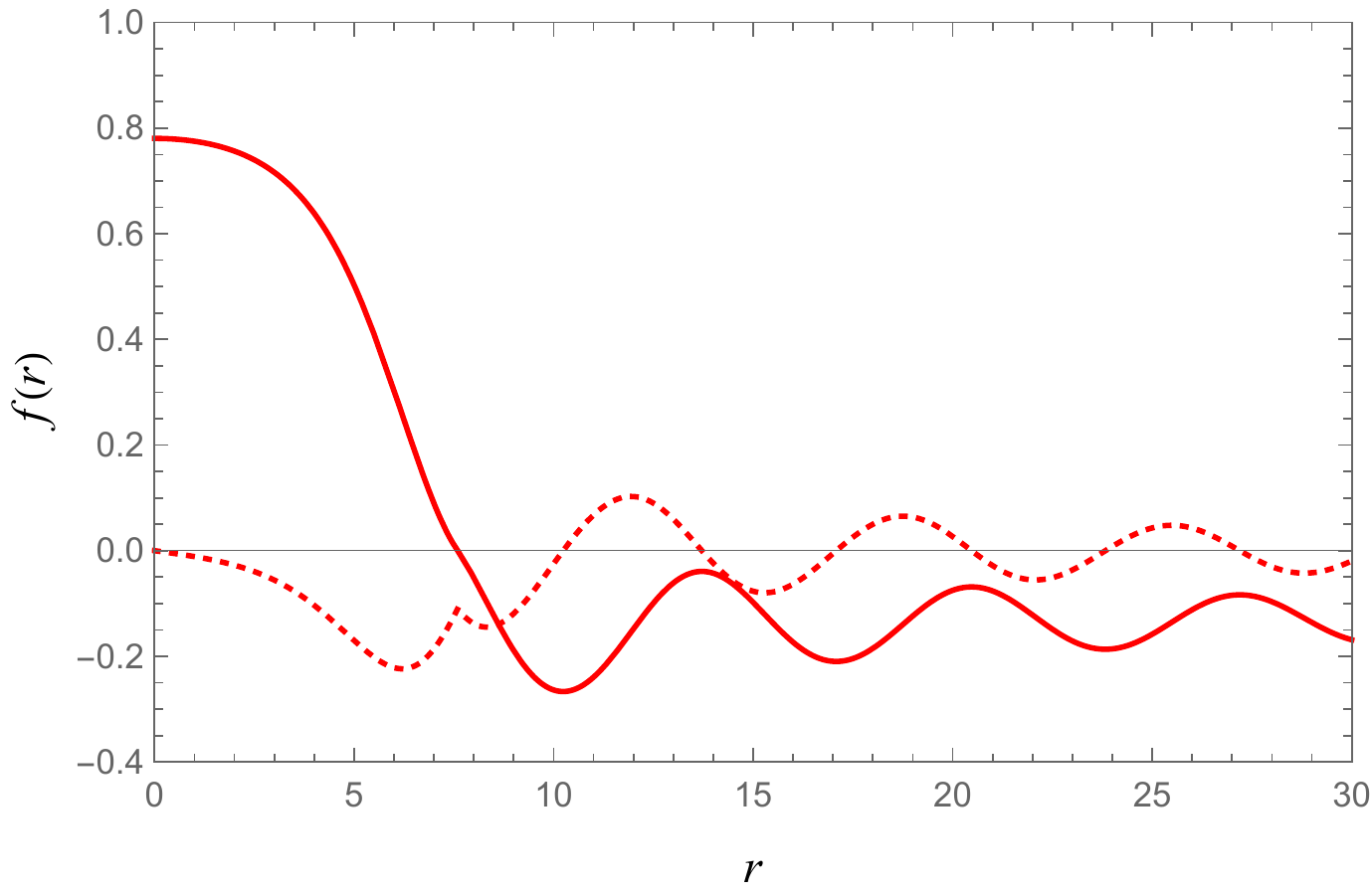}}\hskip0.5cm
\subfigure[]{\includegraphics[width=0.45\textwidth, angle =0]{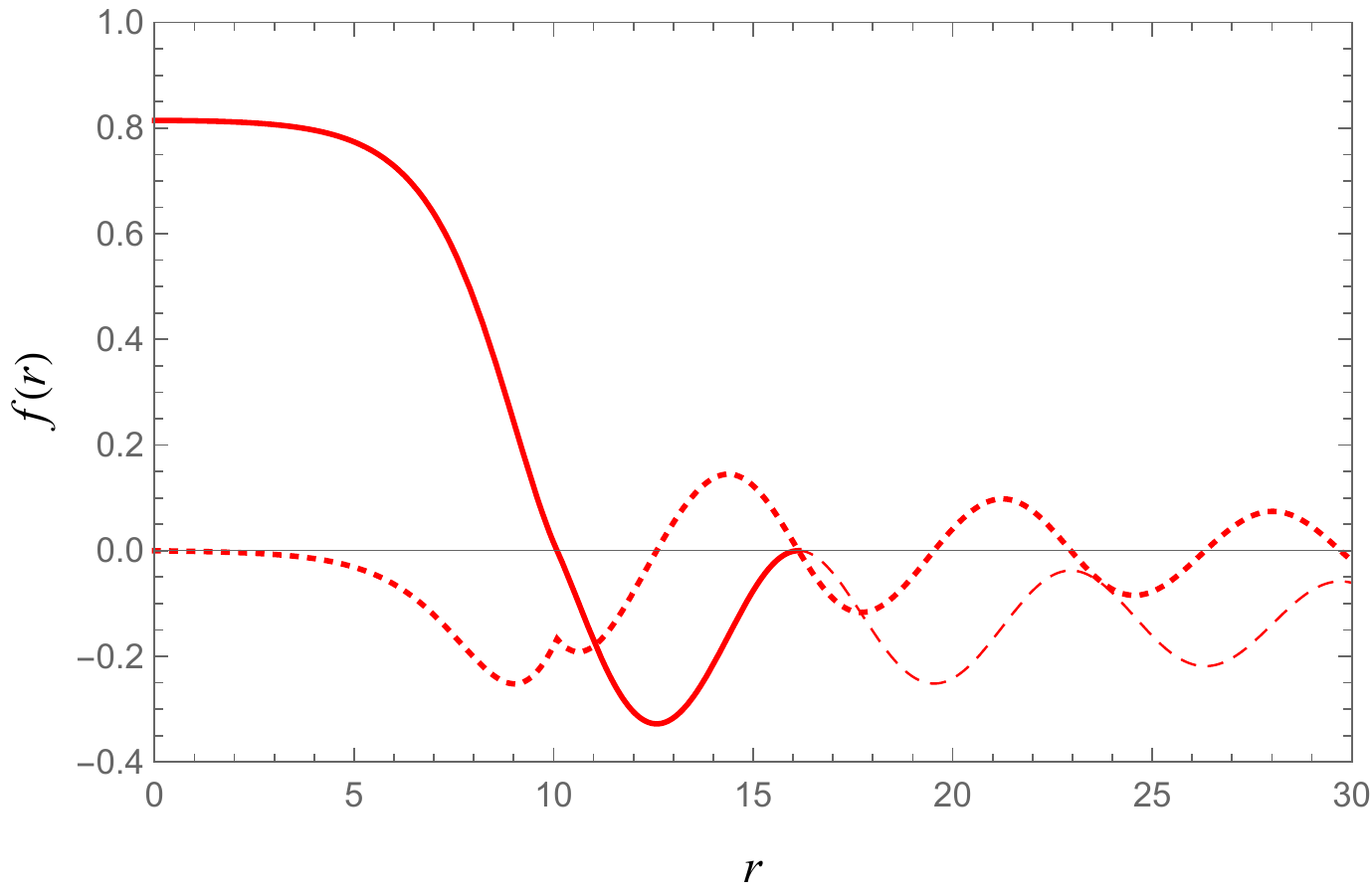}}
   \caption{\label{shoot}  Shooting curves for the $\mathbb{C}P^{1}$ solutions for (a),(b) 0-node and (c),(d) 1-node. 
We determine such a value of the parameter $f(0)$ (the value of the profile function $f(r)$ at the origin) 
that the function $f(r)$ is zero at its first minimum (maximum). The node number corresponds to the number of internal zeros.}
 \end{center}
\end{figure*}

\begin{figure*}[t]
  \begin{center}
\subfigure[]{\includegraphics[width=0.45\textwidth, angle =0]{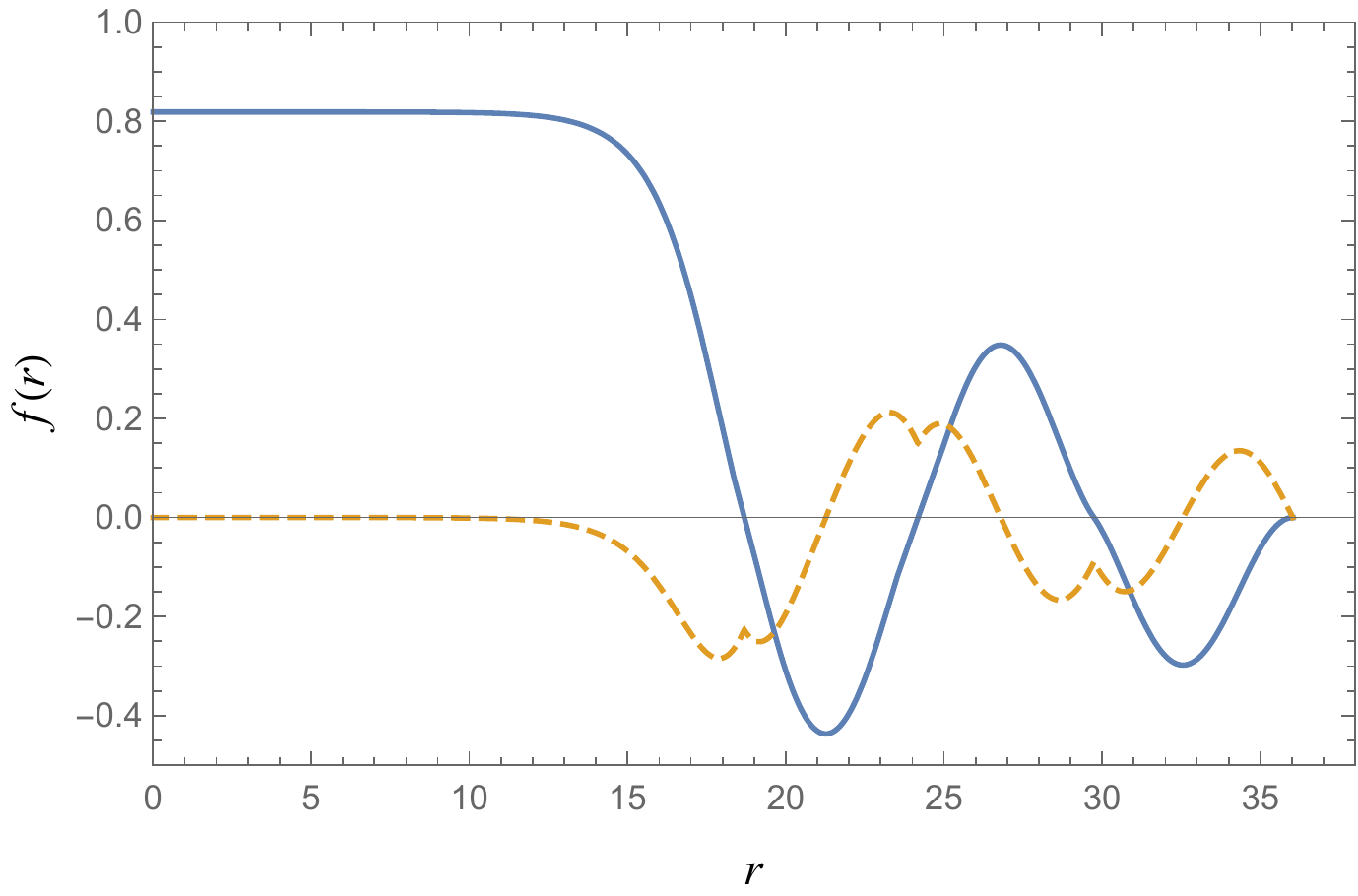}}\hskip0.5cm
\subfigure[]{\includegraphics[width=0.45\textwidth, angle =0]{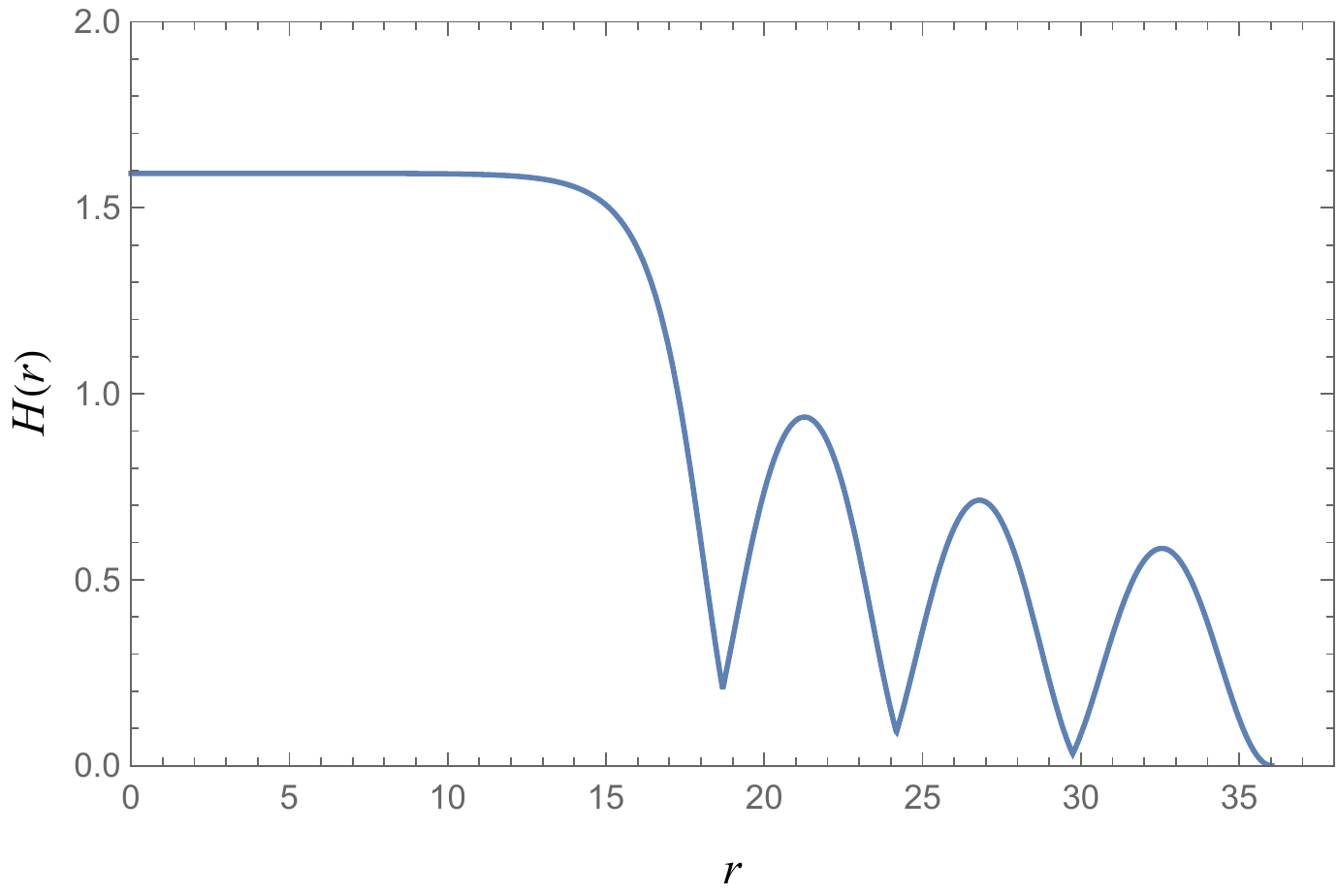}} \\
\subfigure[]{\includegraphics[width=0.45\textwidth, angle =0]{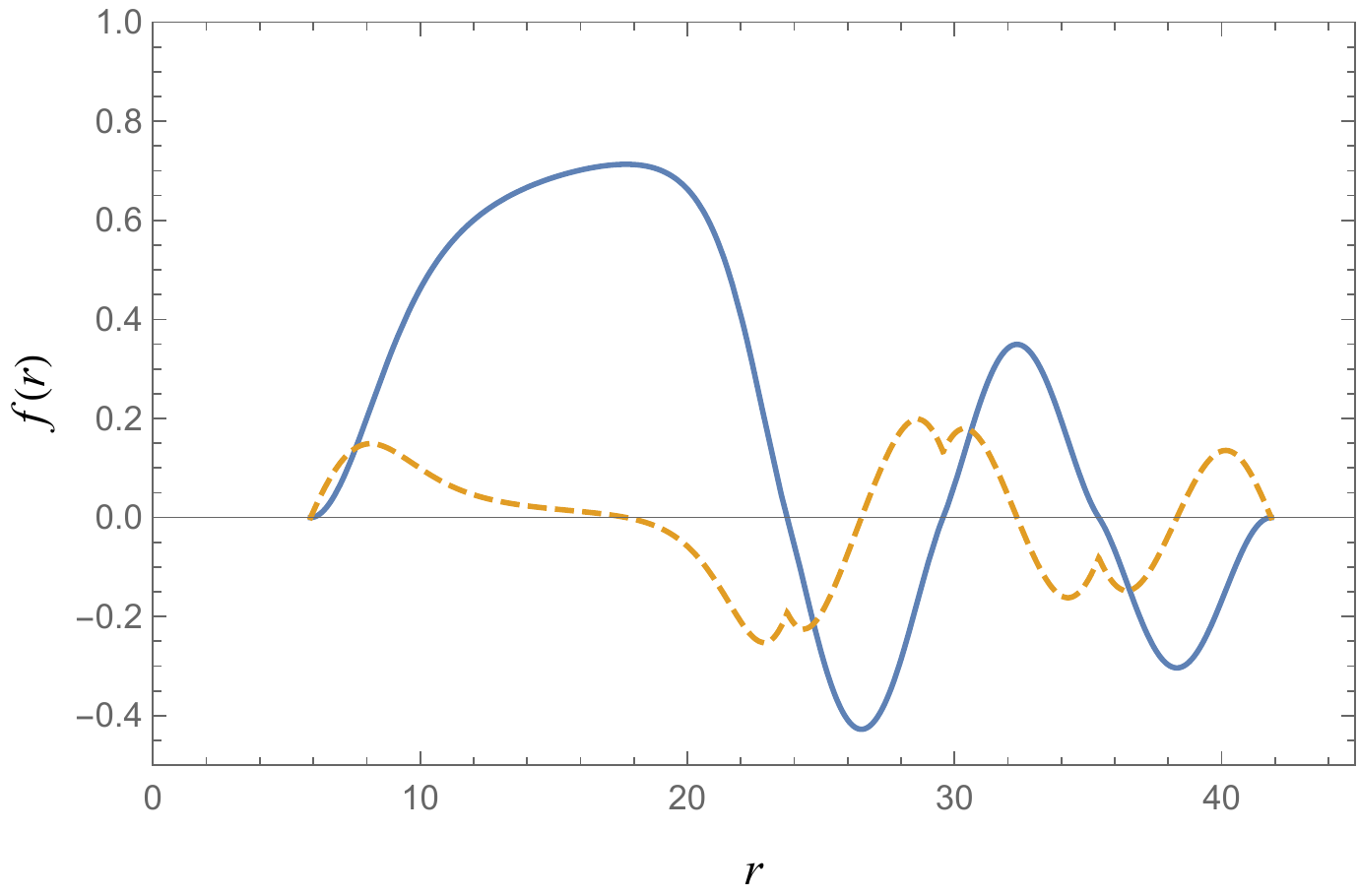}}\hskip0.5cm
\subfigure[]{\includegraphics[width=0.45\textwidth, angle =0]{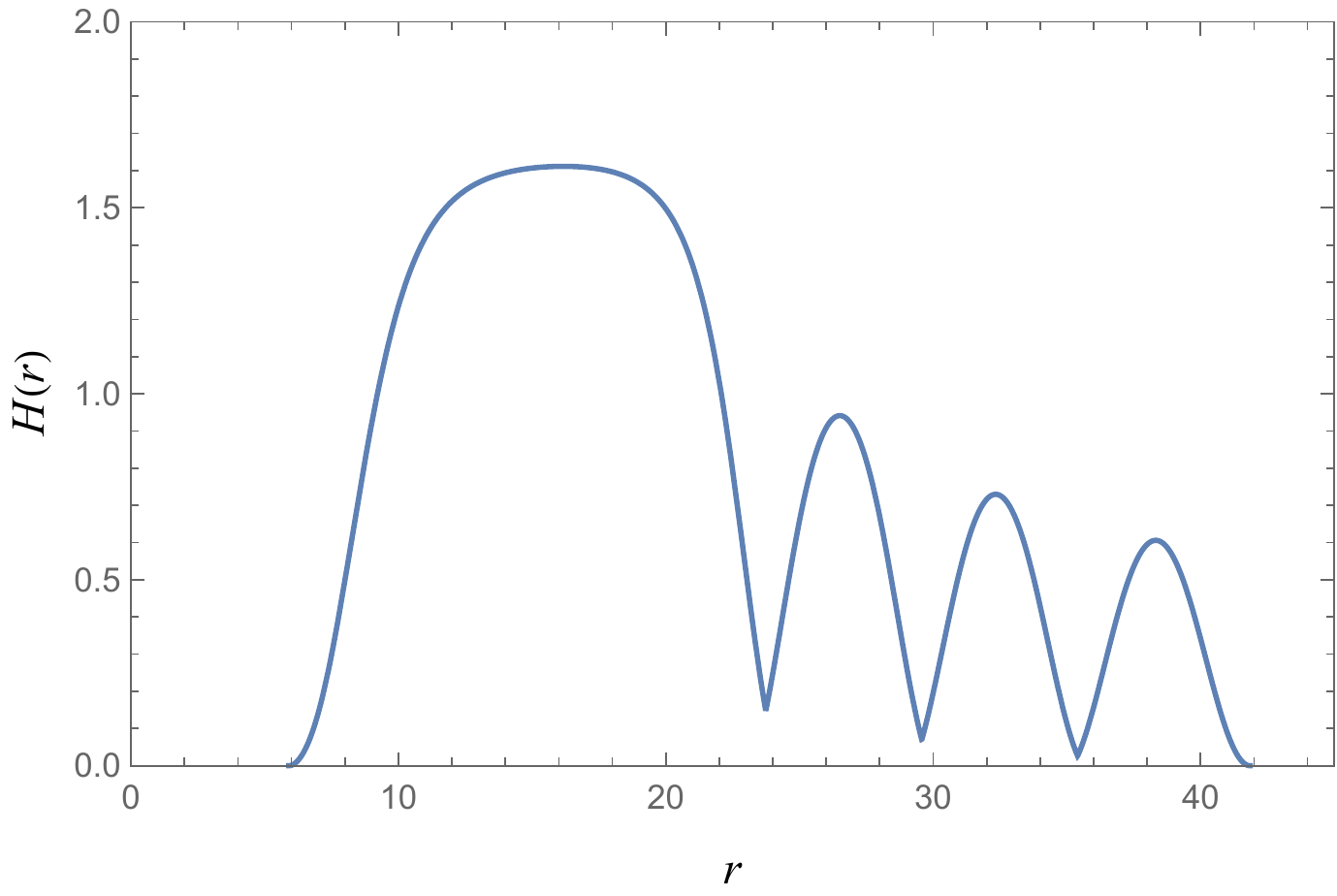}} 

   \caption{\label{nongauge}  The excited compact solutions with $\omega = 1.00$ for (a),(b) $\mathbb{C}P^{1}$ 
and (c),(d) $\mathbb{C}P^{11}$. (a) $f(r)$ for 3-node $Q$-ball, (b) the energy density $H(r)$. 
(c) $f(r)$ for 3-node $Q$-shell, (d) the energy density $H(r)$.}
  \end{center}
\end{figure*}

\begin{figure*}[t]
  \begin{center}
\subfigure[]{\includegraphics[width=0.45\textwidth, angle =0]{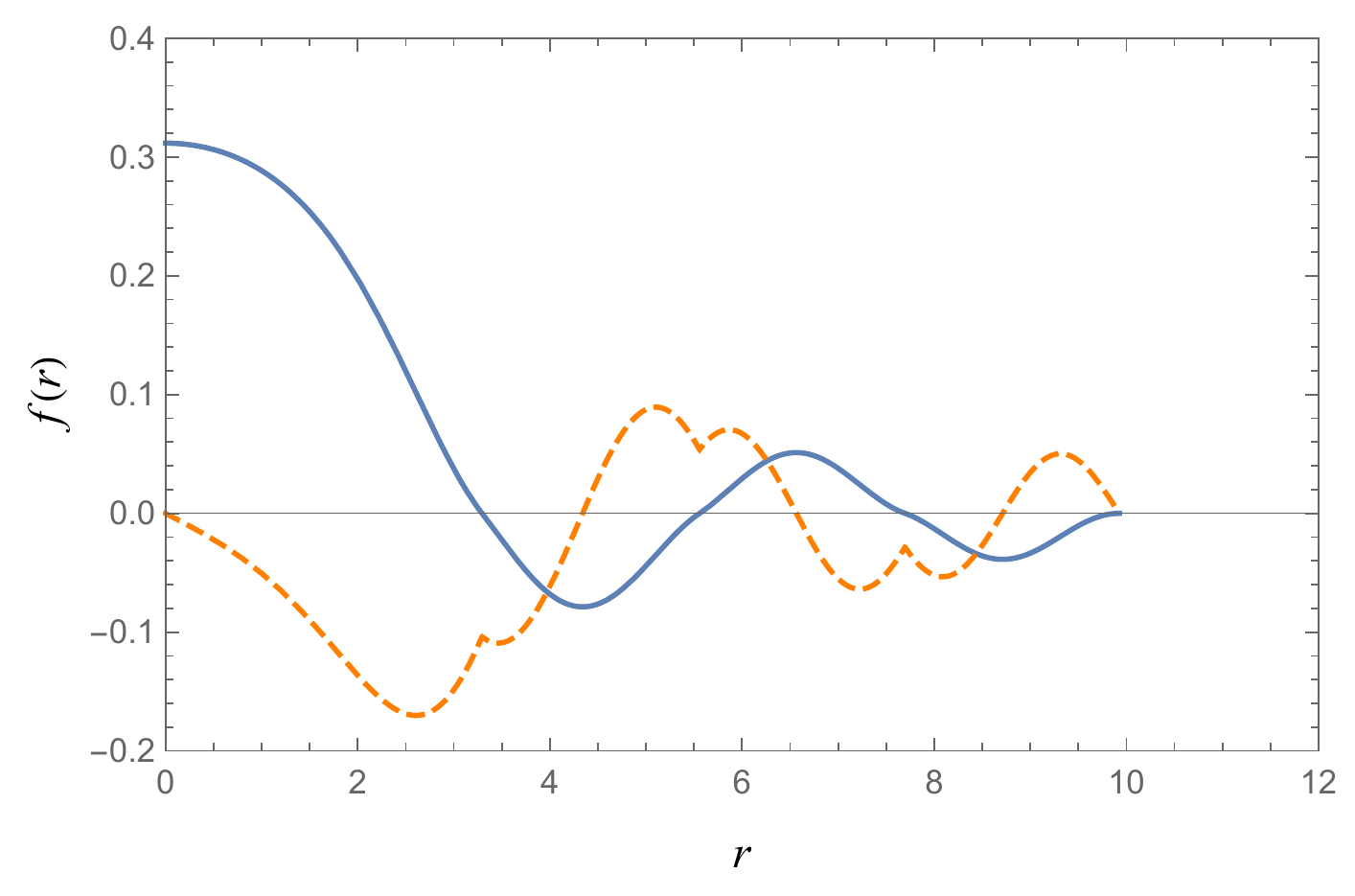}}\hskip0.5cm
\subfigure[]{\includegraphics[width=0.45\textwidth, angle =0]{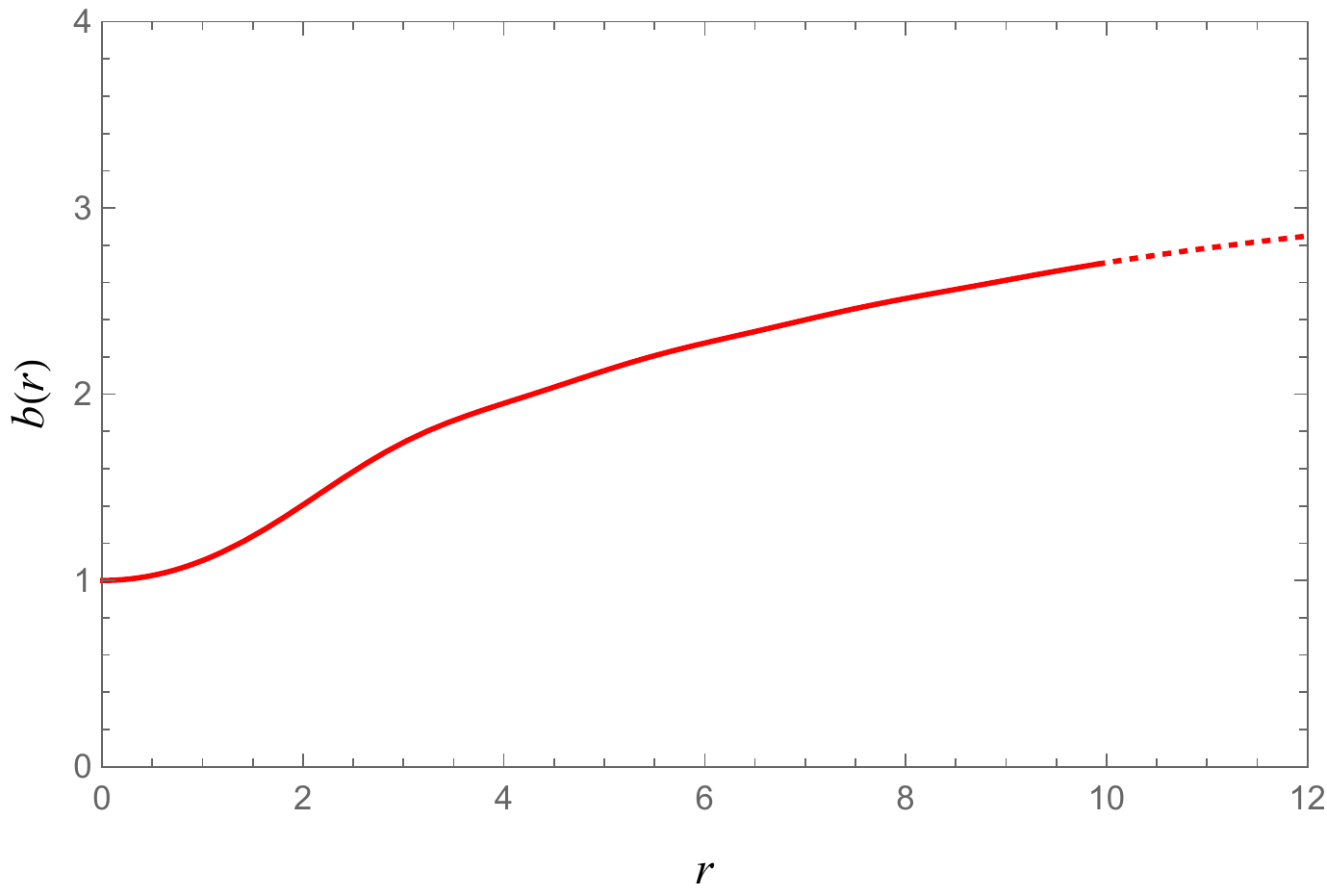}} \\
\subfigure[]{\includegraphics[width=0.45\textwidth, angle =0]{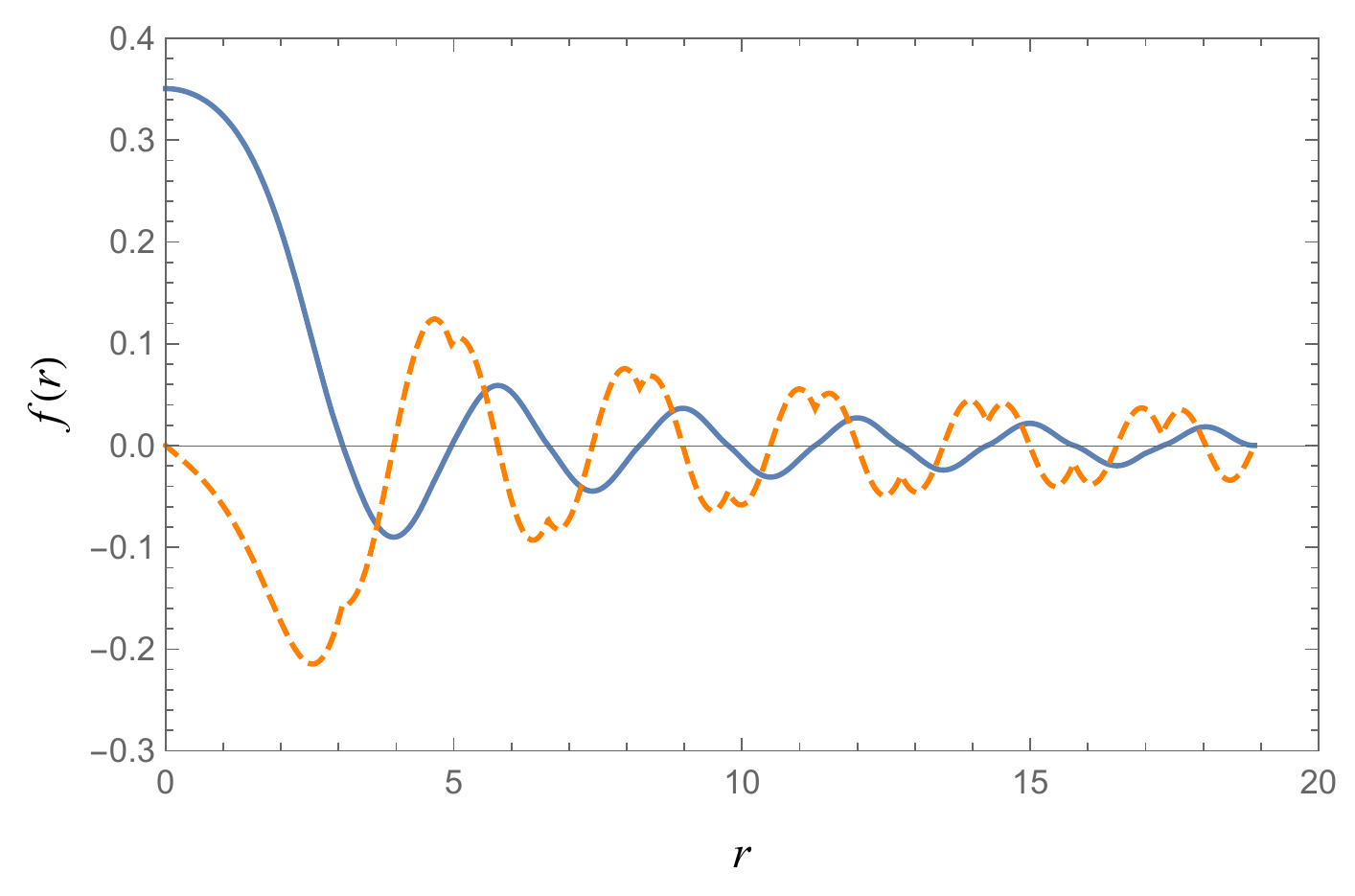}}\hskip0.5cm
\subfigure[]{\includegraphics[width=0.45\textwidth, angle =0]{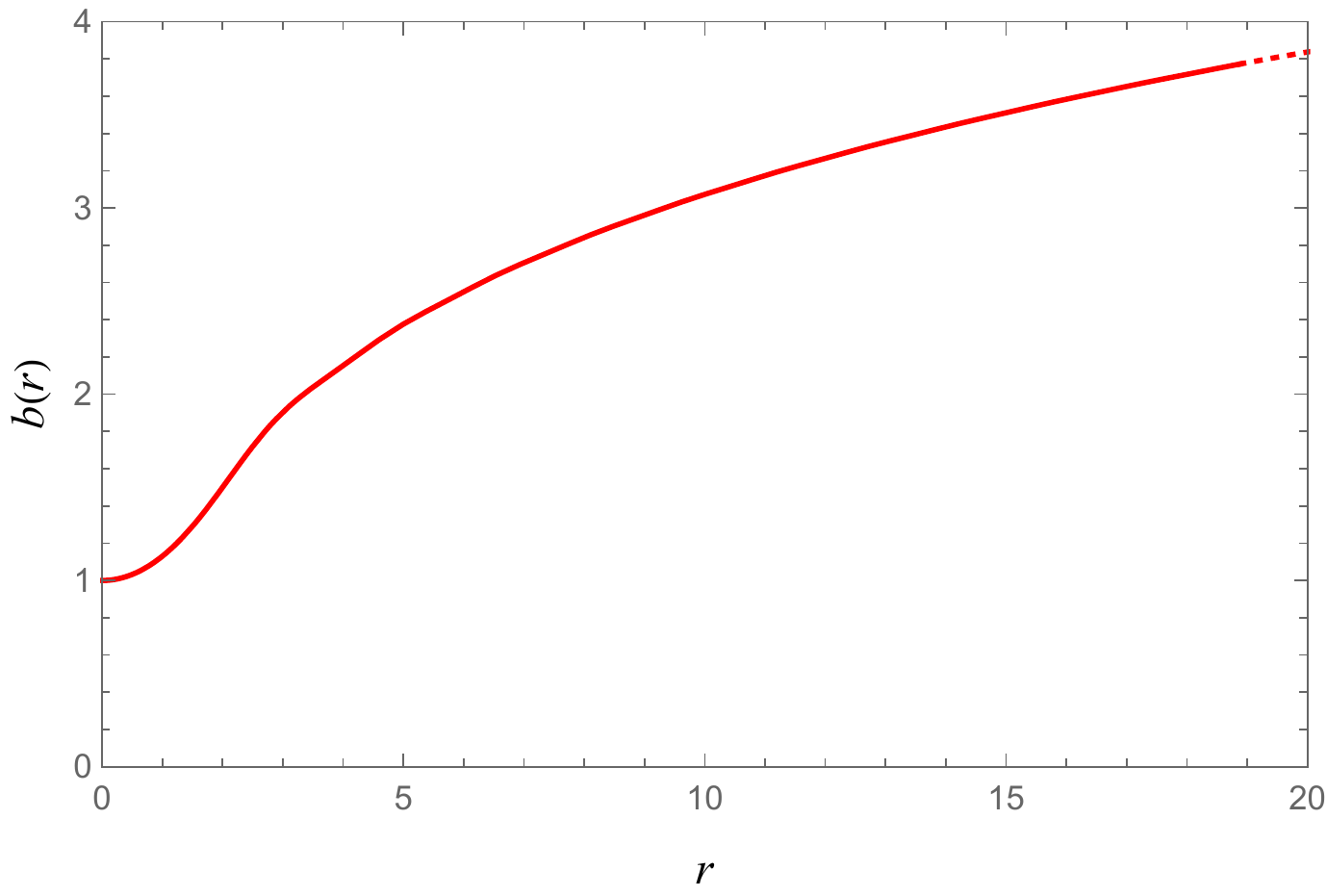}} 

   \caption{\label{fCP1gaugeb}  The excited gauged solution for $\mathbb{C}P^{1}$ $Q$-ball. 
(a) The profile function $f(r)$ for the 3-node. (b) The gauge function $b(r)$ for 3-node. 
(c) The profile $f(r)$ for 10-node. (d) The gauge function $b(r)$ for 10-node. 
Dashed line indicates $f'(r)$ whereas the dotted line represents the gauge function outside the compacton support. }
  \end{center}
\end{figure*}

\begin{figure*}[t]
  \begin{center}
\subfigure[]{\includegraphics[width=0.45\textwidth, angle =0]{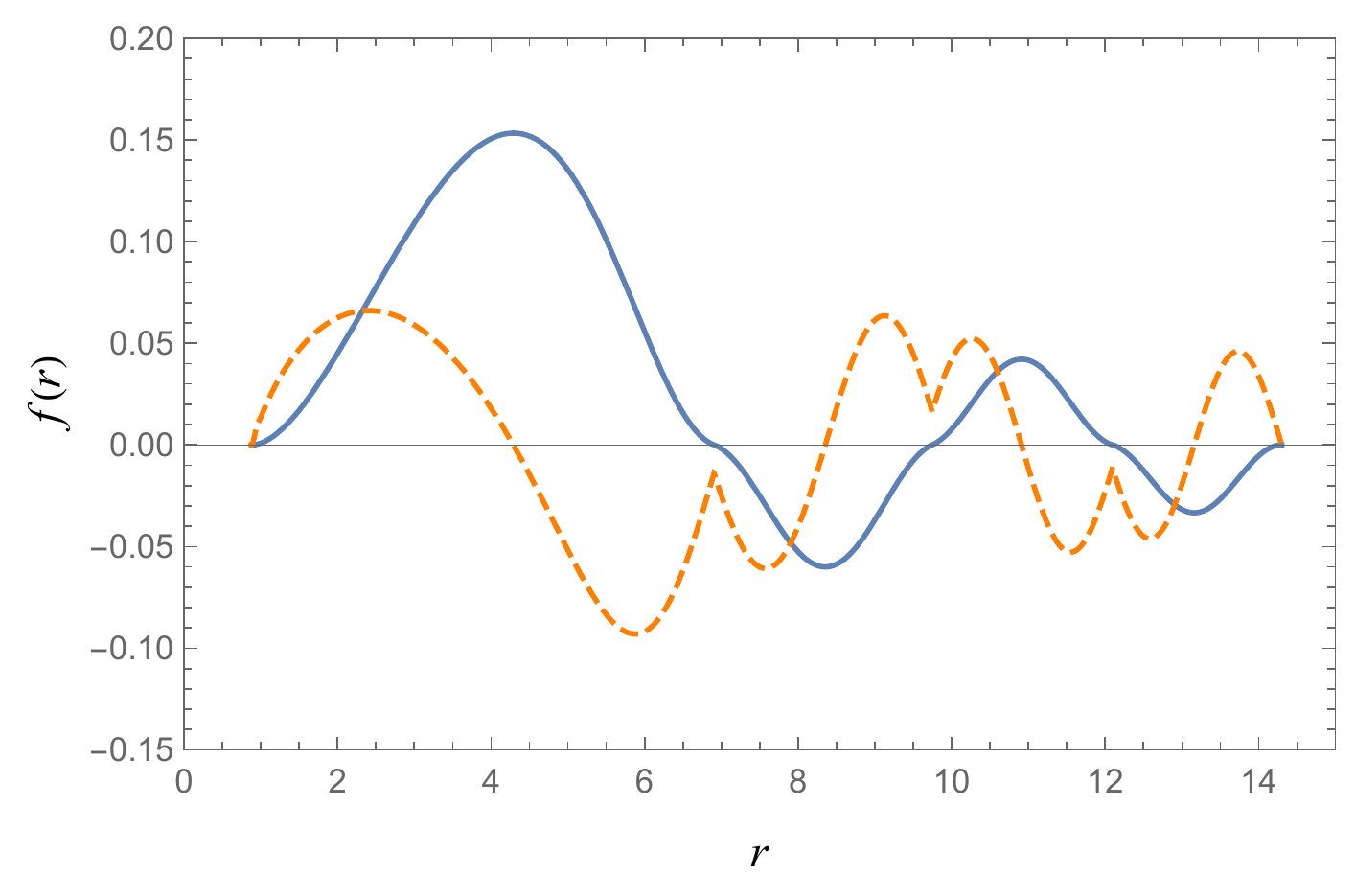}}\hskip0.5cm
\subfigure[]{\includegraphics[width=0.45\textwidth, angle =0]{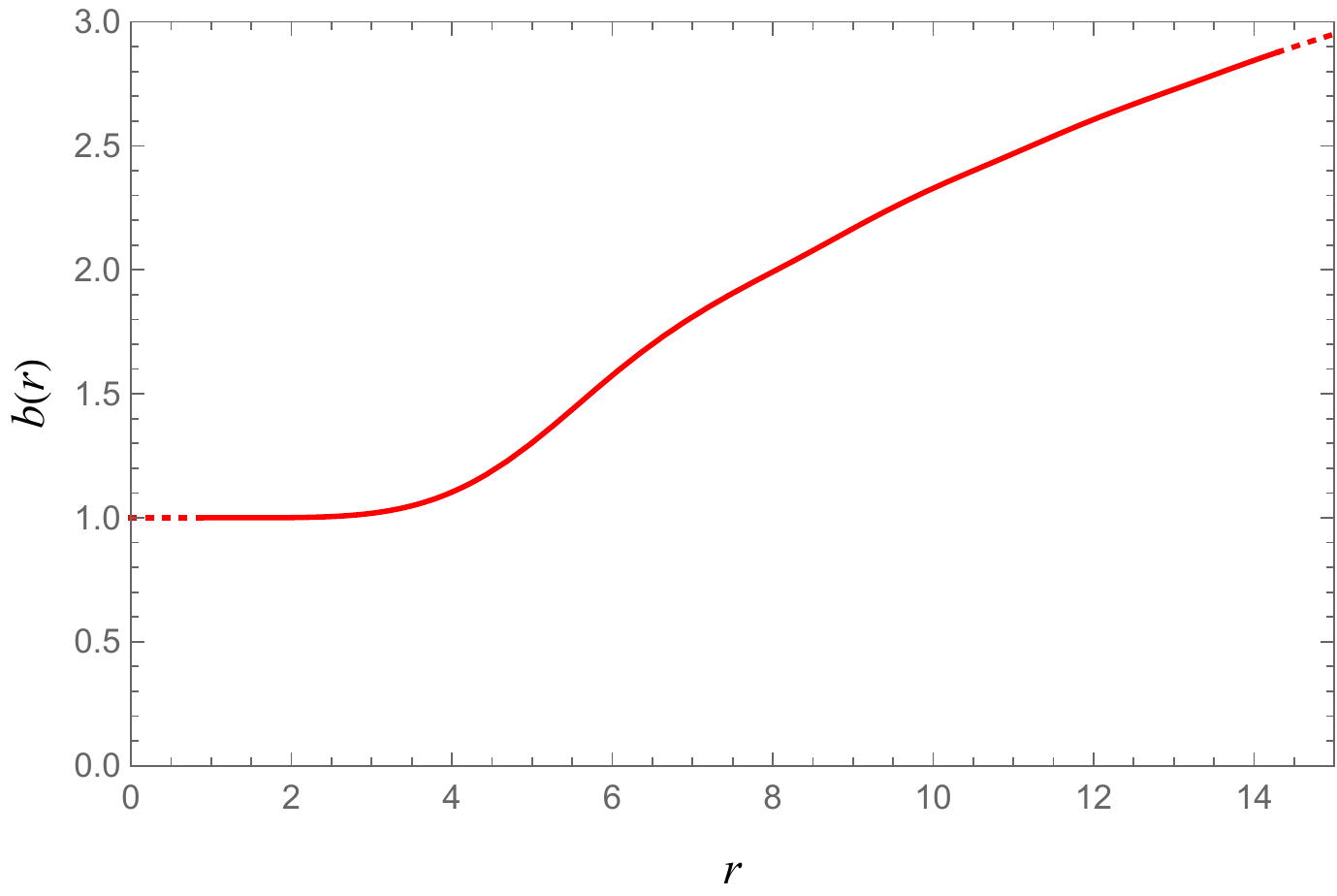}} \\
\subfigure[]{\includegraphics[width=0.45\textwidth, angle =0]{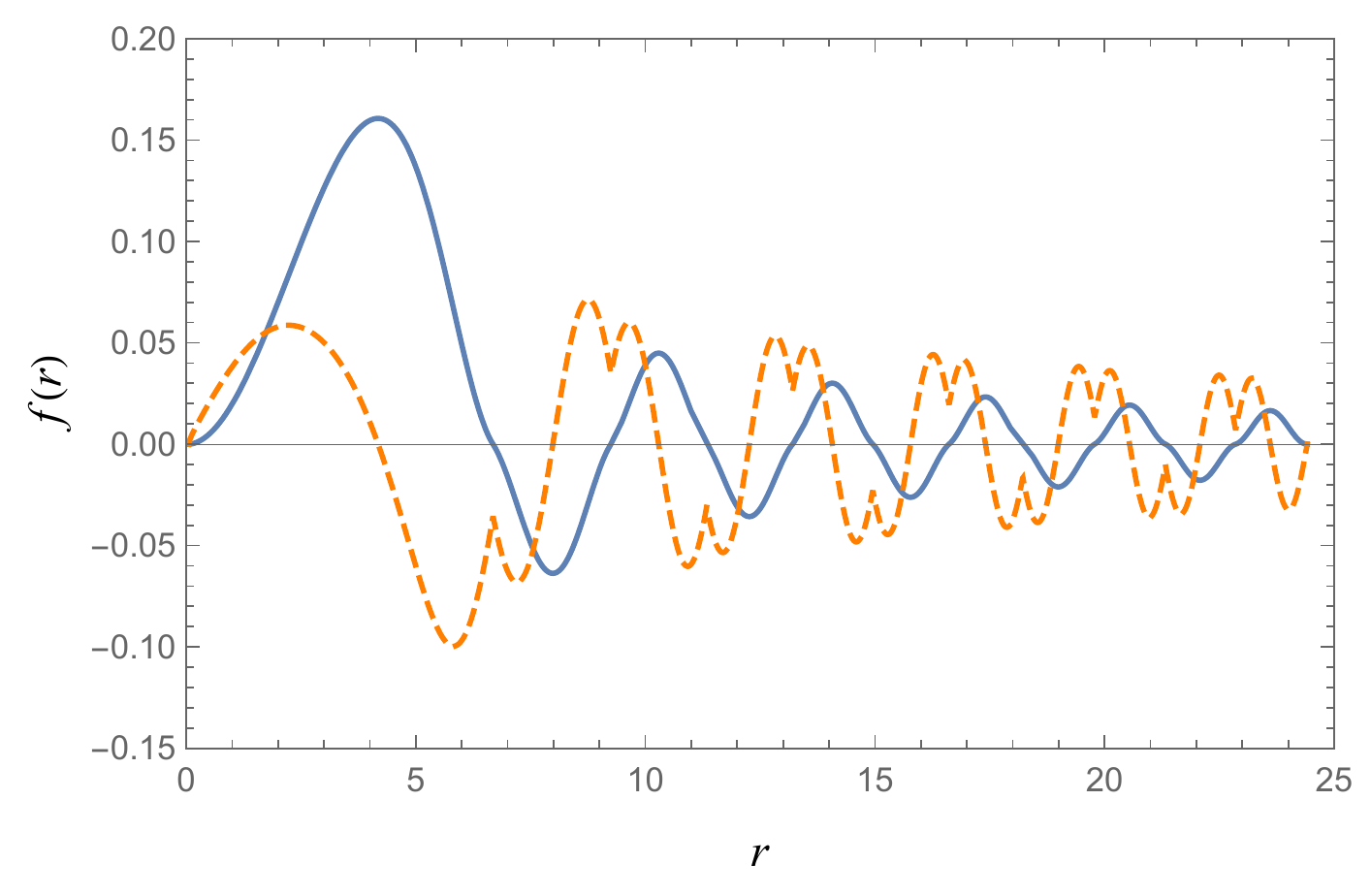}}\hskip0.5cm
\subfigure[]{\includegraphics[width=0.45\textwidth, angle =0]{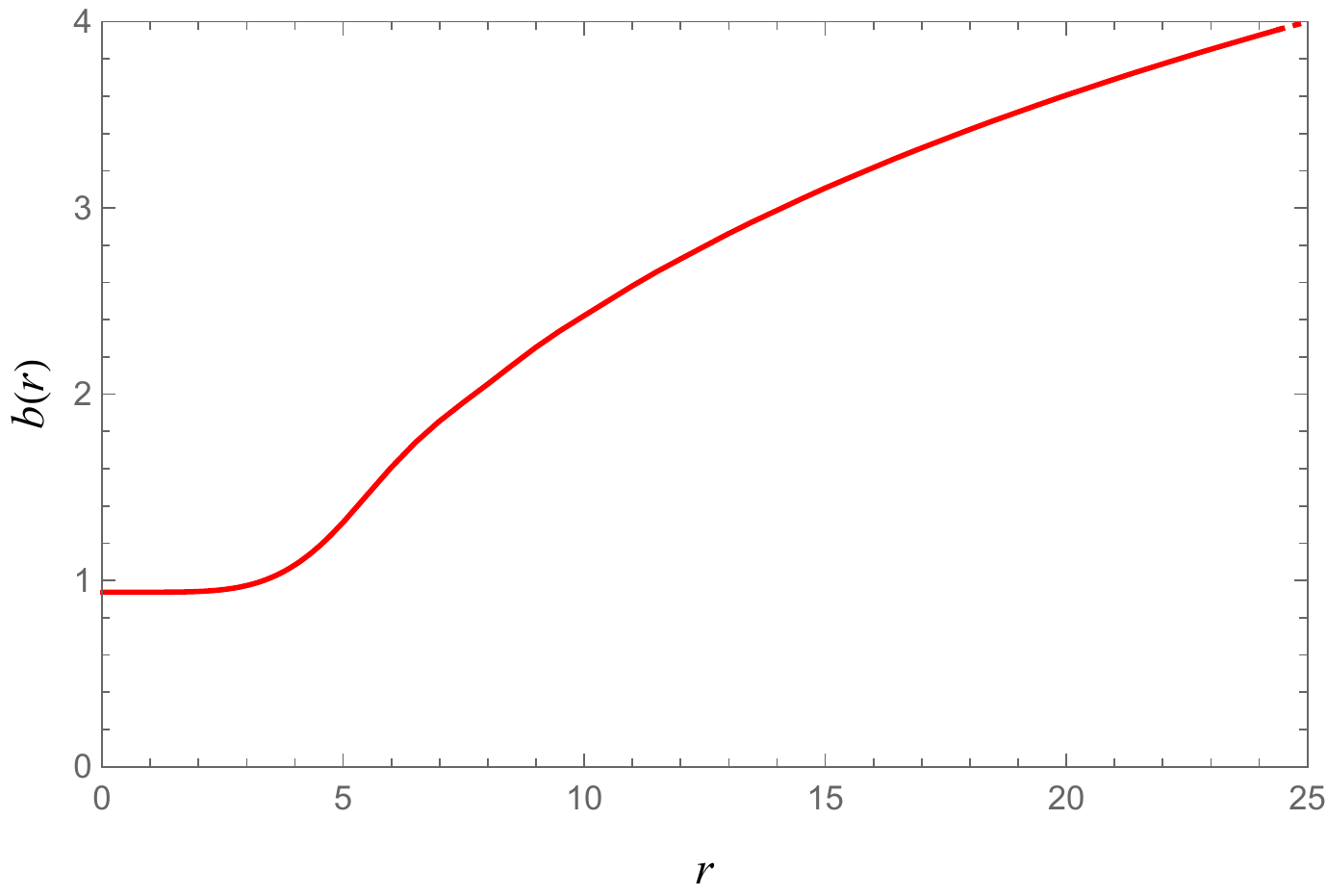}} 

   \caption{\label{fCP1gauges}  The excited gauged solution for $\mathbb{C}P^{1}$ $Q$-shell. 
(a) The profile $f(r)$ function for the 3-node solution. (b) The gauge function $b(r)$ for 3-node. 
(c) The profile $f(r)$ for 10-node. (d) The gauge function $b(r)$ for 10-node. 
The dashed line indicates $f'(r)$ and the dotted line shows the gauge function outside the compacton support. }
  \end{center}
\end{figure*}

\begin{figure*}[t]
  \begin{center}
\subfigure[]{\includegraphics[width=0.45\textwidth, angle =0]{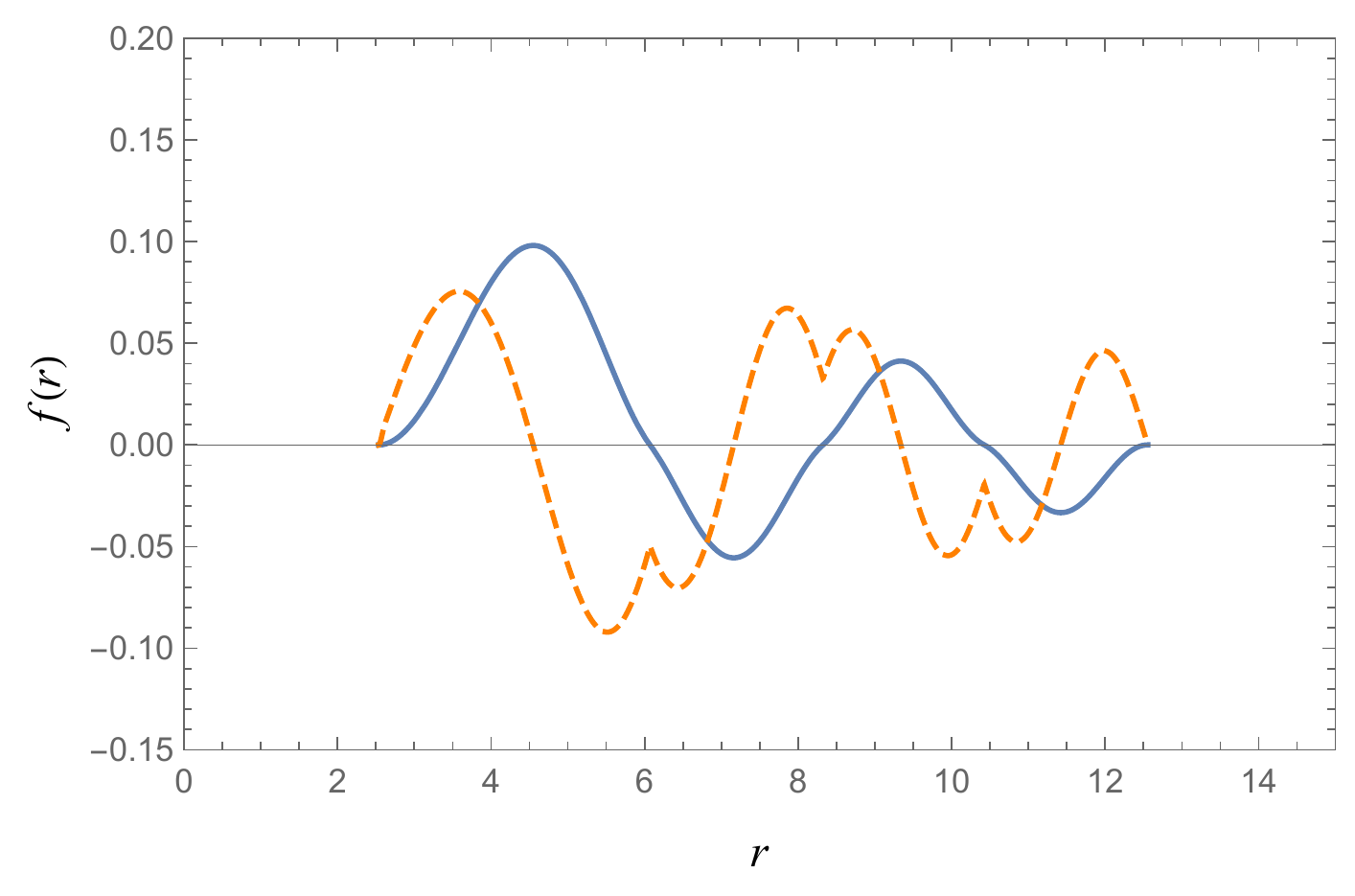}}\hskip0.5cm
\subfigure[]{\includegraphics[width=0.45\textwidth, angle =0]{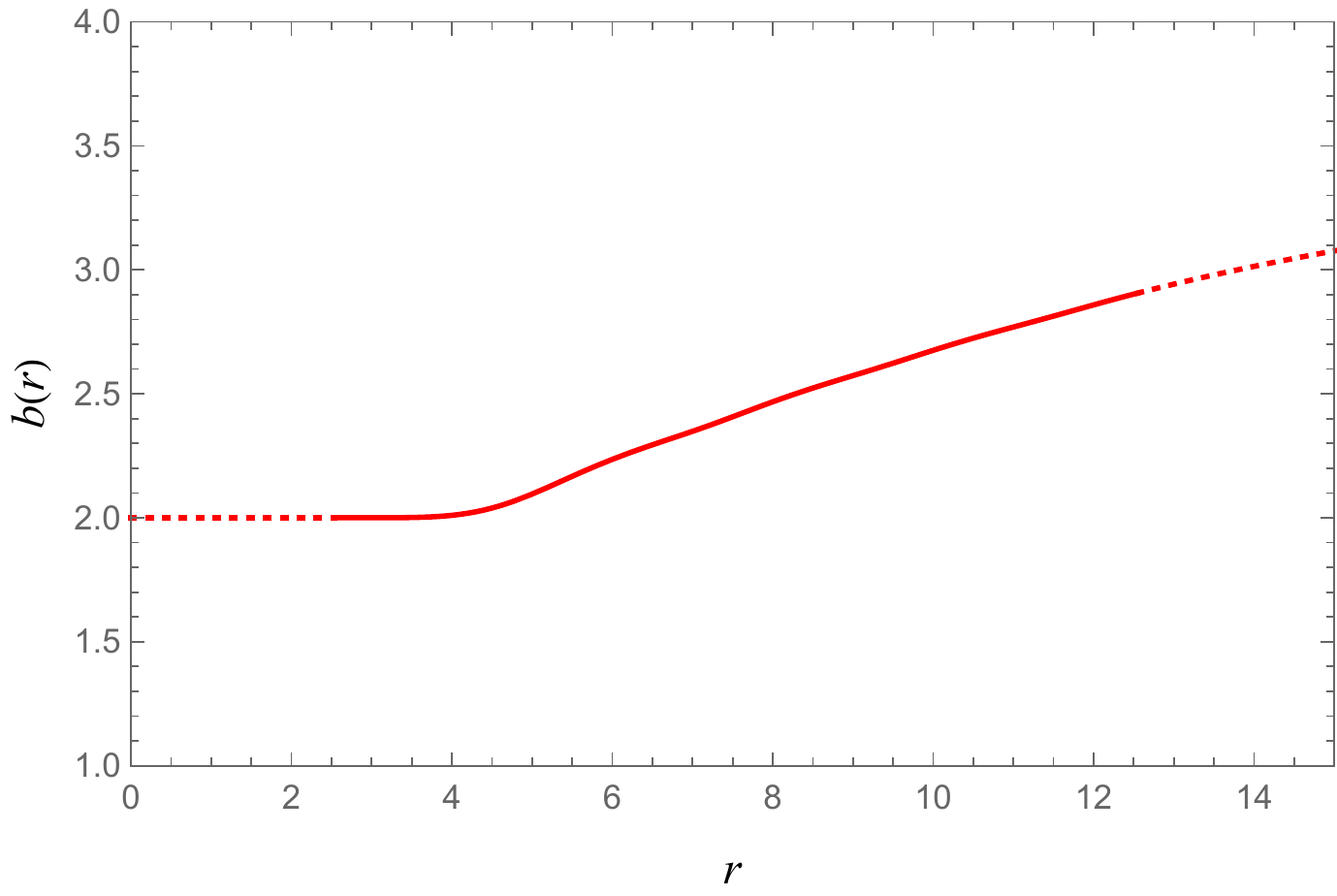}} \\
\subfigure[]{\includegraphics[width=0.45\textwidth, angle =0]{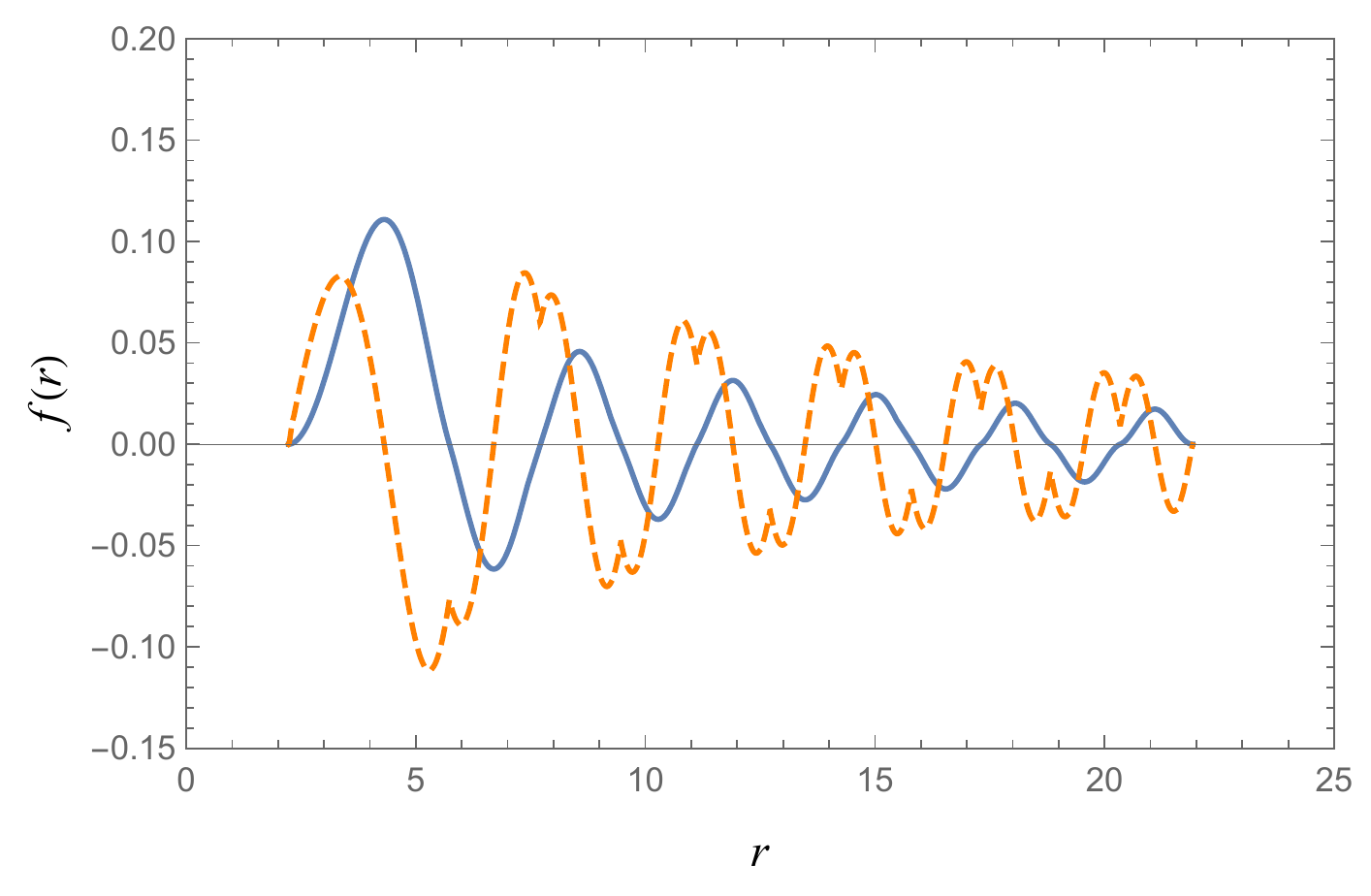}}\hskip0.5cm
\subfigure[]{\includegraphics[width=0.45\textwidth, angle =0]{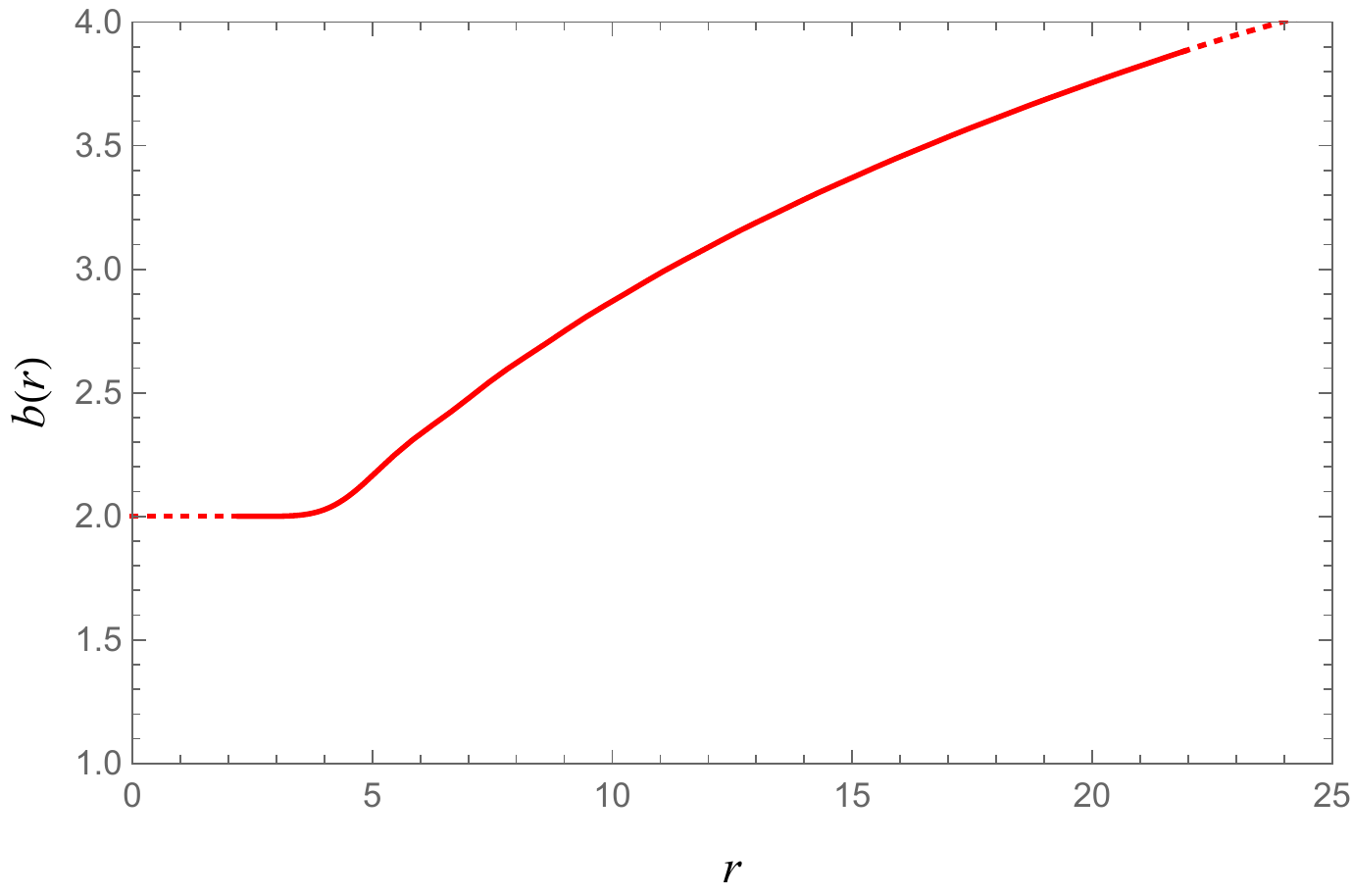}} 

   \caption{\label{fCP11gauges}  The excited gauged solution for $\mathbb{C}P^{11}$ $Q$-shell. 
(a) The profile $f(r)$ function for the 3-node solution. (b) The gauge function $b(r)$ for 3-node. 
(c) The profile $f(r)$ for the 10-node. 
(d) The gauge function $b(r)$ for 10-node. 
The dashed line (orange line) indicates $f'(r)$ and the dotted line (red line) shows the gauge function outside the compacton support.}
  \end{center}
\end{figure*}

\begin{figure*}[t]
  \begin{center}
\subfigure[]{\includegraphics[width=0.45\textwidth, angle =0]{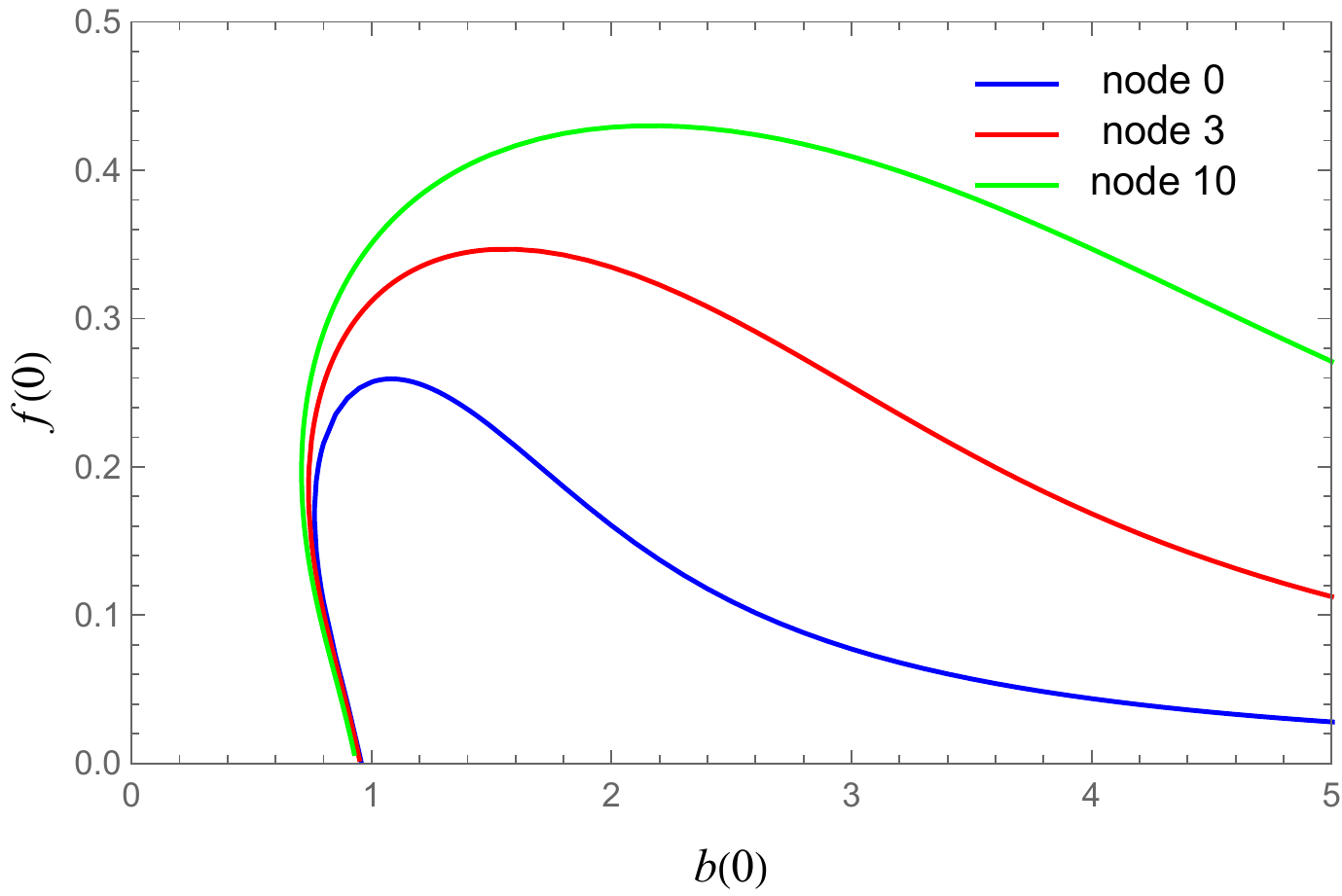}}\hskip0.5cm
\subfigure[]{\includegraphics[width=0.45\textwidth, angle =0]{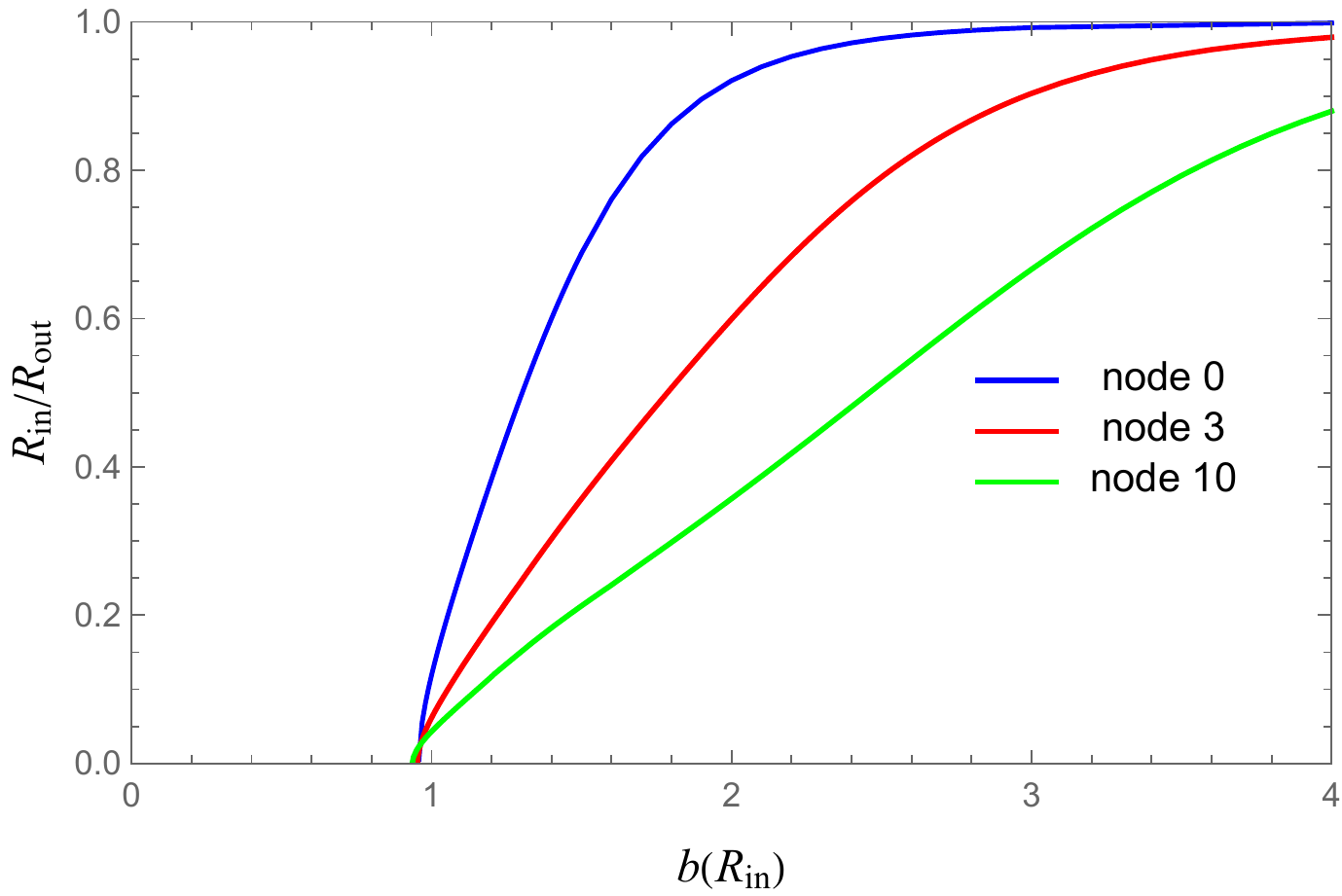}}
 
   \caption{\label{gCP1phase} (a) The phase diagram for $\mathbb{C}P^{1} Q$-balls in space of shooting parameters 
$f(0), b(0)$ that represent the value of the matter profile function $f(r)$ and the gauge field function 
$b(r)$ at the origin $r=0$. The limit $f(0) \rightarrow 0$ indicates that the solution tends to a shell solution. 
(b) The phase diagram for $\mathbb{C}P^{1} Q$-shells with the parameters $R_{\rm in}/R_{\rm out}, b(R_{\rm in})$,
the ratio of the inner radius $R_{\rm in}$ to outer radius $R_{\rm out}$ and the gauge field function $b(r)$ at the inner radius.}
  \end{center}
\end{figure*}

\begin{figure*}[t]
  \begin{center}
\subfigure[]{\includegraphics[width=0.60\textwidth, angle =0]{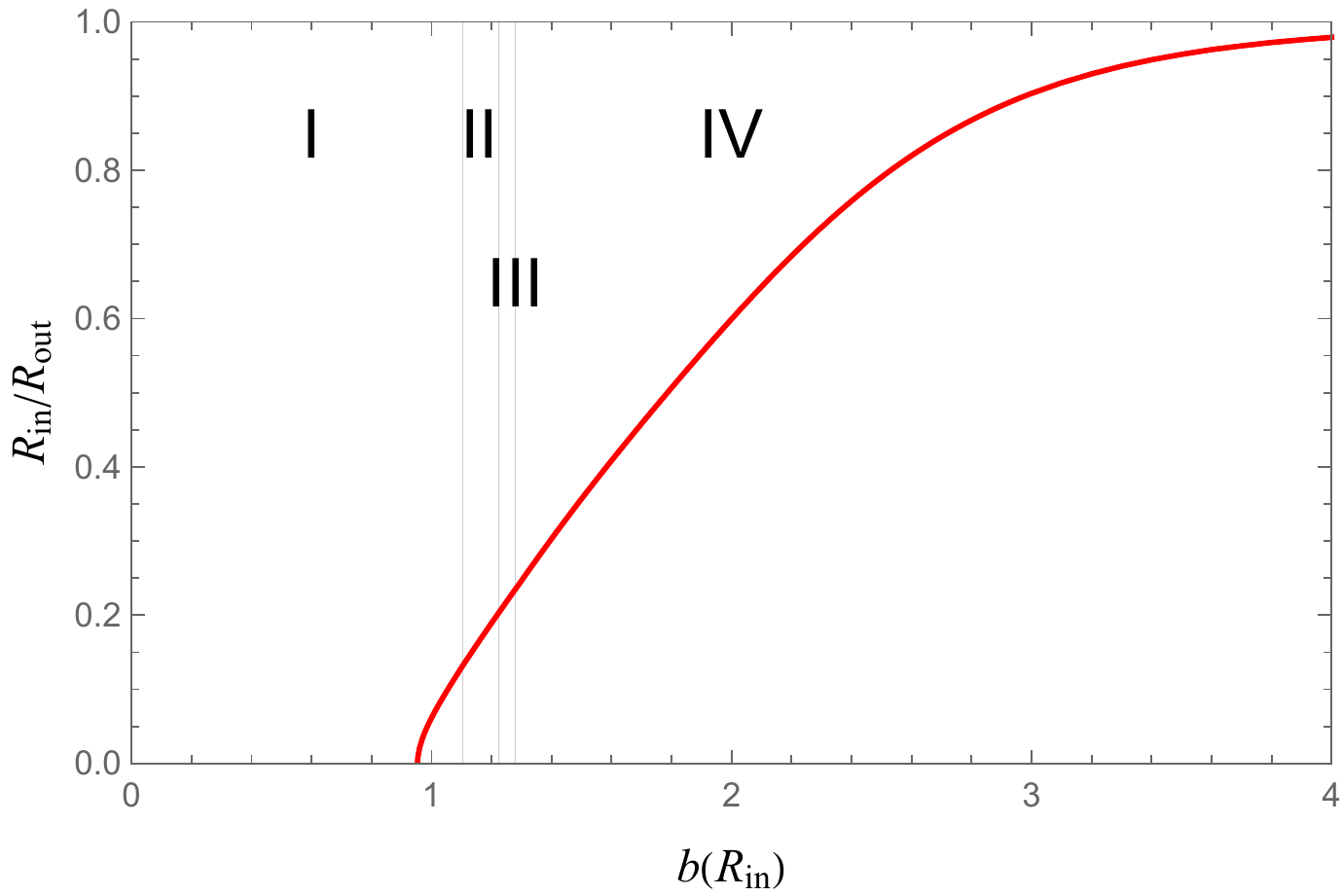}}\hskip0.5cm \\
\subfigure[]{\includegraphics[width=0.45\textwidth, angle =0]{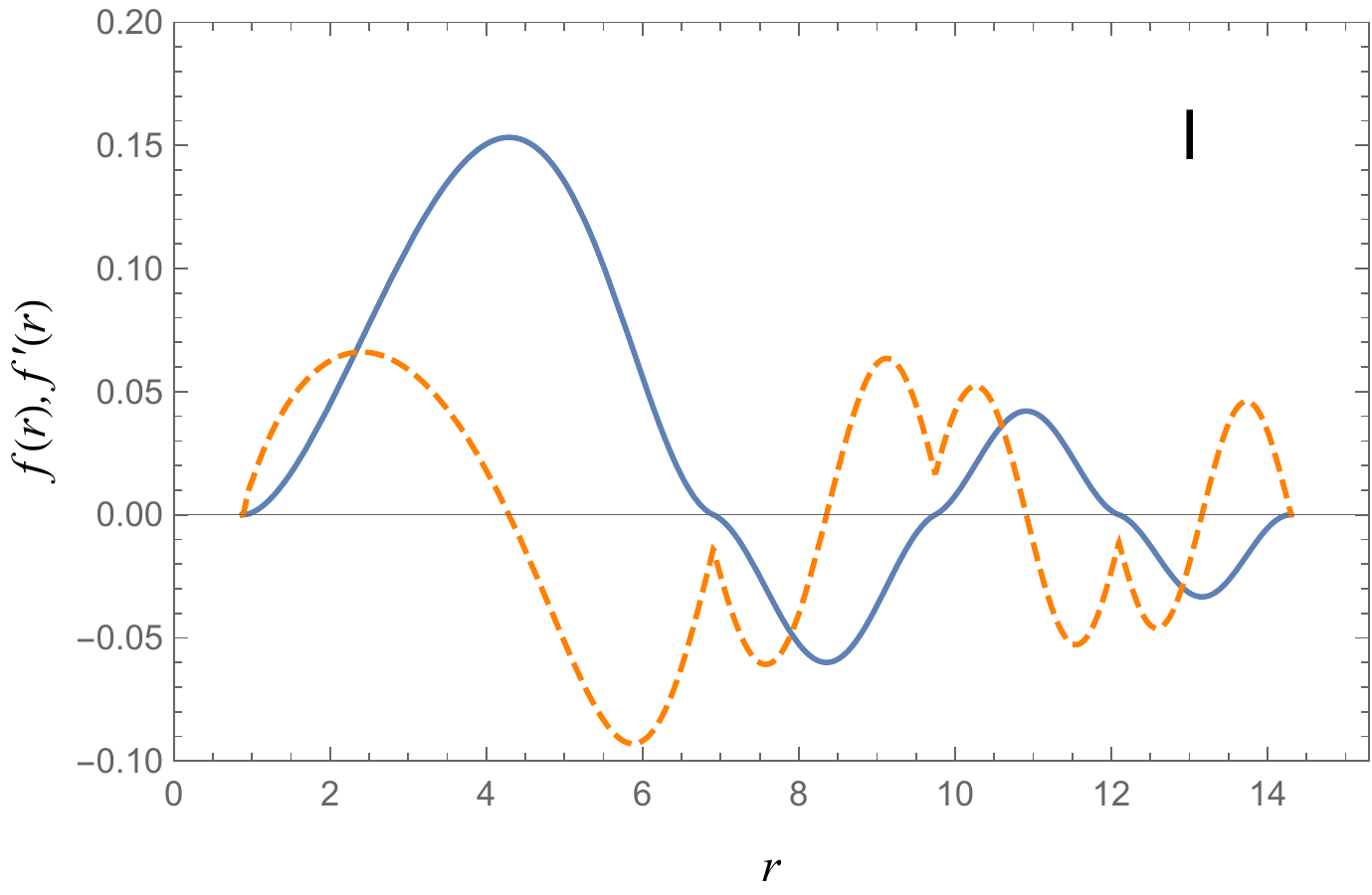}}~~
\subfigure[]{\includegraphics[width=0.45\textwidth, angle =0]{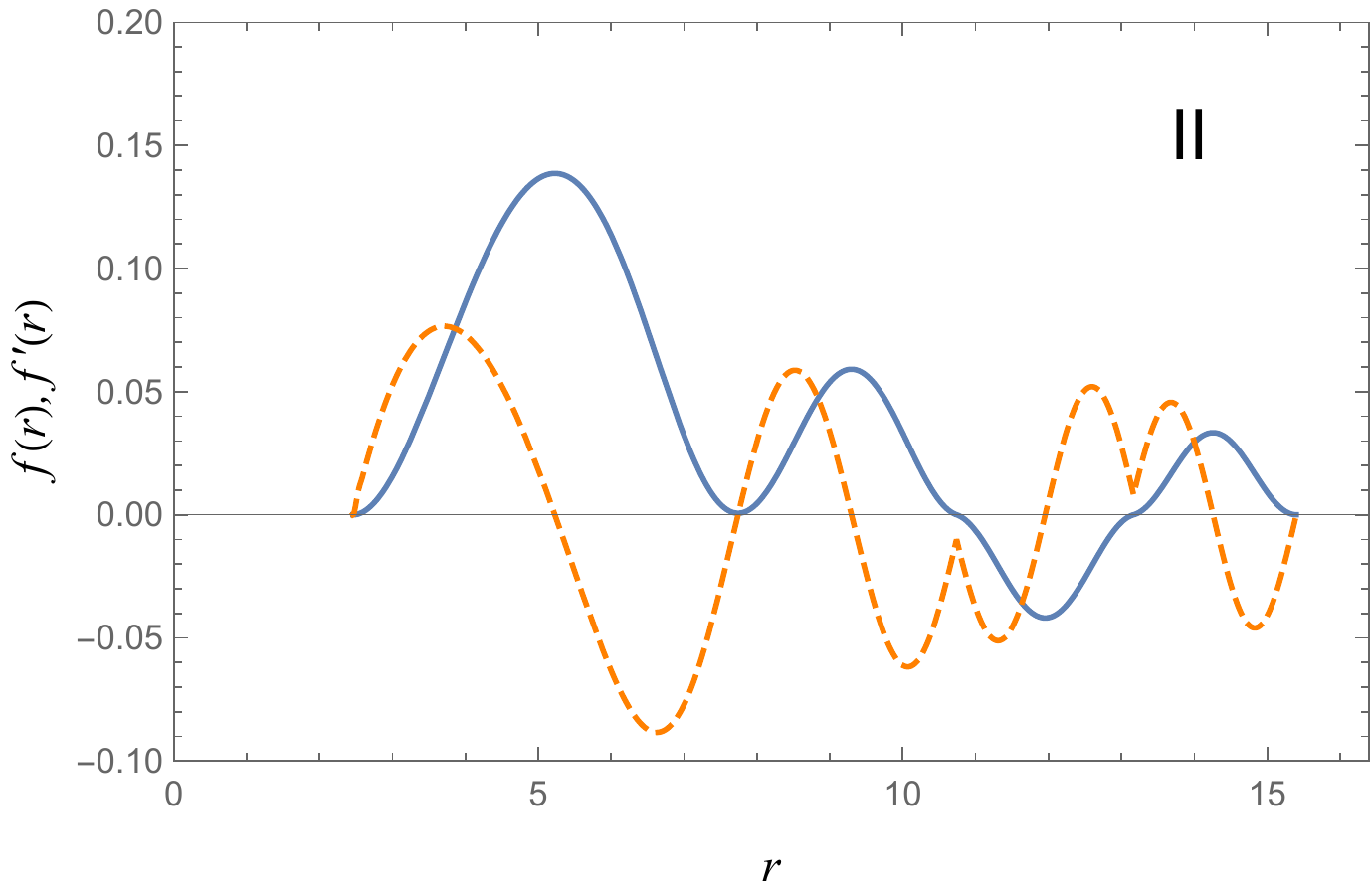}}\hskip0.5cm \\
\subfigure[]{\includegraphics[width=0.45\textwidth, angle =0]{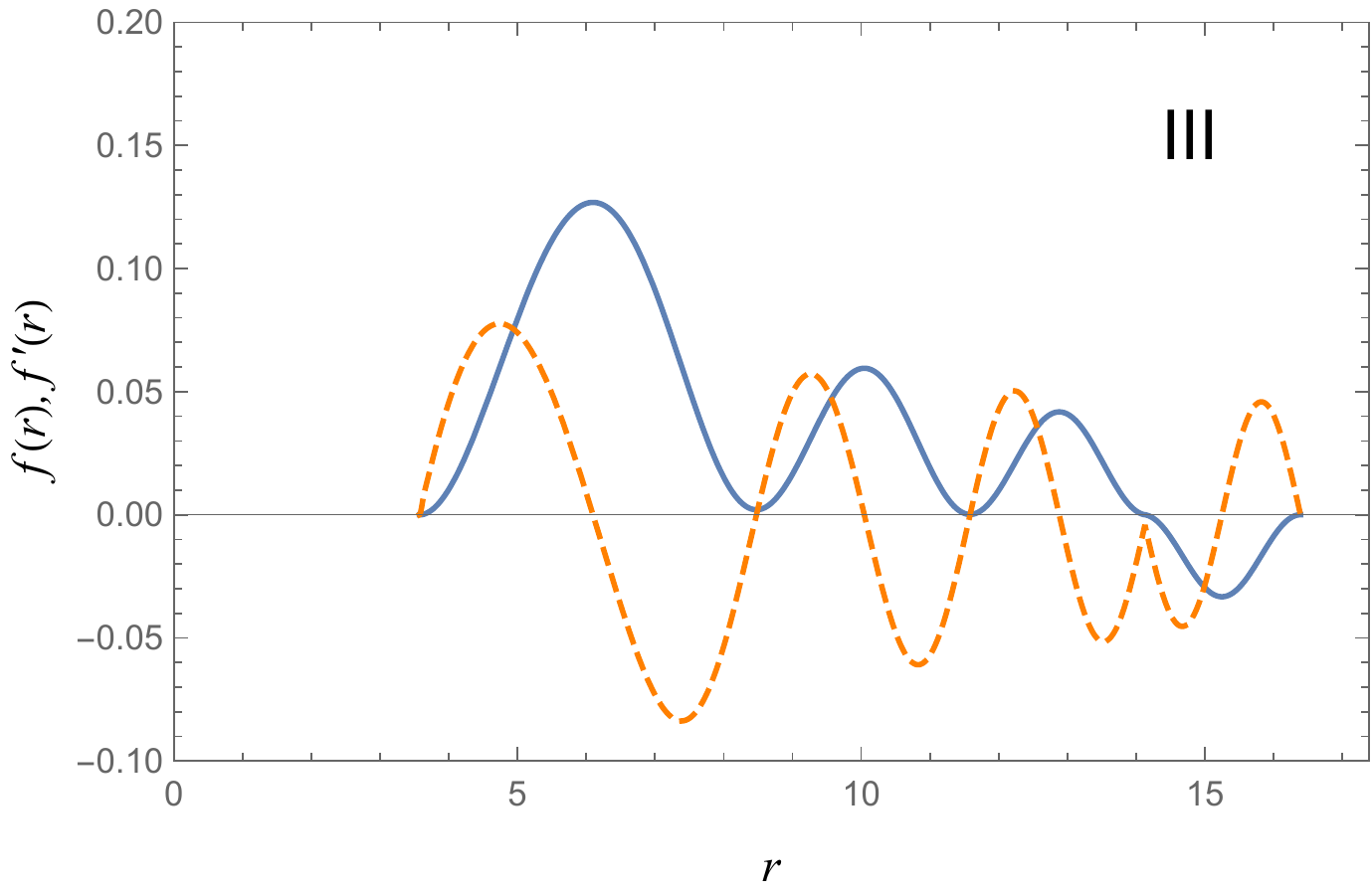}}~~
\subfigure[]{\includegraphics[width=0.45\textwidth, angle =0]{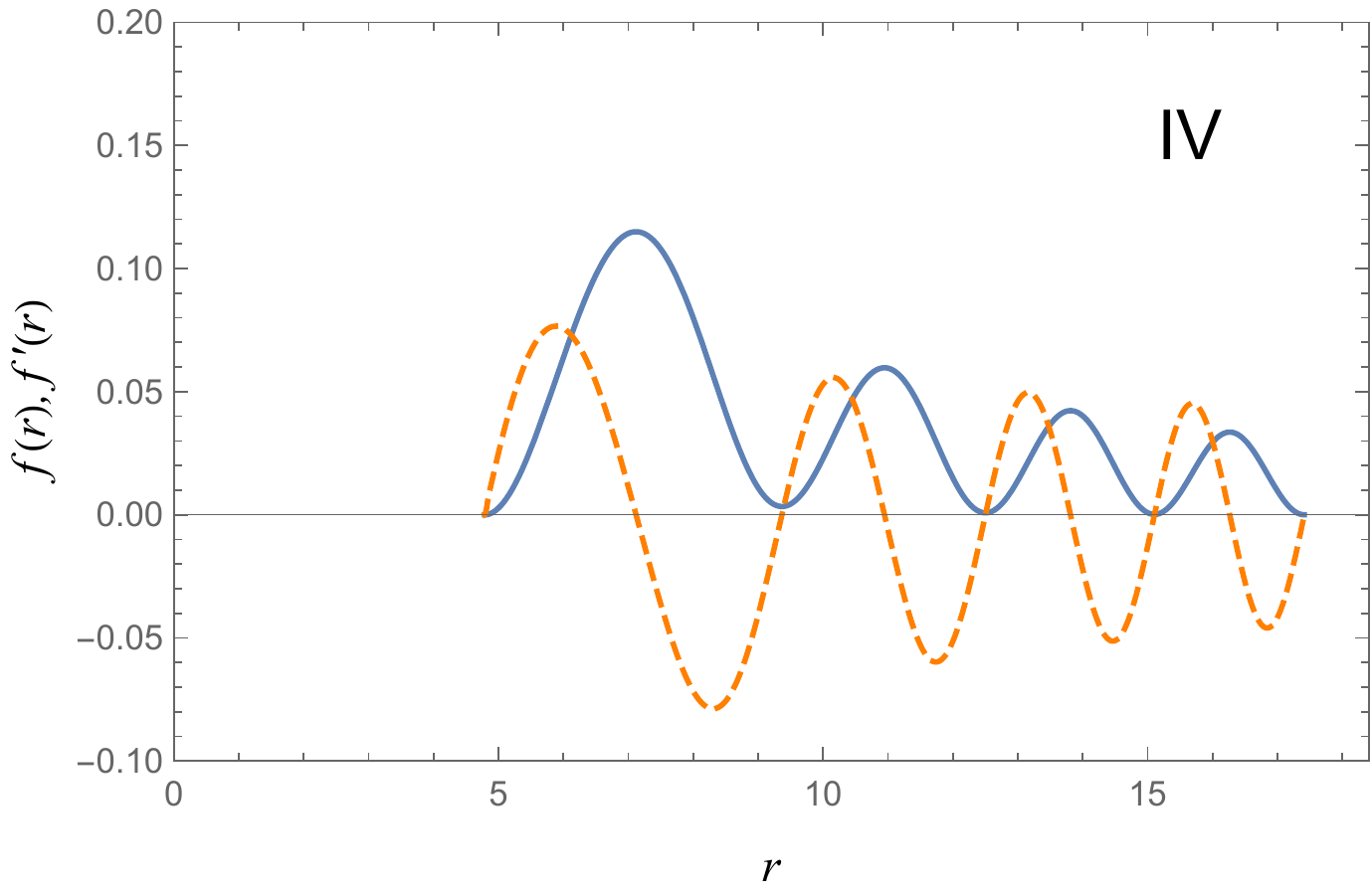}}\hskip0.5cm

   \caption{\label{gCP1Sp}(a) The gauged $\mathbb{C}P^{1}$ shell solutions. 
(a) The phase diagram in the space $R_{\rm in}/R_{\rm out}, b(R_{\rm in})$ with marked four regions.   
The shape of the profile function $f(r)$ changes qualitatively from I to IV as the radius of the shell increases. 
(b) In I $f(r)$ has three nodes,  (c) in II first node is transformed in a local minimum, 
(d) in III the second node is transformed in a local minimum and, finally, in IV the last third node changes into a local minimum. 
The profile function $f(r)$ in IV is always positive and at the last (fourth) minimum it satisfies $f(R)=f'(R)=0$. }
  \end{center}
\end{figure*}

\begin{figure*}[t]
  \begin{center}
\subfigure[]{\includegraphics[width=0.60\textwidth, angle =0]{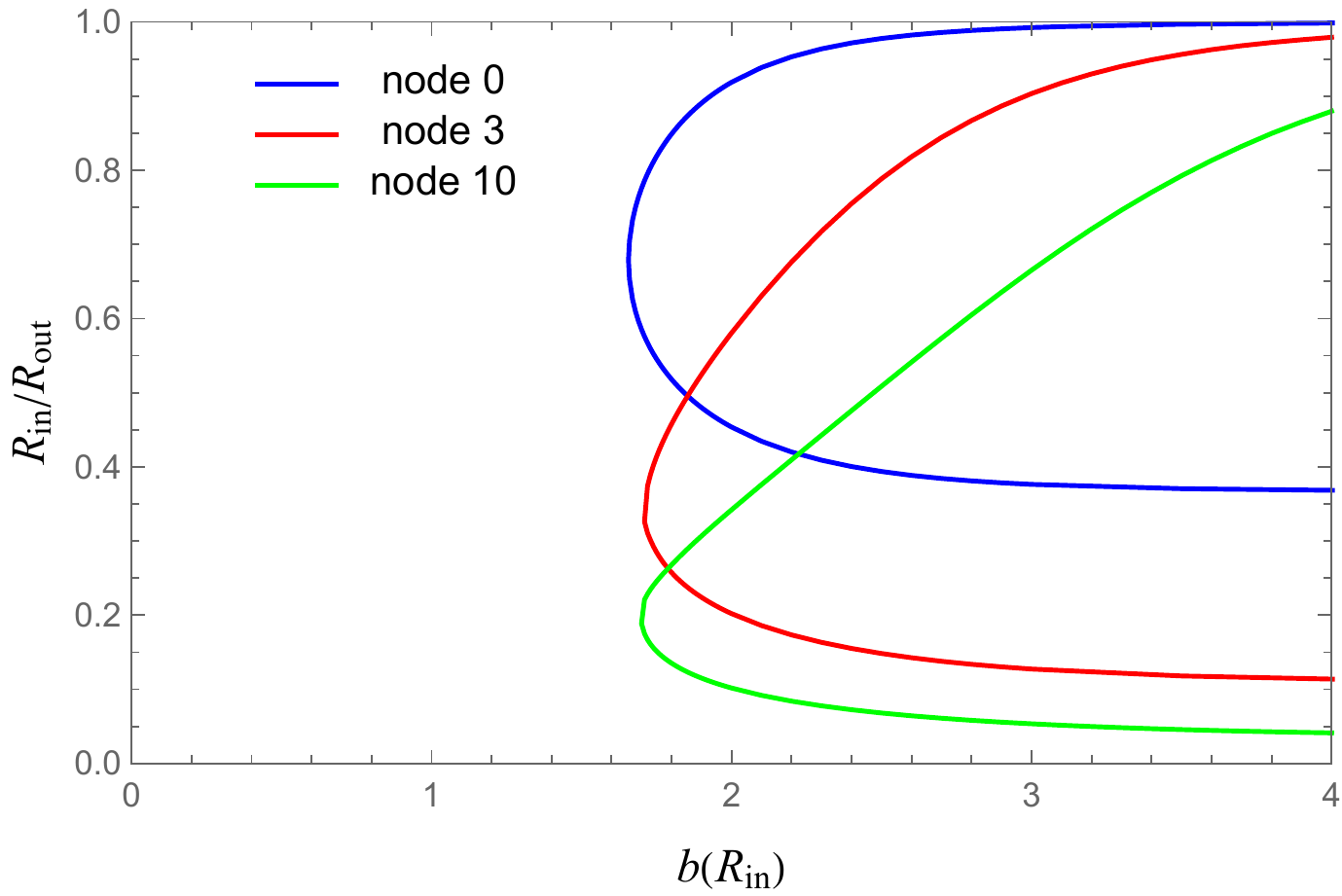}}\hskip0.5cm \\
\subfigure[]{\includegraphics[width=0.45\textwidth, angle =0]{gfCP11n3.pdf}}~~
\subfigure[]{\includegraphics[width=0.45\textwidth, angle =0]{gbCP11n3.pdf}}\hskip0.5cm \\
\subfigure[]{\includegraphics[width=0.45\textwidth, angle =0]{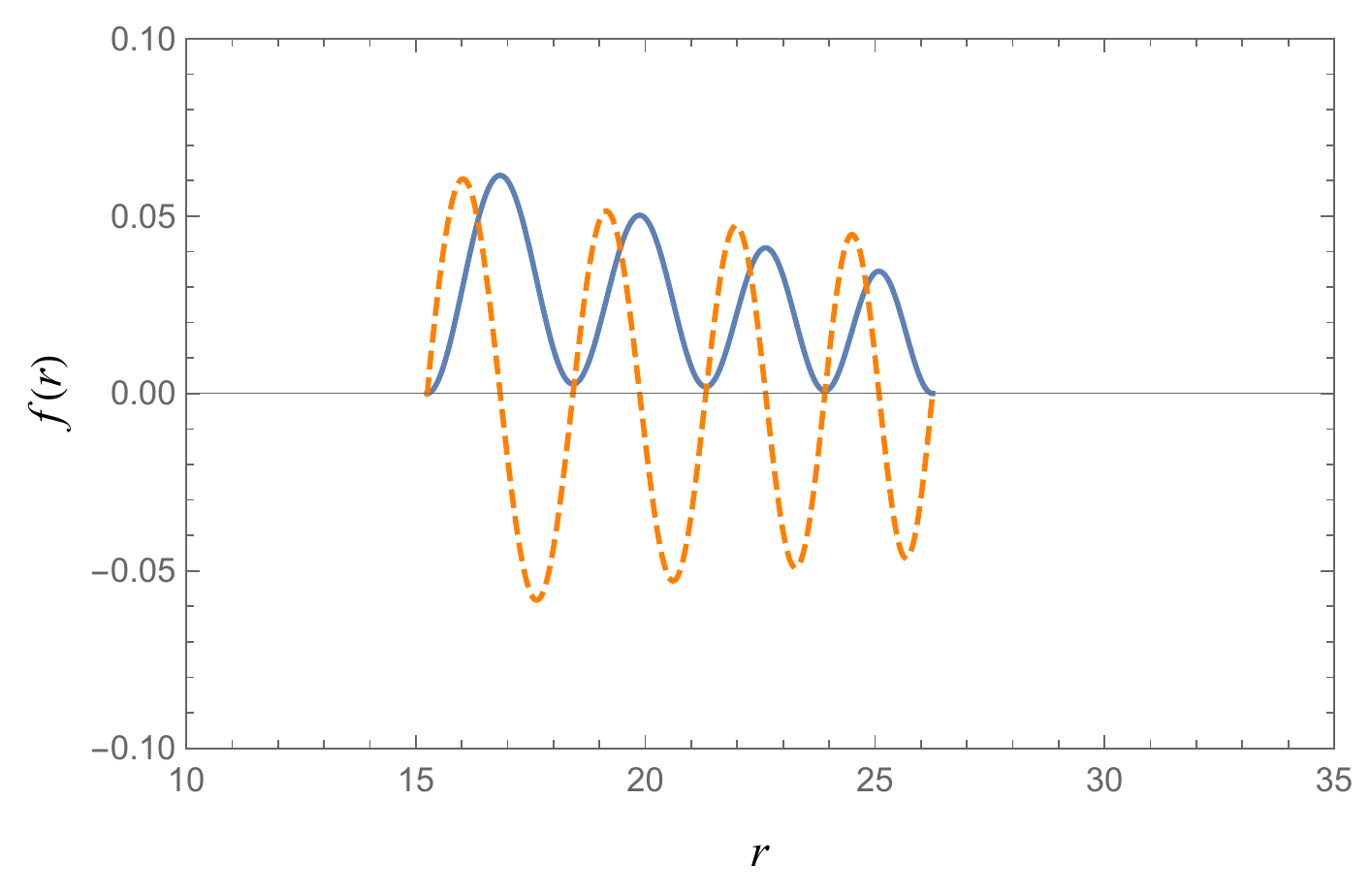}}~~
\subfigure[]{\includegraphics[width=0.45\textwidth, angle =0]{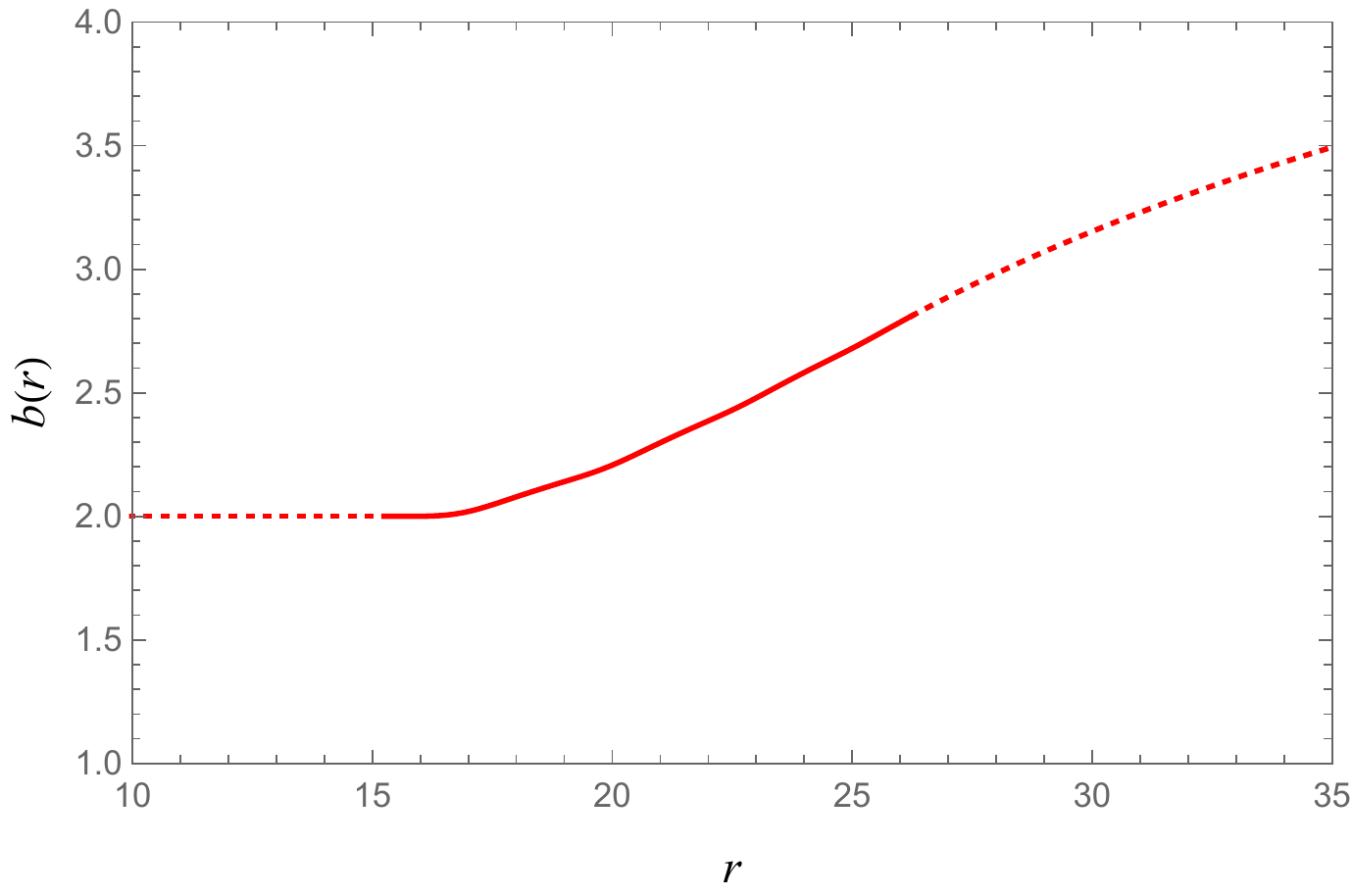}}\hskip0.5cm \\
 
   \caption{\label{gCP11phase} The phase diagram for $\mathbb{C}P^{11}$ $Q$-shells and the profile functions. 
(a) the ratio $R_{\rm in}/R_{\rm out}$ and $b(R_{\rm in})$, where the 
$R_{\rm in}$ is the inner radius and $R_{\rm out}$ is the outer radius. 
The gauge field function i $b(r)$ is taken at the inner compacton radius. 
(b),(c),(d),(e) show the profile functions for $b(R_{\rm in})=2.00$. 
Figures (b) and (c) correspond to the solutions at the lower region of 
(a) with $R_{\rm in}/R_{\rm out}=0.202073$. (d) and (e) shows the solutions for  upper region of  (a) with $R_{\rm in}/R_{\rm out}=0.580547$.}
  \end{center}
\end{figure*}

\begin{figure*}[t]
  \begin{center}
\subfigure[]{\includegraphics[width=0.45\textwidth, angle =0]{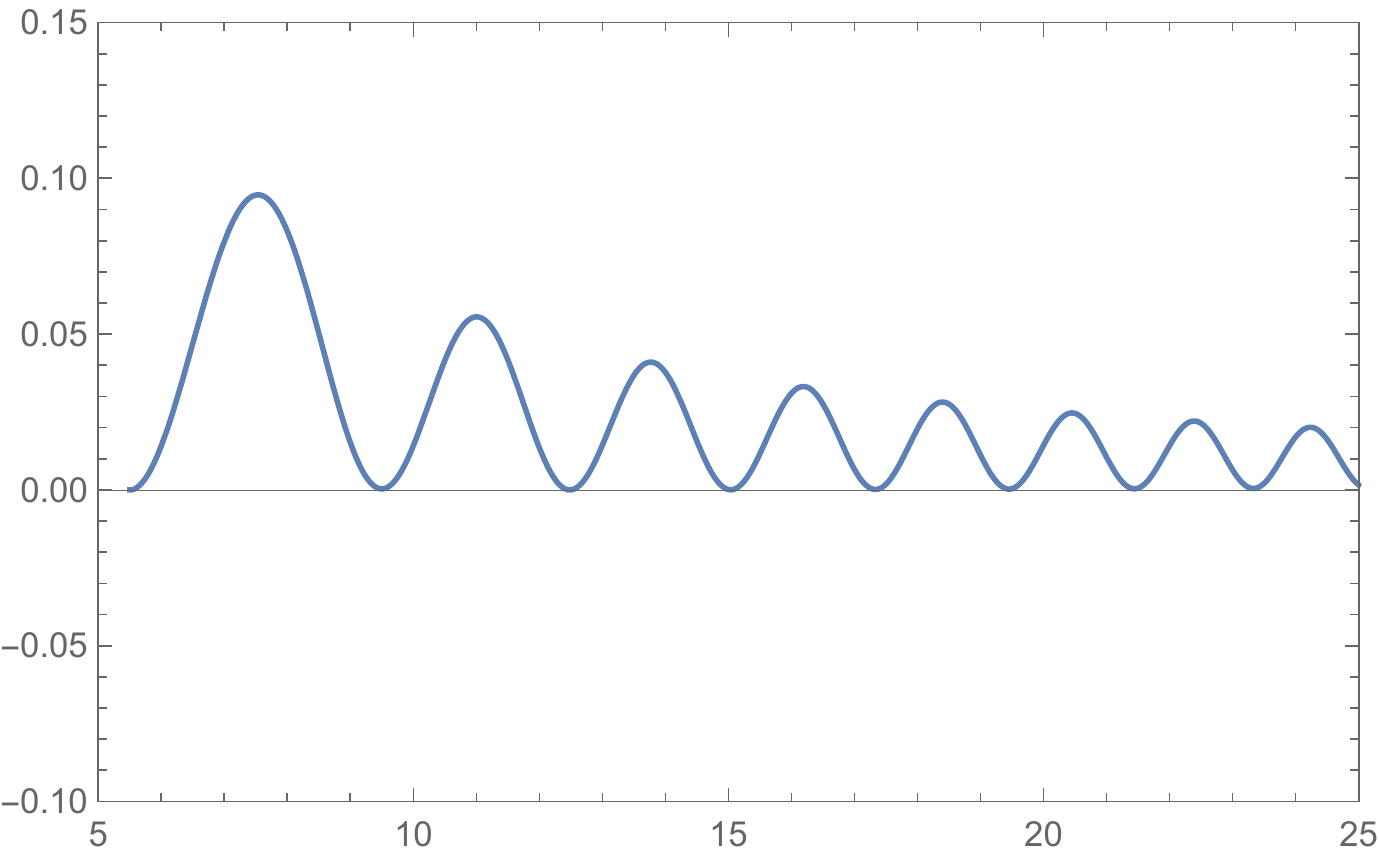}}\hskip0.5cm
\subfigure[]{\includegraphics[width=0.45\textwidth, angle =0]{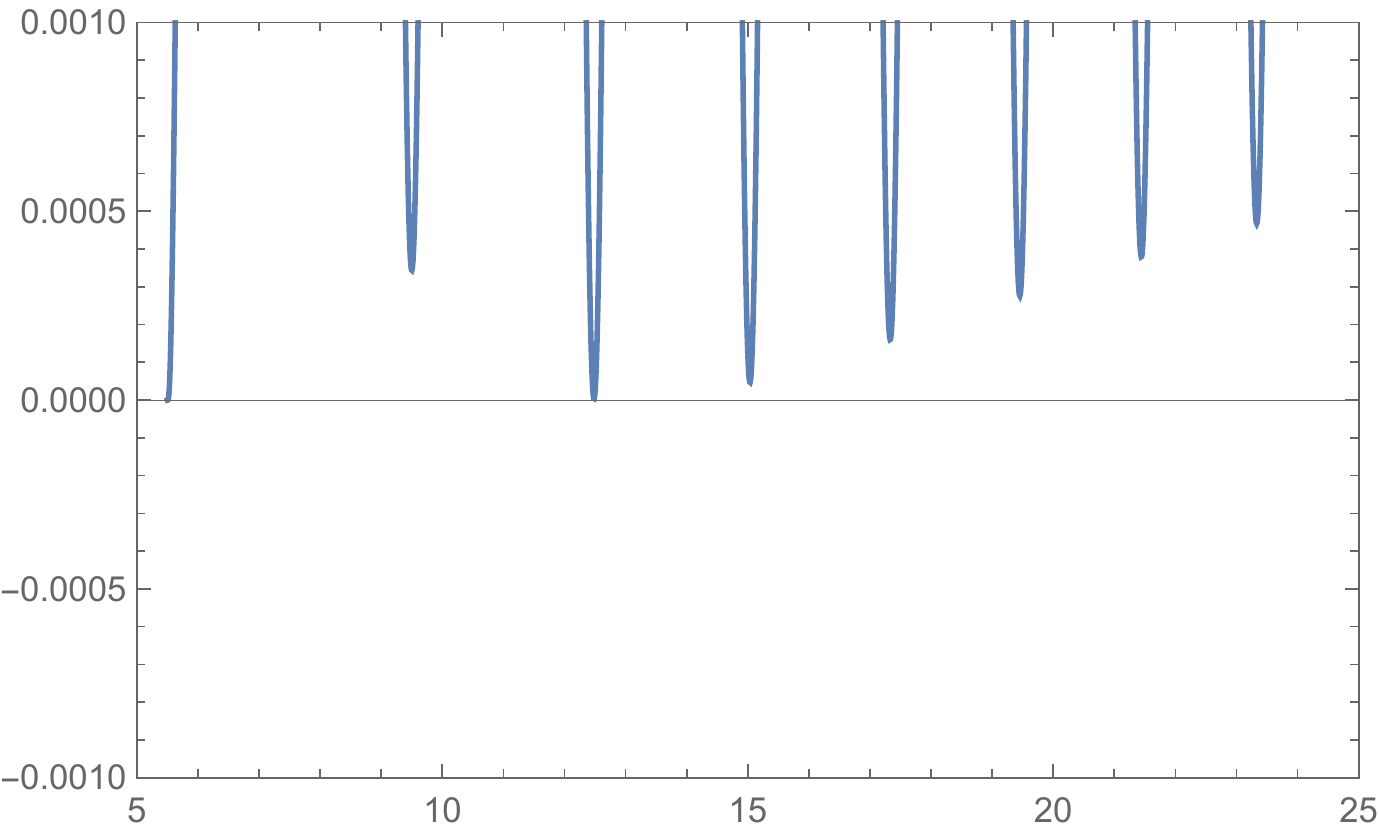}} \\

   \caption{\label{app} (a) The process of searching for the 3-node solution for the gauged $\mathbb{C}P^{11}$ model. 
(b) The blow up of the region containing local minima where $f(r)$ is close to zero. The second minimum is global hence 
one can expect that there is no \textit{2-Local Minima} solution.}
  \end{center}
\end{figure*}

\begin{figure*}[t]
  \begin{center}
\subfigure[]{\includegraphics[width=0.45\textwidth, angle =0]{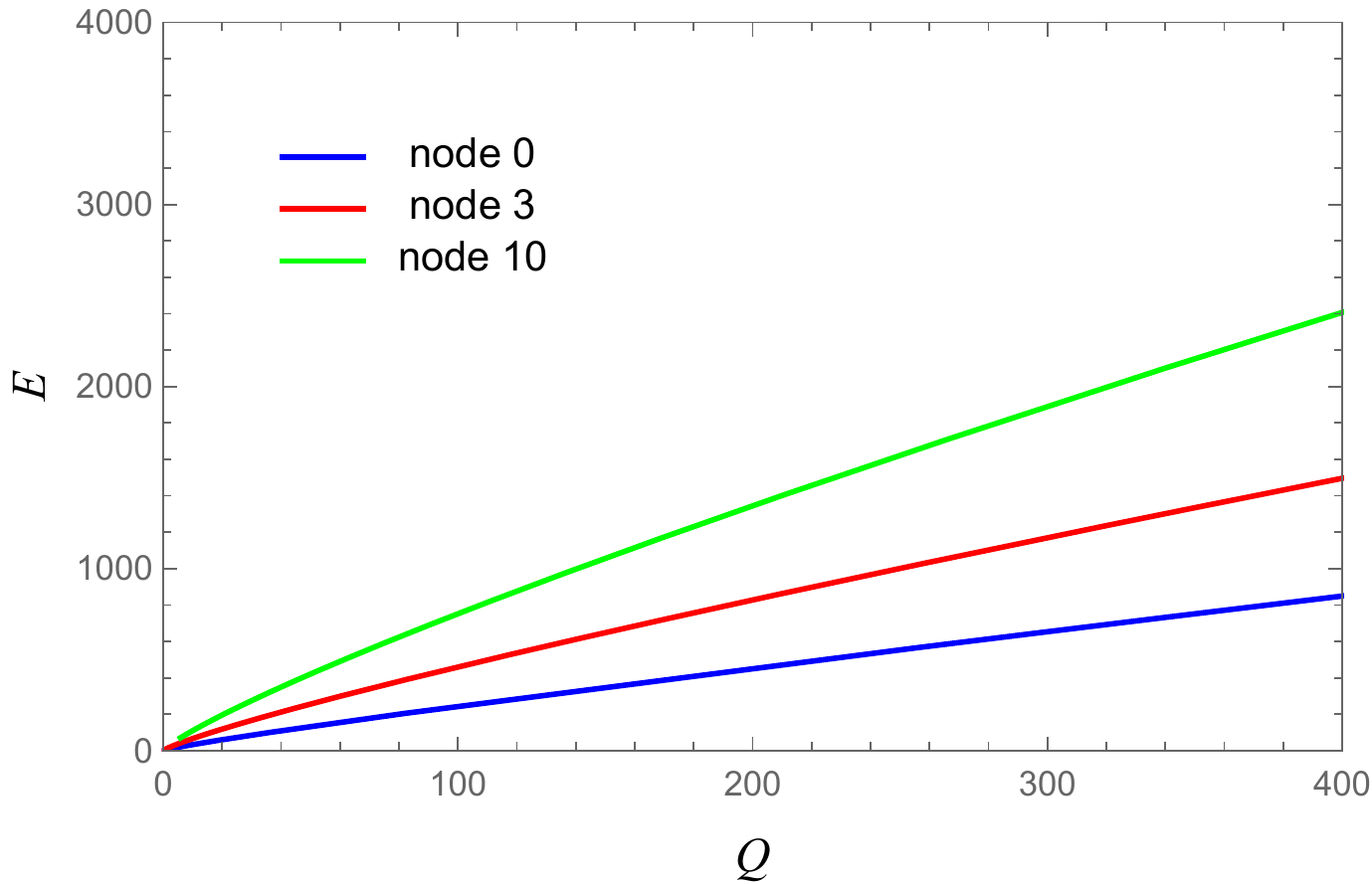}}\hskip0.5cm
\subfigure[]{\includegraphics[width=0.45\textwidth, angle =0]{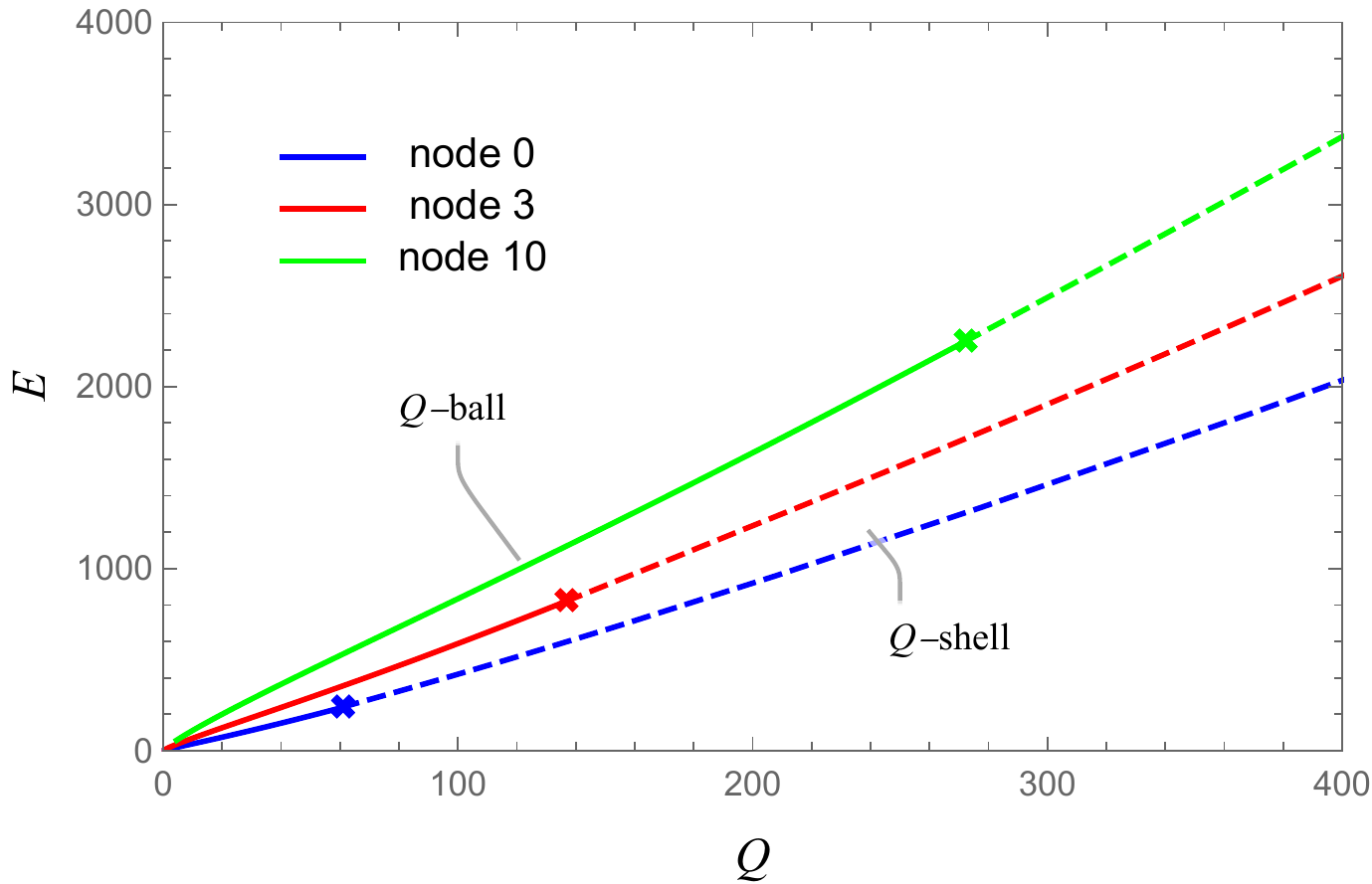}} \\
\subfigure[]{\includegraphics[width=0.45\textwidth, angle =0]{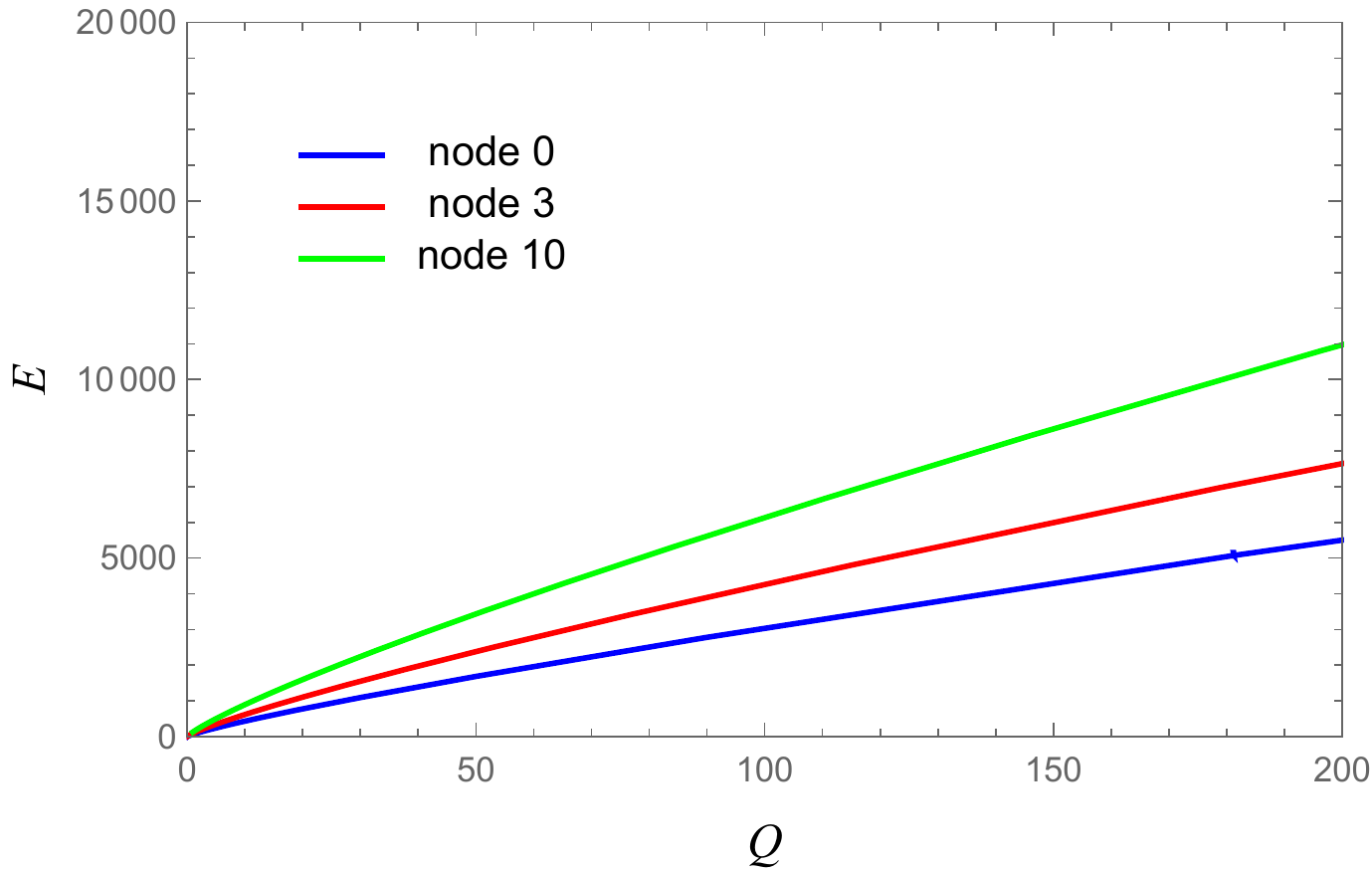}}\hskip0.5cm 
\subfigure[]{\includegraphics[width=0.45\textwidth, angle =0]{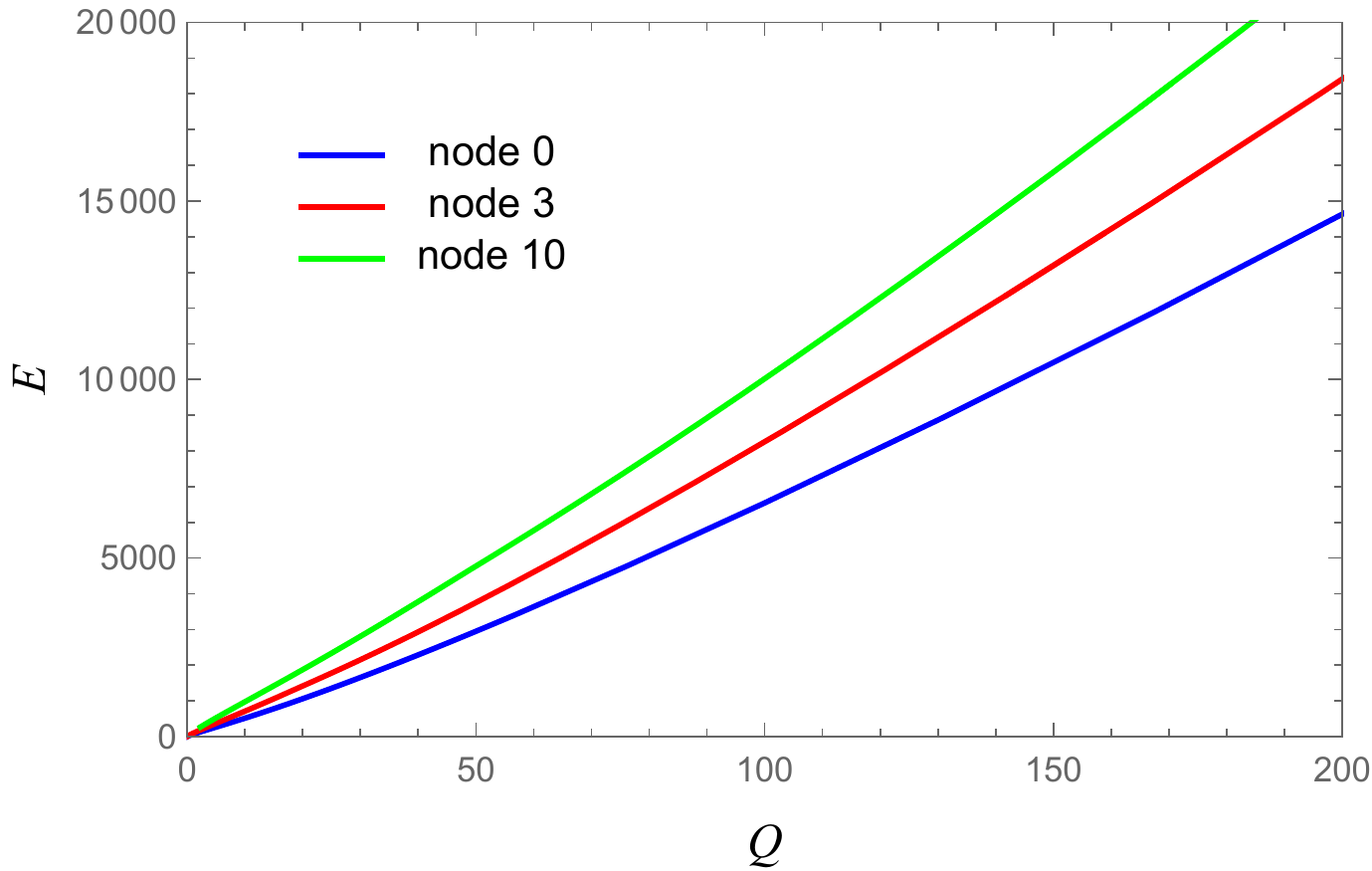}}

   \caption{\label{QE} The relation between $E$ and $Q$ for (a)(b) the $\mathbb{C}P^{1}$ model 
and (c)(d) the $\mathbb{C}P^{11}$ model. (a)(c) correspond with the case of $e=0$ (non-gauged) and 
(b)(d) with the case of $e=1$ (gauged).}
  \end{center}
\end{figure*}

\begin{figure*}[t]
  \begin{center}
\subfigure[]{\includegraphics[width=0.45\textwidth, angle =0]{gfCP11n3.pdf}}~~
\subfigure[]{\includegraphics[width=0.45\textwidth, angle =0]{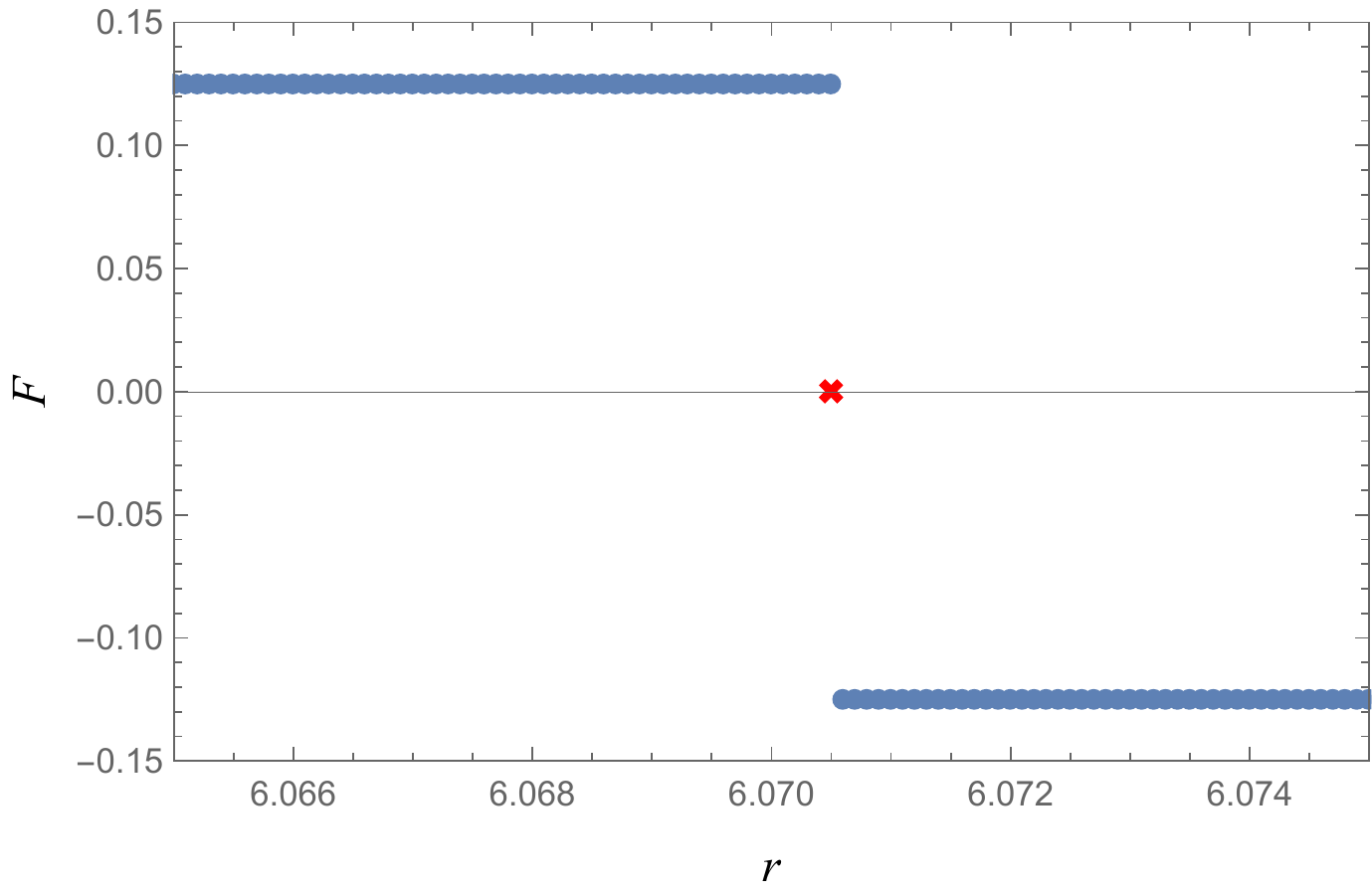}}
 
   \caption{\label{eqconti} (a) The 3-node solution for the $\mathbb{C}P^{11}$ model. 
(b) The (\ref{eq:wop}) evaluated on  the 3-node $\mathbb{C}P^{11}$  solution considered around 
the first nodal point. The values  before and after the nodal point are based on numerical solutions 
whereas the $F=0$, marked by the cross, is imposed.}
  \end{center}
\end{figure*}

\begin{table}[t]
\caption{The value $\alpha$ of fitting $E$ and $Q$ to $E \propto Q^{\alpha}$}
  \label{tab1}
    \centering
\begin{tabular}{lcccc}\hline\hline
node & ~~$\mathbb{C}P^{1}$~~ & ~~gauged $\mathbb{C}P^{1}$~~ & ~~ $\mathbb{C}P^{11}$~~& ~~gauged $\mathbb{C}P^{11}$~~ \\ \hline
0 & 0.895428 & 1.03395(ball)~1.17393(shell) & 0.865907 & 1.16475  \\
3 & 0.850909 & 1.03136(ball)~1.16898(shell) & 0.848827 & 1.16702 \\
10 & 0.840726 & 1.03522(ball)~1.17056(shell) & 0.837085 & 1.15618 \\ \hline\hline
\end{tabular}
\end{table}

\begin{figure}[h]
\tblcaption{The value of minima for each nodal points for the gauged $\mathbb{C}P^{11}$ model.}
\vspace{-0.5cm}
\begin{tabular}{lr}
  \begin{minipage}{.40\textwidth}
    \begin{center}
        \label{}
    \centering
\begin{tabular}{ccc}\hline\hline
node & $r = R_{n}$ & $f(R_{n})$ \\ \hline
0 & 15.73203014 & 0.00348041 \\
1 & 18.63543816 & 0.00306324 \\
2 & 21.24303051 & 0.00218077 \\
3 & 23.61426409 & 0.00147956 \\
4 & 25.80128460 & 0.000988057 \\
5 & 27.84273945 & 0.000650335 \\
6 & 29.76672287 & 0.000417063 \\
7 & 31.59387591 & 0.000254239 \\
8 & 33.33973251 & 0.000139506 \\
9 & 35.01621869 & 0.000058170 \\
10 & 36.63268890 & $4.01175 \times 10^{-7}$ \\ \hline\hline
\end{tabular}
    \end{center}
  \end{minipage}
\hspace{0cm}
  \begin{minipage}{.40\textwidth}
\vspace{4cm}

	\centering
\includegraphics[width=1.0\textwidth, angle =0]{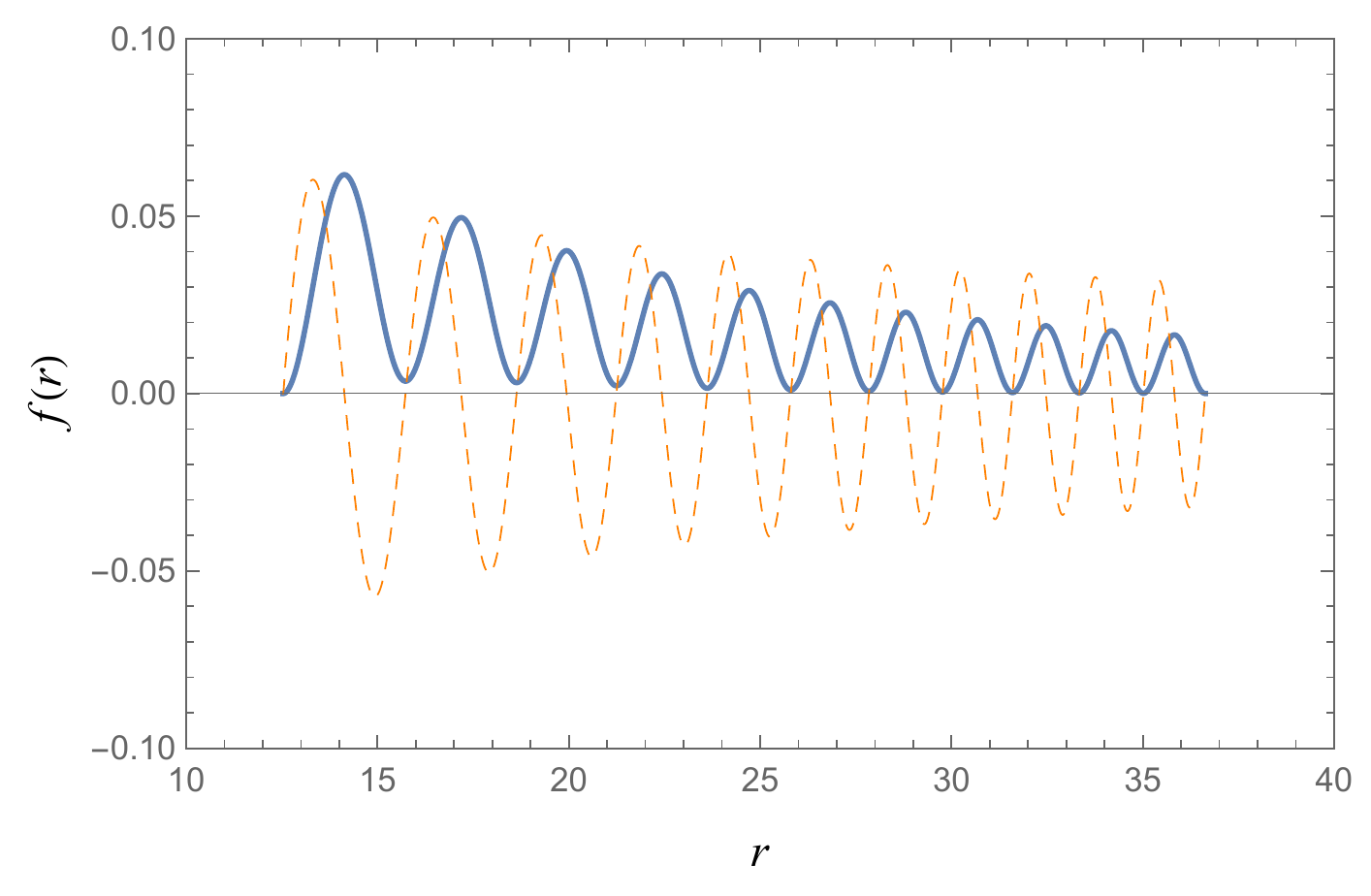}
  \end{minipage}

\end{tabular}
\end{figure}

\section{Solutions}\label{secondmodel}

\subsection{Some analytical solutions in the signum-Gordon model}

A complicated structure of solutions inevitably requires numerical analysis. 
Before proceeding with the numerical analysis, however, it is worth studying some analytical 
solutions of the signum-Gordon model which is related to the non-gauged $\mathbb{C}P^N$ model 
within the small field limit $f,f'\to 0$, see~\cite{Klimas:2017eft}. 
In Fig. \ref{ANw}, we show the convergence of  the solutions of the $\mathbb{C}P^N$ model to analytical  solution. This convergence is obtained by gradual changing of the parameter $\omega$.
For increasing $\omega$, the amplitude of the solutions became smaller and the $\mathbb{C}P^{N}$ solution successively tends to the signum-Gordon solution. 
Here, we examine the nodal, analytical solutions of the signum-Gordon model. 
Applying the ansatz \eqref{ansatzcpn}, one gets the dimensionless radial Euler-Lagrange equation of the signum-Gordon model 
\begin{align}
f''+\frac{2}{r}f'+ \omega^2 f-\frac{n(n+1)}{r^2}f-\lambda\, \textrm{Sign}(f)=0.\label{radialsg}
\end{align}
in order to  simplify the discussion, we consider the case  $n=0$. 
The equation \eqref{radialsg} has the following partial solutions with  constant sign
\begin{align}
f_k(r)=\pm\frac{\lambda}{\omega^2}+A_kj_0(\omega r)+B_kn_0(\omega r).
\end{align}
where $j_0,n_0$ are the zeroth-order spherical Bessel functions. 
The plus/minus sign corresponds to $\textrm{Sign}(f)=\pm 1$ in the equation.
This solution must be regular at $r=0$.
In particular, the nodeless compacton with $f(r)$ restricted to non-negative values is of the form  
\begin{align}
f_0(r) = \lambda \left(1-  \frac{j_0 (\omega r)}{j_0(\omega r_0)}\right).
\label{solutionSG}
\end{align}
The solution satisfies the compacton condition $f(r_0)=0, f'(r_0)=0$, where $\omega r_0\equiv x_1^1\sim 4.49341$.
We denote by  $x_1^1$ the first root of the spherical Bessel function i.e. $j_1(x_1^1)=0$. For {\it the single node solution},
we cover the compacton support  with two partial solutions $(f_0,f_1)$, each with different sign. Such a construction of solutions that consists on  \textit{patches} is a very typical situation in analysis of models with non-analytic potentials.  By assumption, our compacton consists of two nontrivial patches with signs $(\textrm{Sign}(f_0),\textrm{Sign}(f_1))=(+1,-1)$. 
The compacton condition (\ref{compact}) for $f_1$, the free coefficients are as follows
\begin{align}
&A_1=\frac{\lambda}{\omega^2}\frac{n_0'(\omega R)}{W[j_0(\omega R),n_0(\omega R)]}
\nonumber \\
&B_1=-\frac{\lambda}{\omega^2}\frac{j_0'(\omega R)}{W[j_0(\omega R),n_0(\omega R)]}
\end{align}
where the Wronskian $W$ is defined as $W[F,G]\equiv FG'-F'G$. As a result, the solution becomes 
\begin{align}
&f_1(r)=-\frac{\lambda}{\omega^2}\left(1+ \frac{1}{W[j_0(\omega R),n_0(\omega R)]}
\Bigl\{j_0(\omega r)n_0'(\omega R)-j_0'(\omega R)n_0(\omega r)\Bigr\} \right)
\end{align}
The matching point (sphere) $r=r_1$ is neither special (as it would be the light-cone surface) and thus the solution must be continuous and smooth in its first derivative 
\begin{align}
f_0(r_1) = f_1(r_1)=0,~~f_1'(r_1)=f_0'(r_1).
\end{align}
Consequently, the first partial solution $f_0$ has the form of \eqref{solutionSG} with $r_0$ replaced by $r_1$.

The two parameters $r_1, R$ can be fixed by the conditions at the compacton border $f_1(R)=0, f'_1(R)=0$, 
however, due to the complexity of the algebraic equations they can be determined only numerically.  
Note, that in order to solve two equations at the compacton border the parameter $r_1$ must be fine-tuned together with $R$.

The solutions with the higher number of nodes and/or the higher $n$ can be formally obtained in a similar way. 
Certainly, it is a tedious task and effectively only determination of a whole solution made of analytical pieces requires numerical analysis in finding solutions of algebraic equations. We conclude that the full numerical study is a more reasonable choice for a better understanding of the 
whole nature for a large class of the present model. The analytical approach is certainly important for testing numerical solutions in a small amplitude field limit. On the other hand, only numerical solutions can give us an answer what is the form of the profile functions $f(r)$ for a wide range of amplitudes.

\subsection{The $Q$-ball/$Q$-shell}

We first consider the case of non-gauged, excited $Q$-ball/shell solutions, which is obtained by setting $e=0$ and $b(r) = \omega$ for
the gauge function.

Fig.\ref{shoot} shows few shooting curves for the $\mathbb{C}P^{1}$ case. 
We gradually change the free parameter $f_0\equiv f(0)$, which equals to the profile function $f(r)$ at the origin and look for such its value 
that $f(r)$ satisfies the compacton condition (\ref{compact}) at its first local minimum. The shooting curves for zero node solution are sketched in Fig.\ref{shoot}(a),(b) whereas in Fig.\ref{shoot}(c),(d) we show the case of a single node solution. The spike in the first derivative of $f(r)$ is associated with discontinuity of $f''(r)$. This discontinuity is expected because the first derivative of the potential is not continuous at the point where $f(r)$ has its first zero.  According to Fig.\ref{shoot}(d) the solution exists, i.e. it is possible to fine-tune the free parameter in the way that at the next maximum of $f(r)$ the profile function satisfies the compacton condition (\ref{compact}).

In Fig.\ref{nongauge} we plot the three node matter profile function $f(r)$ and corresponding energy density $H(r)$ for $\mathbb{C}P^{1}$ and $\mathbb{C}P^{11}$. The first derivative $f'(r)$ has discontinuities at the points where the profile function $f(r)$ changes its sign, 
because the sign of the potential derivative flips.

\subsection{The gauged $Q$-ball/$Q$-shell}
\label{gECQ}

For the gauged solutions, we set $e=1$. It was pointed out in~\cite{Arodz:2008nm} that for a large charge $Q$ 
the $Q$-ball decays into the $Q$-shell because of electrical repulsion.
We shall show modifications of this behavior in the case of our nodal Q-ball solution.

Here we look at solutions of $\mathbb{C}P^{1}$ model containing three and ten nodes. We plot 
the matter profile function and its derivative $f(r),f'(r)$ as well as the gauge function $b(r)$. 
Fig.\ref{fCP1gaugeb} shows the ball solutions whereas Fig.\ref{fCP1gauges} shows the shell-like solutions. 
The function $f'(r)$ has spikes at the points where the profile function $f(r)$ changes its sign.  On the other hand the gauge function 
is continuous at these points.

We also plot the $\mathbb{C}P^{11}$ $Q$-shell in Fig.\ref{fCP11gauges}.  In this case, only the shell solutions exist. The presented solutions have three and ten nodes. Fig. \ref{gCP1phase}, shows the phase diagram of $\mathbb{C}P^{1}$ solutions. 
Fig. \ref{gCP1phase}(a) represents the phase diagram in the space of parameters $f(0)$ and $b(0)$ that are values of 
the profile function and gauge function at the origin $r=0$. 
In Fig. \ref{gCP1phase}(b), we plot the phase diagram of the ratio of inner and
outer shell radii $R_{\rm in}/R_{\rm out}$ versus the gauge field function $b(r)$ at the inner radius $b(R_{\rm in})$.
This diagram indicates that the nodal Q-ball is smoothly connected to the Q-shell with the same number of nodes.

An interesting feature associated with the form of the profile function is observed in the case of shell solutions.
The three node $\mathbb{C}P^{1}$ solutions for several values of $b(R_{\rm in})$ is presented in the Fig.\ref{gCP1Sp}. 
As increasing the charge (or the radius), the shape of the solutions change from Fig.\ref{gCP1Sp}(b) to (e). In Fig.\ref{gCP1Sp}(a) we show the phase diagram divided in regions with qualitatively different form of the profile function $f(r)$. 
In the region I: Fig.\ref{gCP1Sp}(b), the function $f(r)$ switches its sign three times - it has three internal zeros. At fourth zero the function $f(r)$ satisfies the compacton condition (\ref{compact}).

When $b(R_{\rm in})$ grows then the solution changes from I to II which means that the first nodal point is replaced by a local minimum. 
The profile function at this minimum is strictly positive. We shall call it a solution with one internal local minimum 
\textit{1 local minimum}: $f^{[1\ell m]}(r)$ (the other local minima are zeros - internal and external radius of the compacton).

When increasing further the parameter, the solution changes from type II to type III.  It manifests in the transformation of the second node into a local minimum localized above the axis $r$. The profile function with this property has \textit{2 local minima} and it is denoted by $f^{[2\ell m]}(r)$. The function $f(r)$ satisfies the compacton boundary condition at its last local minimum. 
Finally, for sufficiently high $b(R_{\rm in})$ the solution belongs to the class IV i.e. it has the \textit{3 local minima}: $f^{[3\ell m]}(r)$. Such a solution is strictly positive on the whole support of the compacton except its border where it takes the zero value. We have observed transitions of nodes into local minima only for the gauged model.

In Fig.\ref{gCP11phase} we show the phase diagram for the $\mathbb{C}P^{11}$ model.
We found that there are some solutions with (internal) local minima. When smoothly changing parameters such solutions appear from solutions containing a certain number of nodes. An example of such behavior is shown in
Fig.\ref{gCP11phase} where (a) shows that the phase diagram corresponds with three types of nodal solutions.  
We look in detail at the curve which represents three nodal solutions. Taking for instance  $b(R_{\rm in})=2.0$ we see that there are two points 
 $R_{\rm in}/R_{\rm out}=\{0.202073,0.580547\}$ that corresponds to chosen value of $b(R_{\rm in})$. 
For smaller value, the solution has three nodes (three internal zeros). Such solution is shown in  Fig.\ref{gCP11phase}(b)(c). 
On the other hand, for the bigger value of the ratio the solution has a different form -- it is non-negative and instead of internal zeros 
the profile curve contains local minima that lie above the $r$ axis. This solution is plotted in Fig.\ref{gCP11phase}(d)(e). 

For both $Q$-ball and $Q$-shell type solutions, there are no solutions that have simultaneously the internal local minima and nodes.
In the case of $Q$-shell, the solution with an arbitrary number of local minima may not exist. In Fig.\ref{app} we show the example of such a situation. When searching for three-node solutions for $\mathbb{C}P^{11}$, we see that
the sign flips in a way that the \textit{2 local minima} does not appear. It happens because  
the second nodal point has smaller value than the third one. 
Consequently, there are no $Q$-shells with \textit{2 local minima} and three nodal points for the case $\mathbb{C}P^{11}$.

\subsection{The energy-charge scaling}
\label{Stability}

The energy $E$ and the charge $Q$ in our $\mathbb{C}P^{N}$model obey a certain scaling relation, namely 
 $E \propto Q^{5/6}$ or $E \propto Q^{7/6}$ \cite{Klimas:2017eft, Sawado:2020ncc, Sawado:2021rsc}. 
In this section, we study the stability of the nodal solutions based on these scaling relations.
Fig.\ref{QE} shows the relation between $E$ and $Q$ for the non-gauged ($e=0$) and gauged ($e=1$) solutions in models $\mathbb{C}P^{1}$ and $\mathbb{C}P^{11}$.
For fixed value of the charge $Q$ the energy  $E$ of the solution with the higher number of nodes is bigger than the energy of the solution with a smaller number of nodes. When looking at the energy density we see that there is no significant change of these functions even when the profile function varies significantly as for instance for different types of solutions corresponding listed in Fig.\ref{gCP1Sp}, I-IV. 
This clearly means that the solutions I-IV belong to the same class.

Table.\ref{tab1} shows the results of fitting $E$ and $Q$ to the relation $E \propto Q^{\alpha}$. 
We can see that the more nodes the solutions have, the smaller value  the parameter $\alpha$ takes.

\section{Further discussion}

In this section, we shall give a comment on discontinuities associated with the change of the sign of the profile $f$ in Eq.(\ref{eq:f}). 
In particular, we are interested in the fact whether it has some consequences on the energy density. 
Continuity of the energy density requires that both $f$ and $f'$ are continuous. We do not expect discontinuity in gauge function $b(r)$ because the Maxwell equations do not contain any signum function. In this section, we study how the behavior of the radial function $f(r)$ around the nodal points.

First, we numerically check the continuity condition. From the equation (\ref{eq:f}), we define a function
\begin{align}
F(r) := f''(r) + \frac{2}{r}f'(r) - \frac{n(n+1)f(r)}{r^2} + \frac{(1-f(r)^2)b^2f(r)}{1+f(r)^2}-\frac{2f(r)f'(r)^2}{1+f(r)^2}. \label{eq:wop}
\end{align}
which is equal to l.h.s. of the matter field equation minus the potential derivative. 
Then we substitute a solution into (\ref{eq:wop}) and evaluate $F(r)$. 
In Fig.\ref{eqconti} we plot $F(r)$ in vicinity of the nodal point $f\sim 0$. For the correct solution, 
$F(r)$ should behave like the signum function. This is exactly what we can see in Fig.\ref{eqconti}(b).
At both open segments separated by the nodal point, $F(r)$ is continuous, hence we impose $F(r)=0$ at the nodal point.  With this appropriate definition, one gets the signum function. Note, that from a physical point of view $F(r)=0$ at a certain segment of space is consistent with the vacuum solution $f(r)=0$. This simple analysis proves that $f(r)$ is the solution of (\ref{eq:f}).
Next, we examine the series expansion of solutions at both sides of the nodal point $r=R_{\mathrm{nd}}$
\begin{align}
f(r) = \sum_{k=0}^{\infty}F_{k}(R_{\mathrm{nd}}-r)^k,~~~b(r) = \sum_{k=0}^{\infty}B_{k}(R_{\mathrm{nd}}-r)^k.
\end{align}
Since the value of $f$ is zero at the nodal points, $F_{0}=0$,  then the expansions can be written as
\begin{align}
f_{+}(r) &= F_{1}^{+}(r-R_{\mathrm{nd}}) + \left(\frac{F_{1}^{+}}{R_{\mathrm{nd}}}+\frac{1}{16} \right)(r-R_{\mathrm{nd}})^2 + O((r-R_{\mathrm{nd}})^3) \\
f_{-}(r) &= F_{1}^{-}(r-R_{\mathrm{nd}}) + \left(\frac{F_{1}^{-}}{R_{\mathrm{nd}}}-\frac{1}{16} \right)(r-R_{\mathrm{nd}})^2 + O((r-R_{\mathrm{nd}})^3).
\end{align}
 $f_{+}$ stands for the expansion where $f(r)$ is positive, and $f_{-}$ stands for the expansion where $f(r)$ is negative.
Thus, even when the first-order coefficients match, $F_1^{+} = F_{1}^{-}$, the left and right second-order coefficients (and further)  
are different from each other. 

The point is, therefore, whether the first-order derivatives coincide or not. In the case of their equality, the energy density, which contains the field and its first derivative, becomes continuous (but not necessarily differentiable). 

This is what our numerical results suggest. 
From Eq.(\ref{eq:f}), one can verify indirectly that the profile $f$ is regular (i.e., continuous and differentiable) 
at the point of $f \sim 0$. Equation (\ref{eq:f}) can be cast in the form
\begin{align}
\frac{1}{r^2}\frac{d}{dr}\left( r^2 f' \right) = \frac{n(n+1)f}{r^2} - \frac{(1-f^2)b^2f}{(1+f^2)} + \frac{2ff'^2}{(1+f^2)} + \frac{1}{8}{\mathrm{Sign}}(f) \sqrt{1+f^2}.
\end{align}
Integrating over the segment $[r_{-}, r_{+}]$ that contains the nodal point $r=R_{\mathrm{nd}} : f(r) = 0, r_{-} < r < r_{+}$, 
one gets
\begin{align}
&\Bigl[ r^2 f' \Bigr]_{r_{-}}^{r_{+}} = \int_{r_{-}}^{r_{+}} \Omega (f,f') r^2 dr
\end{align}
Most of  terms in $\Omega$ contain $f$ which means that the integral containing such terms 
vanishes in the limit $r_{\pm} \rightarrow R_{\mathrm{nd}}$.
It is enough to examine the term Sign$(f)$. In  vicinity $r \rightarrow R_{\mathrm{nd}}$, 
the integrand signum function is an odd functional, hence the value of the integral is expected to vanish. 
We conclude that the first derivative $f'$ on the zero crossing point is continuous.

\section{Conclusions}\label{sec6}

We have presented in this paper several nodal compact $Q$-ball ($Q$-shell) solutions in the $\mathbb{C}P^{N}$ nonlinear sigma models. 
There are an infinite number of radii that satisfy the compacton condition such that the field configuration is zero at the compacton radius. They differ by the number of nodes. For a given solution we choose the number of  nodes.
The presence of the signum function in the field equation is a consequence of a V-shaped potential in the Lagrangian. The nodal solutions are such that the profile function changes its sign. Each point where it takes place is a node of the solution. At the last point (radius) the field satisfies the compacton condition. This is the compacton border where the field matches the vacuum solution $f=0$.
We have obtained new solutions with a  given number of nodes for both non-gauged and gauged models.

When smoothly varying parameters in the gauged case we observe that the nodal $Q$-ball smoothly connects the $Q$-shell with the same number of nodes.  
For the gauged solutions we have observed that with increasing the charge (or the radius) the nodal $Q$-shell transforms into the field configuration with some \textit{local minima} which substitute the nodes. Such minima are localized under the $r$ axis. In other words, the profile function has no sign change for such configurations.
We have denoted it by \textit{k local minimum}: $f^{[k-\ell m]}(r)$.

We have looked also at the energy and charge for nodal compactons. Our results show that for fixed Noether charge  $Q$ the energy of the solution $E$ increases together with the number of nodes. Looking at the energy and charge densities of $Q$-balls or $Q$-shells we found that they do not change qualitatively in their form even when the nodal solution transforms into  the solutions with some \textit{local minima}. This clearly means that the $n$-th $Q$-ball,  and $n$-th $Q$-shell with \textit{k-local minima} belong to the same class.

The first derivative of the profile function $f'$ is non-differentiable at the nodal points. It results in the appearance of discontinuity of the second derivative of $f$. We have checked that under continuity of $f'$ at the nodal points the energy and the charge are continuous function, even for nodal solutions. These conditions were investigated both numerically and analytically.

Our new solution has possible applications to boson stars with non-trivial excitations just by implementing gravity into the equation. This solution can be seen as the weak gravity limit of true gravitating boson star solutions.
Extending a class of solutions by the inclusion of excitations we naturally offer a variety of solutions that can be useful for the description of several astronomical phenomena. Alternatively, our results can also be applied to a phenomenon of evaporation of $Q$-balls and to the production of fermions.
These models will be discussed in our subsequent papers in order.

\begin{center}
{\bf Acknowledgment}
\end{center}

The authors would like to thank Yves Brihaye who gave the initial birth of this peculiar setup of the solutions.
We also appreciate Yuki Amari, Luiz Agostinho Ferreira, Atsushi Nakamula and Kouichi Toda for valuable discussions. 
Discussions during the YITP workshop YITP-W-20-03 on ``Strings and Fields 2020'' 
and YITP workshop YITP-W-21-04 on ``Strings and Fields 2021'' 
have been useful to complete this work. 
N.S. was supported in part by JSPS KAKENHI Grant Number JP B20K03278(1).

\bibliography{excitedQballs}

\begin{thebibliography}{30}%
\makeatletter
\providecommand \@ifxundefined [1]{%
 \@ifx{#1\undefined}
}%
\providecommand \@ifnum [1]{%
 \ifnum #1\expandafter \@firstoftwo
 \else \expandafter \@secondoftwo
 \fi
}%
\providecommand \@ifx [1]{%
 \ifx #1\expandafter \@firstoftwo
 \else \expandafter \@secondoftwo
 \fi
}%
\providecommand \natexlab [1]{#1}%
\providecommand \enquote  [1]{``#1''}%
\providecommand \bibnamefont  [1]{#1}%
\providecommand \bibfnamefont [1]{#1}%
\providecommand \citenamefont [1]{#1}%
\providecommand \href@noop [0]{\@secondoftwo}%
\providecommand \href [0]{\begingroup \@sanitize@url \@href}%
\providecommand \@href[1]{\@@startlink{#1}\@@href}%
\providecommand \@@href[1]{\endgroup#1\@@endlink}%
\providecommand \@sanitize@url [0]{\catcode `\\12\catcode `\$12\catcode
  `\&12\catcode `\#12\catcode `\^12\catcode `\_12\catcode `\%12\relax}%
\providecommand \@@startlink[1]{}%
\providecommand \@@endlink[0]{}%
\providecommand \url  [0]{\begingroup\@sanitize@url \@url }%
\providecommand \@url [1]{\endgroup\@href {#1}{\urlprefix }}%
\providecommand \urlprefix  [0]{URL }%
\providecommand \Eprint [0]{\href }%
\providecommand \doibase [0]{http://dx.doi.org/}%
\providecommand \selectlanguage [0]{\@gobble}%
\providecommand \bibinfo  [0]{\@secondoftwo}%
\providecommand \bibfield  [0]{\@secondoftwo}%
\providecommand \translation [1]{[#1]}%
\providecommand \BibitemOpen [0]{}%
\providecommand \bibitemStop [0]{}%
\providecommand \bibitemNoStop [0]{.\EOS\space}%
\providecommand \EOS [0]{\spacefactor3000\relax}%
\providecommand \BibitemShut  [1]{\csname bibitem#1\endcsname}%
\let\auto@bib@innerbib\@empty
\bibitem [{\citenamefont {Arodz}\ and\ \citenamefont
  {Lis}(2008)}]{Arodz:2008jk}%
  \BibitemOpen
  \bibfield  {author} {\bibinfo {author} {\bibfnamefont {H.}~\bibnamefont
  {Arodz}}\ and\ \bibinfo {author} {\bibfnamefont {J.}~\bibnamefont {Lis}},\
  }\bibfield  {title} {\enquote {\bibinfo {title} {{Compact Q-balls in the
  complex signum-Gordon model}},}\ }\href {\doibase 10.1103/PhysRevD.77.107702}
  {\bibfield  {journal} {\bibinfo  {journal} {Phys. Rev. D}\ }\textbf {\bibinfo
  {volume} {77}},\ \bibinfo {pages} {107702} (\bibinfo {year} {2008})},\
  \Eprint {http://arxiv.org/abs/0803.1566} {arXiv:0803.1566 [hep-th]}
  \BibitemShut {NoStop}%
\bibitem [{\citenamefont {Arodz}\ and\ \citenamefont
  {Lis}(2009)}]{Arodz:2008nm}%
  \BibitemOpen
  \bibfield  {author} {\bibinfo {author} {\bibfnamefont {H.}~\bibnamefont
  {Arodz}}\ and\ \bibinfo {author} {\bibfnamefont {J.}~\bibnamefont {Lis}},\
  }\bibfield  {title} {\enquote {\bibinfo {title} {{Compact Q-balls and
  Q-shells in a scalar electrodynamics}},}\ }\href {\doibase
  10.1103/PhysRevD.79.045002} {\bibfield  {journal} {\bibinfo  {journal} {Phys.
  Rev. D}\ }\textbf {\bibinfo {volume} {79}},\ \bibinfo {pages} {045002}
  (\bibinfo {year} {2009})},\ \Eprint {http://arxiv.org/abs/0812.3284}
  {arXiv:0812.3284 [hep-th]} \BibitemShut {NoStop}%
\bibitem [{\citenamefont {Arodz}(2002)}]{Arodz:2002yt}%
  \BibitemOpen
  \bibfield  {author} {\bibinfo {author} {\bibfnamefont {H.}~\bibnamefont
  {Arodz}},\ }\bibfield  {title} {\enquote {\bibinfo {title} {{Topological
  compactons}},}\ }\href@noop {} {\bibfield  {journal} {\bibinfo  {journal}
  {Acta Phys. Polon. B}\ }\textbf {\bibinfo {volume} {33}},\ \bibinfo {pages}
  {1241--1252} (\bibinfo {year} {2002})},\ \Eprint
  {http://arxiv.org/abs/nlin/0201001} {arXiv:nlin/0201001} \BibitemShut
  {NoStop}%
\bibitem [{\citenamefont {Arodz}(2004)}]{Arodz:2003mx}%
  \BibitemOpen
  \bibfield  {author} {\bibinfo {author} {\bibfnamefont {H.}~\bibnamefont
  {Arodz}},\ }\bibfield  {title} {\enquote {\bibinfo {title} {{Symmetry
  breaking transition and appearance of compactons in a mechanical system}},}\
  }\href@noop {} {\bibfield  {journal} {\bibinfo  {journal} {Acta Phys. Polon.
  B}\ }\textbf {\bibinfo {volume} {35}},\ \bibinfo {pages} {625--638} (\bibinfo
  {year} {2004})},\ \Eprint {http://arxiv.org/abs/hep-th/0312036}
  {arXiv:hep-th/0312036} \BibitemShut {NoStop}%
\bibitem [{\citenamefont {Arodz}\ \emph {et~al.}(2005)\citenamefont {Arodz},
  \citenamefont {Klimas},\ and\ \citenamefont {Tyranowski}}]{Arodz:2005gz}%
  \BibitemOpen
  \bibfield  {author} {\bibinfo {author} {\bibfnamefont {H.}~\bibnamefont
  {Arodz}}, \bibinfo {author} {\bibfnamefont {P.}~\bibnamefont {Klimas}}, \
  and\ \bibinfo {author} {\bibfnamefont {T.}~\bibnamefont {Tyranowski}},\
  }\bibfield  {title} {\enquote {\bibinfo {title} {{Field-theoretic models with
  V-shaped potentials}},}\ }\href@noop {} {\bibfield  {journal} {\bibinfo
  {journal} {Acta Phys. Polon. B}\ }\textbf {\bibinfo {volume} {36}},\ \bibinfo
  {pages} {3861--3876} (\bibinfo {year} {2005})},\ \Eprint
  {http://arxiv.org/abs/hep-th/0510204} {arXiv:hep-th/0510204} \BibitemShut
  {NoStop}%
\bibitem [{\citenamefont {Friedberg}\ \emph {et~al.}(1976)\citenamefont
  {Friedberg}, \citenamefont {Lee},\ and\ \citenamefont
  {Sirlin}}]{Friedberg:1976me}%
  \BibitemOpen
  \bibfield  {author} {\bibinfo {author} {\bibfnamefont {R.}~\bibnamefont
  {Friedberg}}, \bibinfo {author} {\bibfnamefont {T.~D.}\ \bibnamefont {Lee}},
  \ and\ \bibinfo {author} {\bibfnamefont {A.}~\bibnamefont {Sirlin}},\
  }\bibfield  {title} {\enquote {\bibinfo {title} {{A Class of Scalar-Field
  Soliton Solutions in Three Space Dimensions}},}\ }\href {\doibase
  10.1103/PhysRevD.13.2739} {\bibfield  {journal} {\bibinfo  {journal} {Phys.
  Rev.}\ }\textbf {\bibinfo {volume} {D13}},\ \bibinfo {pages} {2739--2761}
  (\bibinfo {year} {1976})}\BibitemShut {NoStop}%
\bibitem [{\citenamefont {Coleman}(1985)}]{Coleman:1985ki}%
  \BibitemOpen
  \bibfield  {author} {\bibinfo {author} {\bibfnamefont {Sidney~R.}\
  \bibnamefont {Coleman}},\ }\bibfield  {title} {\enquote {\bibinfo {title} {{Q
  Balls}},}\ }\href {\doibase 10.1016/0550-3213(85)90286-X,
  10.1016/0550-3213(86)90520-1} {\bibfield  {journal} {\bibinfo  {journal}
  {Nucl. Phys.}\ }\textbf {\bibinfo {volume} {B262}},\ \bibinfo {pages} {263}
  (\bibinfo {year} {1985})},\ \bibinfo {note} {[Erratum: Nucl.
  Phys.B269,744(1986)]}\BibitemShut {NoStop}%
\bibitem [{\citenamefont {Friedberg}\ \emph {et~al.}(1987)\citenamefont
  {Friedberg}, \citenamefont {Lee},\ and\ \citenamefont
  {Pang}}]{Friedberg:1986tq}%
  \BibitemOpen
  \bibfield  {author} {\bibinfo {author} {\bibfnamefont {R.}~\bibnamefont
  {Friedberg}}, \bibinfo {author} {\bibfnamefont {T.~D.}\ \bibnamefont {Lee}},
  \ and\ \bibinfo {author} {\bibfnamefont {Y.}~\bibnamefont {Pang}},\
  }\bibfield  {title} {\enquote {\bibinfo {title} {{Scalar Soliton Stars and
  Black Holes}},}\ }\href {\doibase 10.1103/PhysRevD.35.3658} {\bibfield
  {journal} {\bibinfo  {journal} {Phys. Rev.}\ }\textbf {\bibinfo {volume}
  {D35}},\ \bibinfo {pages} {3658} (\bibinfo {year} {1987})}\BibitemShut
  {NoStop}%
\bibitem [{\citenamefont {Lee}(1987)}]{Lee:1986ts}%
  \BibitemOpen
  \bibfield  {author} {\bibinfo {author} {\bibfnamefont {T.~D.}\ \bibnamefont
  {Lee}},\ }\bibfield  {title} {\enquote {\bibinfo {title} {{Soliton Stars and
  the Critical Masses of Black Holes}},}\ }\href {\doibase
  10.1103/PhysRevD.35.3637} {\bibfield  {journal} {\bibinfo  {journal} {Phys.
  Rev.}\ }\textbf {\bibinfo {volume} {D35}},\ \bibinfo {pages} {3637} (\bibinfo
  {year} {1987})}\BibitemShut {NoStop}%
\bibitem [{\citenamefont {Kusenko}(1997)}]{Kusenko:1997zq}%
  \BibitemOpen
  \bibfield  {author} {\bibinfo {author} {\bibfnamefont {Alexander}\
  \bibnamefont {Kusenko}},\ }\bibfield  {title} {\enquote {\bibinfo {title}
  {{Solitons in the supersymmetric extensions of the standard model}},}\ }\href
  {\doibase 10.1016/S0370-2693(97)00584-4} {\bibfield  {journal} {\bibinfo
  {journal} {Phys. Lett.}\ }\textbf {\bibinfo {volume} {B405}},\ \bibinfo
  {pages} {108} (\bibinfo {year} {1997})},\ \Eprint
  {http://arxiv.org/abs/hep-ph/9704273} {arXiv:hep-ph/9704273 [hep-ph]}
  \BibitemShut {NoStop}%
\bibitem [{\citenamefont {Kusenko}\ and\ \citenamefont
  {Shaposhnikov}(1998)}]{Kusenko:1997si}%
  \BibitemOpen
  \bibfield  {author} {\bibinfo {author} {\bibfnamefont {Alexander}\
  \bibnamefont {Kusenko}}\ and\ \bibinfo {author} {\bibfnamefont {Mikhail~E.}\
  \bibnamefont {Shaposhnikov}},\ }\bibfield  {title} {\enquote {\bibinfo
  {title} {{Supersymmetric Q balls as dark matter}},}\ }\href {\doibase
  10.1016/S0370-2693(97)01375-0} {\bibfield  {journal} {\bibinfo  {journal}
  {Phys. Lett.}\ }\textbf {\bibinfo {volume} {B418}},\ \bibinfo {pages}
  {46--54} (\bibinfo {year} {1998})},\ \Eprint
  {http://arxiv.org/abs/hep-ph/9709492} {arXiv:hep-ph/9709492 [hep-ph]}
  \BibitemShut {NoStop}%
\bibitem [{\citenamefont {Kusenko}\ \emph {et~al.}(1998)\citenamefont
  {Kusenko}, \citenamefont {Kuzmin}, \citenamefont {Shaposhnikov},\ and\
  \citenamefont {Tinyakov}}]{Kusenko:1997vp}%
  \BibitemOpen
  \bibfield  {author} {\bibinfo {author} {\bibfnamefont {Alexander}\
  \bibnamefont {Kusenko}}, \bibinfo {author} {\bibfnamefont {Vadim}\
  \bibnamefont {Kuzmin}}, \bibinfo {author} {\bibfnamefont {Mikhail~E.}\
  \bibnamefont {Shaposhnikov}}, \ and\ \bibinfo {author} {\bibfnamefont
  {P.~G.}\ \bibnamefont {Tinyakov}},\ }\bibfield  {title} {\enquote {\bibinfo
  {title} {{Experimental signatures of supersymmetric dark matter Q balls}},}\
  }\href {\doibase 10.1103/PhysRevLett.80.3185} {\bibfield  {journal} {\bibinfo
   {journal} {Phys. Rev. Lett.}\ }\textbf {\bibinfo {volume} {80}},\ \bibinfo
  {pages} {3185--3188} (\bibinfo {year} {1998})},\ \Eprint
  {http://arxiv.org/abs/hep-ph/9712212} {arXiv:hep-ph/9712212 [hep-ph]}
  \BibitemShut {NoStop}%
\bibitem [{\citenamefont {Klimas}\ and\ \citenamefont
  {Livramento}(2017)}]{Klimas:2017eft}%
  \BibitemOpen
  \bibfield  {author} {\bibinfo {author} {\bibfnamefont {P.}~\bibnamefont
  {Klimas}}\ and\ \bibinfo {author} {\bibfnamefont {L.~R.}\ \bibnamefont
  {Livramento}},\ }\bibfield  {title} {\enquote {\bibinfo {title} {{Compact
  Q-balls and Q-shells in CPN type models}},}\ }\href {\doibase
  10.1103/PhysRevD.96.016001} {\bibfield  {journal} {\bibinfo  {journal} {Phys.
  Rev.}\ }\textbf {\bibinfo {volume} {D96}},\ \bibinfo {pages} {016001}
  (\bibinfo {year} {2017})},\ \Eprint {http://arxiv.org/abs/1704.01132}
  {arXiv:1704.01132 [hep-th]} \BibitemShut {NoStop}%
\bibitem [{\citenamefont {Kleihaus}\ \emph {et~al.}(2010)\citenamefont
  {Kleihaus}, \citenamefont {Kunz}, \citenamefont {Lammerzahl},\ and\
  \citenamefont {List}}]{Kleihaus:2010ep}%
  \BibitemOpen
  \bibfield  {author} {\bibinfo {author} {\bibfnamefont {Burkhard}\
  \bibnamefont {Kleihaus}}, \bibinfo {author} {\bibfnamefont {Jutta}\
  \bibnamefont {Kunz}}, \bibinfo {author} {\bibfnamefont {Claus}\ \bibnamefont
  {Lammerzahl}}, \ and\ \bibinfo {author} {\bibfnamefont {Meike}\ \bibnamefont
  {List}},\ }\bibfield  {title} {\enquote {\bibinfo {title} {{Boson Shells
  Harbouring Charged Black Holes}},}\ }\href {\doibase
  10.1103/PhysRevD.82.104050} {\bibfield  {journal} {\bibinfo  {journal} {Phys.
  Rev.}\ }\textbf {\bibinfo {volume} {D82}},\ \bibinfo {pages} {104050}
  (\bibinfo {year} {2010})},\ \Eprint {http://arxiv.org/abs/1007.1630}
  {arXiv:1007.1630 [gr-qc]} \BibitemShut {NoStop}%
\bibitem [{\citenamefont {Kumar}\ \emph {et~al.}(2014)\citenamefont {Kumar},
  \citenamefont {Kulshreshtha},\ and\ \citenamefont
  {Shankar~Kulshreshtha}}]{Kumar:2014kna}%
  \BibitemOpen
  \bibfield  {author} {\bibinfo {author} {\bibfnamefont {Sanjeev}\ \bibnamefont
  {Kumar}}, \bibinfo {author} {\bibfnamefont {Usha}\ \bibnamefont
  {Kulshreshtha}}, \ and\ \bibinfo {author} {\bibfnamefont {Daya}\ \bibnamefont
  {Shankar~Kulshreshtha}},\ }\bibfield  {title} {\enquote {\bibinfo {title}
  {{Boson stars in a theory of complex scalar fields coupled to the U(1) gauge
  field and gravity}},}\ }\href {\doibase 10.1088/0264-9381/31/16/167001}
  {\bibfield  {journal} {\bibinfo  {journal} {Class. Quant. Grav.}\ }\textbf
  {\bibinfo {volume} {31}},\ \bibinfo {pages} {167001} (\bibinfo {year}
  {2014})},\ \Eprint {http://arxiv.org/abs/1605.07210} {arXiv:1605.07210
  [hep-th]} \BibitemShut {NoStop}%
\bibitem [{\citenamefont {Kumar}\ \emph {et~al.}(2015)\citenamefont {Kumar},
  \citenamefont {Kulshreshtha},\ and\ \citenamefont
  {Kulshreshtha}}]{Kumar:2015sia}%
  \BibitemOpen
  \bibfield  {author} {\bibinfo {author} {\bibfnamefont {Sanjeev}\ \bibnamefont
  {Kumar}}, \bibinfo {author} {\bibfnamefont {Usha}\ \bibnamefont
  {Kulshreshtha}}, \ and\ \bibinfo {author} {\bibfnamefont {Daya~Shankar}\
  \bibnamefont {Kulshreshtha}},\ }\bibfield  {title} {\enquote {\bibinfo
  {title} {{Boson stars in a theory of complex scalar field coupled to
  gravity}},}\ }\href {\doibase 10.1007/s10714-015-1918-0} {\bibfield
  {journal} {\bibinfo  {journal} {Gen. Rel. Grav.}\ }\textbf {\bibinfo {volume}
  {47}},\ \bibinfo {pages} {76} (\bibinfo {year} {2015})},\ \Eprint
  {http://arxiv.org/abs/1605.07015} {arXiv:1605.07015 [hep-th]} \BibitemShut
  {NoStop}%
\bibitem [{\citenamefont {Kumar}\ \emph {et~al.}(2016)\citenamefont {Kumar},
  \citenamefont {Kulshreshtha},\ and\ \citenamefont
  {Kulshreshtha}}]{Kumar:2016sxx}%
  \BibitemOpen
  \bibfield  {author} {\bibinfo {author} {\bibfnamefont {Sanjeev}\ \bibnamefont
  {Kumar}}, \bibinfo {author} {\bibfnamefont {Usha}\ \bibnamefont
  {Kulshreshtha}}, \ and\ \bibinfo {author} {\bibfnamefont {Daya~Shankar}\
  \bibnamefont {Kulshreshtha}},\ }\bibfield  {title} {\enquote {\bibinfo
  {title} {{Charged compact boson stars and shells in the presence of a
  cosmological constant}},}\ }\href {\doibase 10.1103/PhysRevD.94.125023}
  {\bibfield  {journal} {\bibinfo  {journal} {Phys. Rev.}\ }\textbf {\bibinfo
  {volume} {D94}},\ \bibinfo {pages} {125023} (\bibinfo {year} {2016})},\
  \Eprint {http://arxiv.org/abs/1709.09449} {arXiv:1709.09449 [hep-th]}
  \BibitemShut {NoStop}%
\bibitem [{\citenamefont {Klimas}\ \emph {et~al.}(2019)\citenamefont {Klimas},
  \citenamefont {Sawado},\ and\ \citenamefont {Yanai}}]{Klimas:2018ywv}%
  \BibitemOpen
  \bibfield  {author} {\bibinfo {author} {\bibfnamefont {Pawe\l{}}\
  \bibnamefont {Klimas}}, \bibinfo {author} {\bibfnamefont {Nobuyuki}\
  \bibnamefont {Sawado}}, \ and\ \bibinfo {author} {\bibfnamefont {Shota}\
  \bibnamefont {Yanai}},\ }\bibfield  {title} {\enquote {\bibinfo {title}
  {{Gravitating compact $Q$-ball and $Q$-shell solutions in the $\mathbb{C}P^N$
  nonlinear sigma model}},}\ }\href {\doibase 10.1103/PhysRevD.99.045015}
  {\bibfield  {journal} {\bibinfo  {journal} {Phys. Rev. D}\ }\textbf {\bibinfo
  {volume} {99}},\ \bibinfo {pages} {045015} (\bibinfo {year} {2019})},\
  \Eprint {http://arxiv.org/abs/1812.08363} {arXiv:1812.08363 [hep-th]}
  \BibitemShut {NoStop}%
\bibitem [{\citenamefont {Yanai}(2019)}]{Yanai:2019wpv}%
  \BibitemOpen
  \bibfield  {author} {\bibinfo {author} {\bibfnamefont {Shota}\ \bibnamefont
  {Yanai}},\ }\bibfield  {title} {\enquote {\bibinfo {title} {{Q-balls, -shells
  of a nonlinear sigma model with finite cosmological constants}},}\ }\href
  {\doibase 10.1088/1742-6596/1194/1/012114} {\bibfield  {journal} {\bibinfo
  {journal} {J. Phys. Conf. Ser.}\ }\textbf {\bibinfo {volume} {1194}},\
  \bibinfo {pages} {012114} (\bibinfo {year} {2019})}\BibitemShut {NoStop}%
\bibitem [{\citenamefont {Sawado}\ and\ \citenamefont
  {Yanai}(2020)}]{Sawado:2020ncc}%
  \BibitemOpen
  \bibfield  {author} {\bibinfo {author} {\bibfnamefont {Nobuyuki}\
  \bibnamefont {Sawado}}\ and\ \bibinfo {author} {\bibfnamefont {Shota}\
  \bibnamefont {Yanai}},\ }\bibfield  {title} {\enquote {\bibinfo {title}
  {{Compact, charged boson stars and shells in the $\mathbb{C}P^N$ gravitating
  nonlinear sigma model}},}\ }\href {\doibase 10.1103/PhysRevD.102.045007}
  {\bibfield  {journal} {\bibinfo  {journal} {Phys. Rev. D}\ }\textbf {\bibinfo
  {volume} {102}},\ \bibinfo {pages} {045007} (\bibinfo {year} {2020})},\
  \Eprint {http://arxiv.org/abs/2006.03424} {arXiv:2006.03424 [hep-th]}
  \BibitemShut {NoStop}%
\bibitem [{\citenamefont {Sawado}\ and\ \citenamefont
  {Yanai}(2021)}]{Sawado:2021rsc}%
  \BibitemOpen
  \bibfield  {author} {\bibinfo {author} {\bibfnamefont {Nobuyuki}\
  \bibnamefont {Sawado}}\ and\ \bibinfo {author} {\bibfnamefont {Shota}\
  \bibnamefont {Yanai}},\ }\bibfield  {title} {\enquote {\bibinfo {title}
  {{Phase analyses for compact, charged boson stars and shells harboring black
  holes in the $\mathbb{C}P^N$ nonlinear sigma model}},}\ }\href {\doibase
  10.1103/PhysRevD.103.125018} {\bibfield  {journal} {\bibinfo  {journal}
  {Phys. Rev. D}\ }\textbf {\bibinfo {volume} {103}},\ \bibinfo {pages}
  {125018} (\bibinfo {year} {2021})},\ \Eprint
  {http://arxiv.org/abs/2103.05877} {arXiv:2103.05877 [hep-th]} \BibitemShut
  {NoStop}%
\bibitem [{\citenamefont {Klimas}\ \emph {et~al.}(2021)\citenamefont {Klimas},
  \citenamefont {Kubaski}, \citenamefont {Sawado},\ and\ \citenamefont
  {Yanai}}]{Klimas:2021eue}%
  \BibitemOpen
  \bibfield  {author} {\bibinfo {author} {\bibfnamefont {P.}~\bibnamefont
  {Klimas}}, \bibinfo {author} {\bibfnamefont {L.~C.}\ \bibnamefont {Kubaski}},
  \bibinfo {author} {\bibfnamefont {N.}~\bibnamefont {Sawado}}, \ and\ \bibinfo
  {author} {\bibfnamefont {S.}~\bibnamefont {Yanai}},\ }\bibfield  {title}
  {\enquote {\bibinfo {title} {{Compact Q-balls and Q-shells in a
  multi-component $\mathbb{C}P^N$ model}},}\ }\href {\doibase
  10.1007/JHEP09(2021)084} {\  (\bibinfo {year} {2021}),\
  10.1007/JHEP09(2021)084},\ \Eprint {http://arxiv.org/abs/2107.09831}
  {arXiv:2107.09831 [hep-th]} \BibitemShut {NoStop}%
\bibitem [{\citenamefont {Brihaye}\ and\ \citenamefont
  {Hartmann}(2009)}]{Brihaye:2008cg}%
  \BibitemOpen
  \bibfield  {author} {\bibinfo {author} {\bibfnamefont {Yves}\ \bibnamefont
  {Brihaye}}\ and\ \bibinfo {author} {\bibfnamefont {Betti}\ \bibnamefont
  {Hartmann}},\ }\bibfield  {title} {\enquote {\bibinfo {title} {{Angularly
  excited and interacting boson stars and Q-balls}},}\ }\href {\doibase
  10.1103/PhysRevD.79.064013} {\bibfield  {journal} {\bibinfo  {journal} {Phys.
  Rev.}\ }\textbf {\bibinfo {volume} {D79}},\ \bibinfo {pages} {064013}
  (\bibinfo {year} {2009})},\ \Eprint {http://arxiv.org/abs/0812.3968}
  {arXiv:0812.3968 [hep-ph]} \BibitemShut {NoStop}%
\bibitem [{\citenamefont {Bernal}\ \emph {et~al.}(2010)\citenamefont {Bernal},
  \citenamefont {Barranco}, \citenamefont {Alic},\ and\ \citenamefont
  {Palenzuela}}]{Bernal:2009zy}%
  \BibitemOpen
  \bibfield  {author} {\bibinfo {author} {\bibfnamefont {Argelia}\ \bibnamefont
  {Bernal}}, \bibinfo {author} {\bibfnamefont {Juan}\ \bibnamefont {Barranco}},
  \bibinfo {author} {\bibfnamefont {Daniela}\ \bibnamefont {Alic}}, \ and\
  \bibinfo {author} {\bibfnamefont {Carlos}\ \bibnamefont {Palenzuela}},\
  }\bibfield  {title} {\enquote {\bibinfo {title} {{Multi-state Boson
  Stars}},}\ }\href {\doibase 10.1103/PhysRevD.81.044031} {\bibfield  {journal}
  {\bibinfo  {journal} {Phys. Rev.}\ }\textbf {\bibinfo {volume} {D81}},\
  \bibinfo {pages} {044031} (\bibinfo {year} {2010})},\ \Eprint
  {http://arxiv.org/abs/0908.2435} {arXiv:0908.2435 [gr-qc]} \BibitemShut
  {NoStop}%
\bibitem [{\citenamefont {Collodel}\ \emph {et~al.}(2017)\citenamefont
  {Collodel}, \citenamefont {Kleihaus},\ and\ \citenamefont
  {Kunz}}]{Collodel:2017biu}%
  \BibitemOpen
  \bibfield  {author} {\bibinfo {author} {\bibfnamefont {Lucas~G.}\
  \bibnamefont {Collodel}}, \bibinfo {author} {\bibfnamefont {Burkhard}\
  \bibnamefont {Kleihaus}}, \ and\ \bibinfo {author} {\bibfnamefont {Jutta}\
  \bibnamefont {Kunz}},\ }\bibfield  {title} {\enquote {\bibinfo {title}
  {{Excited Boson Stars}},}\ }\href {\doibase 10.1103/PhysRevD.96.084066}
  {\bibfield  {journal} {\bibinfo  {journal} {Phys. Rev.}\ }\textbf {\bibinfo
  {volume} {D96}},\ \bibinfo {pages} {084066} (\bibinfo {year} {2017})},\
  \Eprint {http://arxiv.org/abs/1708.02057} {arXiv:1708.02057 [gr-qc]}
  \BibitemShut {NoStop}%
\bibitem [{\citenamefont {Alcubierre}\ \emph {et~al.}(2018)\citenamefont
  {Alcubierre}, \citenamefont {Barranco}, \citenamefont {Bernal}, \citenamefont
  {Degollado}, \citenamefont {Diez-Tejedor}, \citenamefont {Megevand},
  \citenamefont {Nunez},\ and\ \citenamefont {Sarbach}}]{Alcubierre:2018ahf}%
  \BibitemOpen
  \bibfield  {author} {\bibinfo {author} {\bibfnamefont {Miguel}\ \bibnamefont
  {Alcubierre}}, \bibinfo {author} {\bibfnamefont {Juan}\ \bibnamefont
  {Barranco}}, \bibinfo {author} {\bibfnamefont {Argelia}\ \bibnamefont
  {Bernal}}, \bibinfo {author} {\bibfnamefont {Juan~Carlos}\ \bibnamefont
  {Degollado}}, \bibinfo {author} {\bibfnamefont {Alberto}\ \bibnamefont
  {Diez-Tejedor}}, \bibinfo {author} {\bibfnamefont {Miguel}\ \bibnamefont
  {Megevand}}, \bibinfo {author} {\bibfnamefont {Dario}\ \bibnamefont {Nunez}},
  \ and\ \bibinfo {author} {\bibfnamefont {Olivier}\ \bibnamefont {Sarbach}},\
  }\bibfield  {title} {\enquote {\bibinfo {title} {{$\ell$-Boson stars}},}\
  }\href {\doibase 10.1088/1361-6382/aadcb6} {\bibfield  {journal} {\bibinfo
  {journal} {Class. Quant. Grav.}\ }\textbf {\bibinfo {volume} {35}},\ \bibinfo
  {pages} {19LT01} (\bibinfo {year} {2018})},\ \Eprint
  {http://arxiv.org/abs/1805.11488} {arXiv:1805.11488 [gr-qc]} \BibitemShut
  {NoStop}%
\bibitem [{\citenamefont {Alcubierre}\ \emph {et~al.}(2019)\citenamefont
  {Alcubierre}, \citenamefont {Barranco}, \citenamefont {Bernal}, \citenamefont
  {Degollado}, \citenamefont {Diez-Tejedor}, \citenamefont {Megevand},
  \citenamefont {Nunez},\ and\ \citenamefont {Sarbach}}]{Alcubierre:2019qnh}%
  \BibitemOpen
  \bibfield  {author} {\bibinfo {author} {\bibfnamefont {Miguel}\ \bibnamefont
  {Alcubierre}}, \bibinfo {author} {\bibfnamefont {Juan}\ \bibnamefont
  {Barranco}}, \bibinfo {author} {\bibfnamefont {Argelia}\ \bibnamefont
  {Bernal}}, \bibinfo {author} {\bibfnamefont {Juan~Carlos}\ \bibnamefont
  {Degollado}}, \bibinfo {author} {\bibfnamefont {Alberto}\ \bibnamefont
  {Diez-Tejedor}}, \bibinfo {author} {\bibfnamefont {Miguel}\ \bibnamefont
  {Megevand}}, \bibinfo {author} {\bibfnamefont {Dario}\ \bibnamefont {Nunez}},
  \ and\ \bibinfo {author} {\bibfnamefont {Olivier}\ \bibnamefont {Sarbach}},\
  }\bibfield  {title} {\enquote {\bibinfo {title} {{Dynamical evolutions of
  $\ell$-boson stars in spherical symmetry}},}\ }\href {\doibase
  10.1088/1361-6382/ab4726} {\bibfield  {journal} {\bibinfo  {journal} {Class.
  Quant. Grav.}\ }\textbf {\bibinfo {volume} {36}},\ \bibinfo {pages} {215013}
  (\bibinfo {year} {2019})},\ \Eprint {http://arxiv.org/abs/1906.08959}
  {arXiv:1906.08959 [gr-qc]} \BibitemShut {NoStop}%
\bibitem [{\citenamefont {Loginov}\ and\ \citenamefont
  {Gauzshtein}(2020{\natexlab{a}})}]{Loginov:2020xoj}%
  \BibitemOpen
  \bibfield  {author} {\bibinfo {author} {\bibfnamefont {A.~Yu.}\ \bibnamefont
  {Loginov}}\ and\ \bibinfo {author} {\bibfnamefont {V.~V.}\ \bibnamefont
  {Gauzshtein}},\ }\bibfield  {title} {\enquote {\bibinfo {title} {{Radially
  excited $U\left(1\right)$ gauged $Q$-balls}},}\ }\href {\doibase
  10.1103/PhysRevD.102.025010} {\bibfield  {journal} {\bibinfo  {journal}
  {Phys. Rev. D}\ }\textbf {\bibinfo {volume} {102}},\ \bibinfo {pages}
  {025010} (\bibinfo {year} {2020}{\natexlab{a}})},\ \Eprint
  {http://arxiv.org/abs/2004.03446} {arXiv:2004.03446 [hep-th]} \BibitemShut
  {NoStop}%
\bibitem [{\citenamefont {Loginov}\ and\ \citenamefont
  {Gauzshtein}(2020{\natexlab{b}})}]{Loginov:2020lwg}%
  \BibitemOpen
  \bibfield  {author} {\bibinfo {author} {\bibfnamefont {A.~Yu}\ \bibnamefont
  {Loginov}}\ and\ \bibinfo {author} {\bibfnamefont {V.~V.}\ \bibnamefont
  {Gauzshtein}},\ }\bibfield  {title} {\enquote {\bibinfo {title} {{Radially
  and azimuthally excited states of a soliton system of vortex and Q-ball}},}\
  }\href {\doibase 10.1140/epjc/s10052-020-08715-z} {\bibfield  {journal}
  {\bibinfo  {journal} {Eur. Phys. J. C}\ }\textbf {\bibinfo {volume} {80}},\
  \bibinfo {pages} {1123} (\bibinfo {year} {2020}{\natexlab{b}})},\ \Eprint
  {http://arxiv.org/abs/2009.12818} {arXiv:2009.12818 [hep-th]} \BibitemShut
  {NoStop}%
\bibitem [{\citenamefont {Almumin}\ \emph {et~al.}(2021)\citenamefont
  {Almumin}, \citenamefont {Heeck}, \citenamefont {Rajaraman},\ and\
  \citenamefont {Verhaaren}}]{Almumin:2021gax}%
  \BibitemOpen
  \bibfield  {author} {\bibinfo {author} {\bibfnamefont {Yahya}\ \bibnamefont
  {Almumin}}, \bibinfo {author} {\bibfnamefont {Julian}\ \bibnamefont {Heeck}},
  \bibinfo {author} {\bibfnamefont {Arvind}\ \bibnamefont {Rajaraman}}, \ and\
  \bibinfo {author} {\bibfnamefont {Christopher~B.}\ \bibnamefont
  {Verhaaren}},\ }\bibfield  {title} {\enquote {\bibinfo {title} {{Excited
  Q-Balls}},}\ }\href@noop {} {\  (\bibinfo {year} {2021})},\ \Eprint
  {http://arxiv.org/abs/2112.00657} {arXiv:2112.00657 [hep-th]} \BibitemShut
  {NoStop}%
\end{thebibliography}%

\end{document}